\pdfoutput=1
\documentclass[cernpreprint, atlasdraft=false, UKenglish, texmf, orcidlogo]{atlasdoc}
 
\usepackage[subcaption]{atlaspackage}
\usepackage{atlasbiblatex}
 
\usepackage{atlasphysics}

\graphicspath{{logos/}{figures/}}
 
\usepackage{ANA-DAPR-2021-02-PAPER-defs}
 
\usepackage{placeins}
\usepackage{gensymb}

\AtlasTitle{Beam-induced backgrounds measured in the ATLAS detector during local gas injection into the LHC beam vacuum}

\AtlasAbstract{
Inelastic beam-gas collisions at the Large Hadron Collider (LHC), within a few hundred metres of the
ATLAS experiment, are known to give the dominant contribution to beam backgrounds.
These are monitored by ATLAS with a dedicated Beam Conditions Monitor (BCM) and with the
rate of fake jets in the calorimeters. These two methods are complementary since the BCM
probes backgrounds just around the beam pipe while fake jets are observed at radii of up to
several metres.
In order to quantify the correlation between the residual gas density in the LHC
beam vacuum  and the experimental backgrounds recorded by ATLAS, several dedicated tests
were performed during LHC Run 2.
Local pressure bumps, with a gas density several orders of magnitude higher than
during normal operation, were introduced at different locations. The changes
of beam-related backgrounds, seen in ATLAS, are correlated with the local pressure
variation. In addition the rates of beam-gas events are estimated from
the pressure measurements and pressure bump profiles obtained from calculations. Using these rates,
the efficiency of the ATLAS beam background monitors to detect beam-gas events is derived
as a function of distance from the interaction point.
These efficiencies and characteristic distributions of fake jets from the beam backgrounds
are found to be in good agreement with results of beam-gas simulations performed with the
\textsc{Fluka} Monte Carlo programme.
}

\AtlasRefCode{DAPR-2021-02}
 
\PreprintIdNumber{CERN-EP-2024-113}

\AtlasJournalRef{JINST 19 (2024) P06014}
\AtlasDOI{10.1088/1748-0221/19/06/P06014}

\hypersetup{pdftitle={ATLAS document},pdfauthor={The ATLAS Collaboration}}
 
\begin{document}
 
\maketitle
 
\tableofcontents

\section{Introduction}
 
The Large Hadron Collider (LHC)\,\cite{Evans:2008zzb} is an accelerator of 27\,km circumference where proton bunches of a
clockwise \beam{1} and a counter-clockwise \beam{2} are colliding at four interaction regions (IR).
During LHC \run{2}, between 2015 and 2018, the proton energy was 6.5\,\TeV. Interesting physics events have to be extracted
from $\sim 10^9$\,Hz of minimum-bias collisions overlaid with some beam induced background (BIB). The intensity of BIB is small
compared to the minimum-bias background but some BIB events can mimic high transverse momentum (\pt) objects that are characteristic for
exotic new physics\,\cite{EXOT-2019-05,EXOT-2019-23,EXOT-2019-24}.
 
At the LHC BIB arises from proton collisions with residual gas in the beam pipe (beam-gas interactions)
or from beam halo protons\footnote{Beam halo is composed by a small fraction of particles surrounding the dense beam
core\,\protect\cite{PhysRevSTAB.17.081004}.} impacting on limiting apertures, e.g. collimators\,\cite{nima381:531,nikolai1,BRUCE2013825}.
The rate of beam-gas events (\bgr) depends on the beam intensity and the density and molecular
composition of the residual gas in the beam pipe. Beam halo losses have a more complicated dependence on the setting of the
various collimators and on the beam quality.
 
The two principal BIB observables monitored in the  ATLAS experiment\,\cite{PERF-2007-01} are  hits in the dedicated
Beam Conditions Monitor (BCM)\,\cite{bcmdetref} and fake jets. Energetic particles emitted at very small angles from the
collision of a high-energy proton with a stationary nucleus can initiate particle showers in the accelerator elements.
These showers are mostly confined to small radii just around the beam pipe where they can hit the BCM detector modules and
produce a trigger signature, indicative of BIB.
Fake jets arise predominantly from high-energy muons produced in kaon and pion decays. When traversing
the ATLAS calorimeters they have a small probability to deposit a substantial fraction of their energy through radiative
processes. If the energy deposit of such an event is large enough it can be reconstructed as a (fake) jet.
 
Simulations\,\cite{DAPR-2017-01}, which agree well with measurements of total BIB rates, indicate that most of the background
seen by the BCM originates from beam-gas collisions at less than 60\,m from the interaction point (IP). Fake jets, on the other hand,
are due to beam losses far upstream -- beyond $\sim$50\,m from the IP.
So far there has been no direct experimental verification of these simulation predictions, which was the main motivation
for the study described in this paper.
 
BIB monitoring and detailed analysis of BIB events collected by the ATLAS detector during LHC \run{1}\,\cite{DAPR-2012-01, DAPR-2014-01}
and \run{2}, in particular during dedicated loss-map tests\,\cite{PhysRevSTAB.17.081004}, have shown that beam-halo losses are
a minor contribution\,\cite{PhysRevAccelBeams.22.021004} and most of the BIB observed by ATLAS originates from inelastic beam-gas collisions.
 
In order to quantify the sensitivity of ATLAS to beam-gas events at various well-defined locations, a series of
dedicated pressure bump tests was performed during LHC \run{2}. In these tests individual non-evaporable
getter (NEG) cartridges\,\cite{neg-overview} were heated such that some of the gas they had absorbed was re-injected into the LHC beam
vacuum. By this method pressure bumps, with a typical extent of $\sim$1\,m FWHM along the beam direction, were created at four different
$|z$|-locations\footnote{In the
right-handed ATLAS coordinate system the origin is at the nominal
IP and the azimuthal angle, $\phi$, is measured relative to the $x$-axis, which points towards
the centre of the LHC ring. The polar angle, $\theta$, is defined with respect to the $z$-axis,
which points along the \beam{2} direction. The pseudorapidity is given by $\eta=-\ln\tan(\theta/2)$.}
between 19\,m and 151\,m. In these bumps the pressure was up to four orders of magnitude higher than during normal beam operation.
Given that the BIB observed in ATLAS, during physics data taking, is dominated by beam-gas collisions distributed over a typical length of
the order of 100\,m, these local pressure bumps resulted in a detectable increase of the BIB. Since the efficiency of ATLAS to
record a beam-gas event depends on the location of that event, the additional BIB rates from some pressure bumps exceeded the normal level by
orders of magnitude, while for some others an excess was barely resolvable.
 
These pressure bump tests provide detailed data for benchmarking the BIB simulations, and the results also are useful
input for designing LHC beam vacuum upgrades such that the beam-gas background in the experiments
is minimised.
 
\subsection{Short overview of the test procedure}
 
\begin{figure}
\centering
\includegraphics[width=0.95\textwidth]{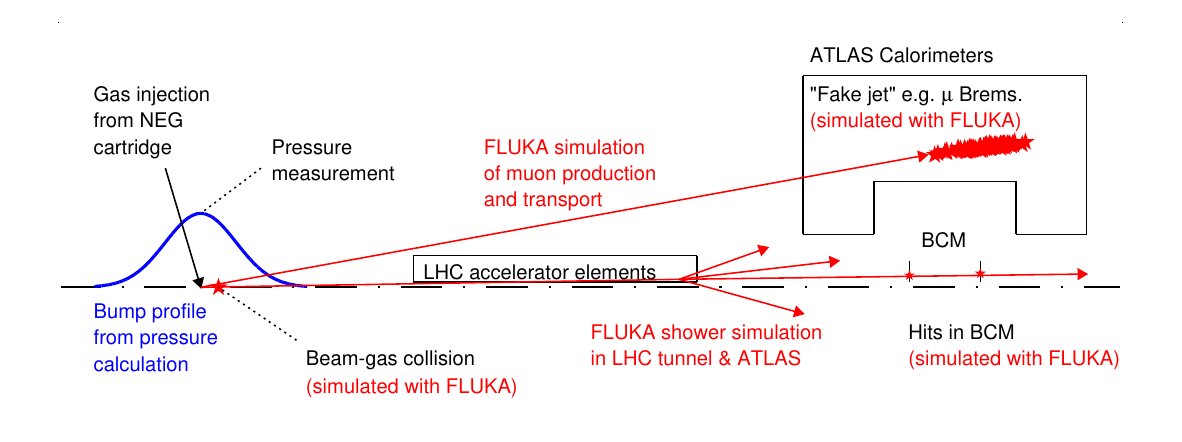}
\caption{Schematic illustration of the pressure bump test. The red arrows and stars indicate particle collisions and
trajectories.
The blue text and profile indicate the estimates of the pressure bump profiles, which are discussed in Section\,\protect\ref{sect:profiles}.
The red text refers to quantities and processes simulated with \fluka, as described in Section\,\protect\ref{sect:fluka}.
Black text is used for detector and accelerator components, real events, actions and measured observables.}
\label{fig:analysis}
\end{figure}
 
Figure\,\ref{fig:analysis} shows a schematic summary of the measurement, analysis and simulation procedures. The creation of the pressure bump
and the associated pressure measurement are described in Sections\,\ref{sect:injection} and \ref{sect:press-meas}, respectively.
 
Two definitions of the relation between the BIB seen by ATLAS and the pressure increase at the bump location can be considered.
\begin{enumerate}
\item The ratio, $\omega(z_i)$, between the additional background rate $\Delta B$, observed by a given ATLAS detector, and the additional pressure,
$\Delta P(z_i)$, measured by a vacuum gauge at the bump location $z_i$, i.e.
\begin{equation}
\omega(z_i) = \frac{\Delta B}{\Delta P(z_i)}.
\label{eq:omegadef}
\end{equation}
\item The ratio, $\varepsilon(z)$, between the additional background rate observed by a given ATLAS detector and the rate of additional
beam gas events, $\Delta$\bgr, due to the introduced pressure bump. Here $\Delta$\bgr is the change of the integral over the gas density in
the bump area, i.e.
\begin{equation}
\varepsilon(z) = \frac{\Delta B}{\Delta R_{\mathrm{BG}}} = \frac{\Delta B}{f\frac{\Delta P(z_i)}{P_{\mathrm{calc}}(z_i)} \int \rho_{\mathrm{calc}}(z)dz},
\label{eq:epsdef}
\end{equation}
where $\rho_{\mathrm{calc}}(z)$ denotes the pressure profile that is obtained from calculations described in Section\,\ref{sect:profiles}
and $P_{\mathrm{calc}}(z_i)$ is the corresponding calculated pressure at the vacuum gauge location.
The ratio of the measured $\Delta P(z_i)$ and the calculated value is used to renormalise the profile such that it matches the measured
pressure at the gauge location. The factor $f$ is related to the interaction cross section of the protons and the beam intensity,
as described in Section\,\ref{sect:injection}.
\end{enumerate}
 
The determination of $\omega(z_i)$ is described in Section\,\ref{sect:BkgdVsP}.
Since $\omega(z_i)$ is entirely based on measured data, it is not subject to any of the assumptions and calculations that have to be made in order
to derive $\varepsilon(z)$. Thus it is a useful quantity to  verify the consistency of the results for the same bump location in different LHC fills.
Also the $|z|$-dependence of the ratio of BCM and fake-jet BIB rates can be assessed from the measured data by using the respective $\omega(z_i)$-values.
The downside of the $\omega(z_i)$-parameter is that it does not accurately describe the sensitivity of ATLAS to beam-gas background because a single pressure measurement does
not provide enough information to determine the number of beam-gas collisions.
Furthermore, the relation between $P(z_i)$ and the integrated density profile along the beam
trajectory, $\int\!\rho(z)dz$, depends on the local vacuum equipment configuration, e.g. the distance between the point of gas injection and the
vacuum gauge. Since these configurations are different at all bump locations, $\omega(z_i)$ provides only an approximate measure of the
relative sensitivity between the four pressure bump locations.
 
The $\varepsilon(z)$ represents the real efficiency, i.e. the ratio between observed and produced events,
which does not depend on the local vacuum equipment configuration or the beam conditions. Its determination needs, however, information about
the gas density, $\rho(z)$, traversed by the beam in order to determine the rate of beam-gas interactions. Because the density profile
cannot be measured by the LHC instrumentation, $\varepsilon(z)$ cannot be derived from data alone and is subject
to the uncertainties of the dedicated pressure calculations, described in Section\,\ref{sect:profiles}.
 
The $\omega(z_i)$ and $\varepsilon(z)$ are derived in Sections\,\ref{sect:BkgdVsP} and \ref{sect:abseff}, respectively and their uncertainties
are discussed in Section\,\ref{sect:syserrors}.
BIB simulations with the \fluka code\,\cite{flukaref1,flukaref2} are described in Section\,\ref{sect:fluka}.
These simulations compute the efficiency per event and can therefore be directly compared with $\varepsilon(z)$, which is done
in Section\,\ref{sect:bibdeteff}. The azimuthal and pseudorapidity distributions of the simulated fake jets
are compared with the measurements for different pressure bump locations in Section\,\ref{sect:jet-dist}.
 
Section\,\ref{sect:aftertest} gives a brief report of an unexpected increase of BIB levels after the last pressure bump test
and Section\,\ref{sect:concl} provides the conclusions.
 
\section{Beam background measurements in ATLAS}
 
\subsection{The ATLAS experiment}
 
ATLAS\,\cite{PERF-2007-01} is one of the two general purpose experiments, designed to fully exploit the proton--proton
physics at the highest luminosities achievable at the LHC.
 
The inner detector (ID) provides the tracking and momentum measurement of charged particles up to $|\eta|=2.5$.
From the smallest to the largest radii it comprises a pixel detector, a
silicon micro-strip tracker and a transition radiation detector. All of these are subdivided into
barrel and endcap parts, embedded in a solenoid, which creates a magnetic field of 2\,T along the
$z$-axis.
 
The BCM is housed inside the ID and consists of four small diamond sensors on both sides of the IP at $|z|=184$\,cm. Following the
ATLAS convention of labelling the sides of the experiment as A and C, the BCM-A and the BCM-C modules are on the incoming
\beam{1} and \beam{2} sides, respectively.
The sensors, with an active area of $8\times 8$\,mm$^2$, are arranged as a cross at an average radius of 55\,mm from the
centre of the beam line. In order to increase the track length within the sensitive volume, and thereby the
signal of a  BIB-particle, the modules are tilted by 45\,degrees with respect to the $z$-axis\,\cite{bcmdetref}.
 
The ID is surrounded by an electromagnetic calorimeter using lead as the absorber and liquid argon (LAr)
as the active medium. The barrel part has a half-length of $\sim$3\,m and extends radially from $r=1.5$\,m to
2.0\,m. At central rapidities the hadron energy is measured by a steel/scintillator tile
calorimeter, extending from $r=2.3$ to 4.3\,m and having a half-length of $\sim$6\,m.
In the endcaps both calorimeters are based on LAr technology and provide coverage up to $|\eta|=3.2$.
These calorimeters are complemented by a forward calorimeter (FCAL), which reaches up to $|\eta|=4.9$.
All calorimeters provide nanosecond timing resolution.
 
The calorimeters are surrounded by a muon spectrometer based on three large air-core superconducting
toroidal magnets with eight coils each: one barrel toroid and two endcap toroids positioned inside the
barrel at both ends of the central solenoid.
 
The very forward region, behind the FCAL, is filled with massive shielding in order to reduce the
radiation levels in the muon spectrometer. This shielding also serves as a plug of the tunnel entry
such that the BIB  -- except the high-energy muon component -- entering the ATLAS cavern is efficiently suppressed.
 
Potentially interesting events are selected with a two-stage trigger\,\cite{TRIG-2016-01}, where the first stage (\Lvl{1}) is hardware-based
and fires on simple physics objects.
The \Lvl{1} selection  is followed by a software high-level trigger, which further reduces the rate before the selected
events are stored for offline analysis.
An extensive software suite\,\cite{ATL-SOFT-PUB-2021-001} is used in the reconstruction of data, in detector operations and in the
trigger and data acquisition systems of the experiment.
 
\subsection{ATLAS beam background triggers}
 
The measurement of BIB utilises bunches that do not collide with bunches of the opposite beam at the ATLAS IP. In the \Lvl{1} trigger
these unpaired bunches form their own bunch group. This grouping isolates the background trigger items from
those used for physics and allows trigger prescales to be applied independently. This is particularly important for the low-\pt
single-jet trigger, which is used for fake-jet monitoring. In colliding bunches low-\pt jets appear at overwhelming rates and the
only possibility to collect a sizeable sample of fake-jet events is to trigger independently on the unpaired bunches without the
need reduce the rate by a prescale.
 
ATLAS data-taking is subdivided into luminosity blocks (LBs) with a typical duration of $\sim$60\,s. This time is
short enough to consider variations of beam conditions to be negligible.\footnote{This assumption is not true
for fast losses due to sudden beam instabilities, but these are not a subject of the present study.} Although individual events have a
time-stamp with nanosecond accuracy the LBs provide convenient time-bins for trend monitoring. This also ensures that
\Lvl{1} trigger rates (which are always averaged over a LB), offline data and information received from the LHC
can be analysed within a common time binning.
 
\begin{figure}[t]
\centering
\includegraphics[width=0.95\textwidth]{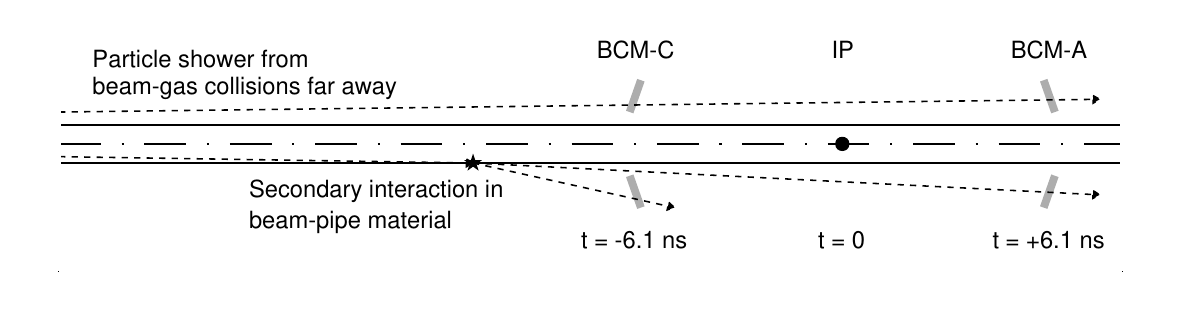}
\caption{Illustration of the BCM background trigger signature for \beam{2}. The vertical dimensions
are not to scale. In particular, the size of the BCM sensors is exaggerated and their tilt is indicated only schematically.
The dot-dashed line is the beam line and the two solid lines illustrate the wall of the beam pipe.
The dashed lines represent the trajectories of particles produced in proton collisions with residual gas or accelerator elements.
The trigger can be fired either by a single particle traversing a module on both sides of the IP or by two different particles
as illustrated on the upper and lower parts of the figure, respectively.
The only requirement is that there are early and in-time hits in any of the four modules on opposite sides of the IP.
}
\label{fig:bcmsignal}
\end{figure}

The signature of BIB in the BCM, shown schematically in Figure\,\ref{fig:bcmsignal}, is formed by one
early hit in any of the four sensors on one side and an in-time hit in any sensor on the other side of the IP.
An early hit is defined to be in a 5.46\,ns wide time-window,
centred $\sim$6\,ns before the nominal collision time. An in-time hit has the same window width, but centred
$\sim$6\,ns after the nominal collision time, i.e. aligned in time with the collision products from the IP. Due to its
location the BCM is particularly sensitive to BIB showers developing at small radii, just outside the beam pipe.
 
Fake jets are almost exclusively due to radiative energy losses of high-energy muons traversing the detector, in particular
bremsstrahlung, resulting in electromagnetic showers with a relatively small extent and negligible hadronic activity.
They are selected in the range $|\eta|<3.2$ by a single jet trigger\,\cite{TRIG-2012-01}, which fires on a transverse energy deposition above \mbox{\et\,$>$\,12\,\GeV},
calibrated at the electromagnetic scale. Since, however, the fake jet is not associated to any object with \pt defined at the IP,
the \et should be considered an apparent \et, which is assigned by the reconstruction algorithm on the basis of the position and magnitude
of the energy deposition in the calorimeter.

The jet trigger is fully efficient up to $|\eta|=2.8$ and has reduced sensitivity up to $|\eta|=3.2$.
At various stages of the trigger chain the recorded events are subject to dead time, which is evaluated for
each LB. A dead-time correction is applied by scaling the recorded event rate correspondingly.
The trigger efficiency starts to rise gradually above the \ET-threshold of 12\,\GeV, and full efficiency is
reached only around 20\,\GeV.
For most of the results in this paper the trigger efficiency can be considered an intrinsic property of the jet-observable that is
used to detect a BIB event by the fake-jet monitoring. Thus, whatever the efficiency to detect the fake jet is, it is the same in the
pressure bump test and during normal physics operation.
In order to maximise the number of fake-jet events for the analysis, all jets that have a reconstructed transverse energy of 12\,\GeV
and for which the corresponding \Lvl{1} jet trigger has fired are retained.
However, the reduced efficiency below 20\,\GeV becomes significant when comparing measured and simulated fake-jet rates. For these
comparisons only jets at the efficiency plateau, i.e. with \ET$>$20\,\GeV, are considered.
 
\subsection{Fake-jet reconstruction and cleaning}
 
Even in unpaired bunches the jet sample has a non-negligible contribution from collisions of protons in the
unpaired bunch with ghost charge in the other beam, i.e. recaptured protons that have left their original radio-frequency bucket\,\cite{DAPR-2014-01}.
These ghost collisions take place at the IP but the luminosity of these encounters is typically $\sim$3 orders of magnitude below
that of two colliding nominal bunches. This contribution can
be removed with high efficiency by applying a vertex veto, i.e. requiring that the event where the jet is observed has no
reconstructed primary vertex\,\cite{ATL-PHYS-PUB-2015-026}.
 
The jets are reconstructed with the anti-$k_t$ jet algorithm\,\cite{Cacciari:2008gp,Fastjet} from clusters at the electromagnetic
scale\,\cite{PERF-2014-07}, using a radius parameter $R=0.4$. No further calibrations are applied since these would not be appropriate for
the fake jets that do not originate from the IP, are not subject to event pile-up, and have a negligible hadronic contribution.
Standard jet quality criteria\,\cite{LARG-2013-01}, are applied in order to remove instrumental backgrounds,
e.g. effects of calorimeter noise.
 
In the barrel calorimeters, the axis of a BIB-induced shower is parallel to the beam direction, instead of pointing to the IP.
Previous studies\,\cite{DAPR-2017-01} have shown that \fluka simulations, implementing only a rudimentary clustering of
the total deposited energy, reproduce the fake jet rates reconstructed from the data. This indicates that, even though the energy
deposition is elongated in the wrong direction with respect to the IP, the cluster in the data is not split by the reconstruction.

\subsection{BCM efficiency and purity}
 
The threshold for the BCM to record a hit corresponds to an energy deposition of around 250\,\keV, above which the
efficiency is assumed to be close to 100\%. After a hit, however, an individual module has a dead time with a typical
duration of the order of 10\,ns, depending on the deposited energy.
Thus a hit before the trigger window
can mask another that would have contributed to the trigger signature. This potential efficiency drop has to be
taken into account when comparing data with simulations.
 
Since the BCM background trigger involves an early hit, its sensitivity to ghost collisions is strongly suppressed.
Less than 4\% of the triggered events would be rejected by a vertex veto. Thus the number of events collected was
maximised and issues of dead-time corrections were mitigated by basing the analysis on the raw \Lvl{1} trigger rates,
recorded from the unpaired bunches.

\section{The pressure bump tests}
 
\subsection{The IR1 region around ATLAS}
 
Earlier BIB simulation studies, e.g. those presented in Refs.\,\cite{BRUCE2013825,nima381:531, DAPR-2017-01}, predict that BIB can be observed from
beam-gas interactions in the entire long straight section (LSS) and part of the LHC arc, up to a distance of about 500\,m from the IP.
 
\begin{figure}[t]
\centering
\includegraphics[width=0.95\textwidth]{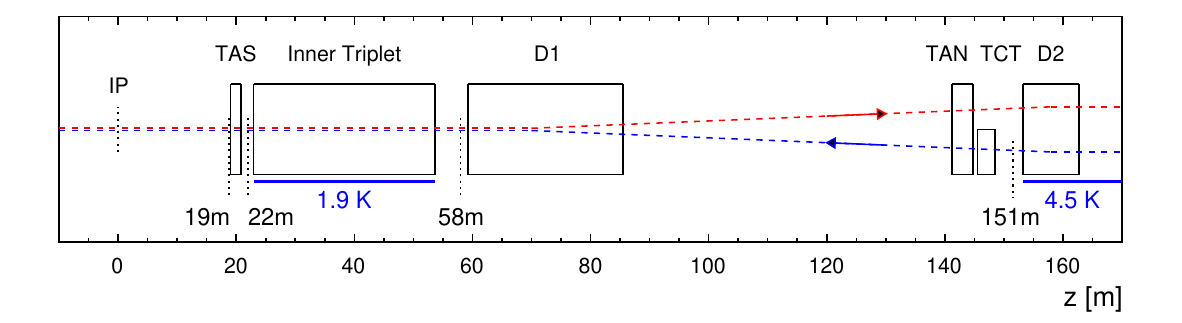}
\caption{Schematic illustration of the LSS up to the D2 dipole, i.e. the part relevant for this study. The pressure bump
locations at 19\,m, 22\,m, 58\,m and 151\,m are indicated, as well as the temperatures of the cryogenic elements in the area. The vertical dimensions
are not to scale.}
\label{fig:lss1-schema}
\end{figure}
 
The main accelerator components in the part of the LSS that is relevant for the pressure bump tests are illustrated in Figure\,\ref{fig:lss1-schema}.
Between the IP and $z=18.5$\,m, the radius of the beam pipe varies between 28\,mm and 120\,mm.
The cryogenic inner triplet of quadrupole magnets extends from $|z|=23$\,m up to $|z|=55$\,m and inside of it a perforated beam screen (BS), kept
at a temperature of $\sim$20\,K, reduces the synchrotron radiation load on the beam-pipe wall. Residual gas in this
cryogenic section can pass through the perforations and condense on the cold bore, kept at 1.9\,K.
The triplet is protected by the secondaries absorber (TAS) from the collision
debris created at the IP. The TAS devices, one on either side of the experiment, starting at $|z|=19$\,m, are 1.8\,m long copper blocks
with an aperture of 17\,mm radius for the passing beams. Vacuum stations with ion pumps and NEG cartridges, placed on either
side of the TAS, provide efficient local pumping.
 
From $|z|=55$\,m to $|z|=152.5$\,m the beam pipe is at room temperature. One of the few warm magnets of
the LHC, the D1 separation dipole,
is located in this region. Extending from 59\,m to 83\,m, its role is to separate the beams from the common pipe at the IP into the
two separate pipes in the LHC arc. Other beam-line elements that are potentially significant for this study are the set of
tertiary collimators (TCT) and the neutral absorber (TAN).
The TCTs are located in the region $|z|=145$\,m$-148$\,m, and with apertures of 18.4\,mm in the horizontal
and 11.6\,mm in the vertical plane\footnote{Corresponding to 2016 collimator settings.} they are likely to
intercept some of the secondaries produced in the most distant pressure bump at 151\,m. The role of the TAN is to
intercept neutral particles produced at the IP and that are not deflected by the D1 dipole, before they would impinge on the D2 dipole, which
is operated at 4.5\,K.

The room-temperature sections of the LHC beam pipe are coated with a TiZrV NEG layer\,\cite{neg-benvenuti}, which provides distributed
pumping. Especially in the region |z|<18\,m this NEG coating and a pair of low-capacity annular ion pumps, located at
$z\approx \pm 3.7$\,m, ensure a good vacuum around the IP.
 
In the cold arc of the LHC, starting at $|z|\approx 270$\,m, a good vacuum is ensured by the cryogenic surfaces of the
magnets, which constitute very efficient pumps. No gauges or other active vacuum equipment exists in the cold regions,
so the pressure bump study is limited to four points in the LSS, $|z|=19$\,m, 22\,m, 58\,m and 151\,m, which are indicated in
Figure\,\ref{fig:lss1-schema}.

\subsection{Beam conditions}
 
During \run{2} of the LHC the colliding protons were grouped in bunches with a typical
population of $n_b=10^{11}$. A bunch train was composed of a varying number of bunches, spaced by 25\,ns. Larger
gaps, of up to 800\,ns, were left between the trains to account for the rise-times of injection kickers. One wide
gap of $3.1\,{\mathrm \mu}$s is needed in the bunch pattern in order to allow for a clean beam abort.
 
Typically twelve unpaired bunches per beam are included in an LHC fill pattern. These are usually grouped in short trains
immediately after the abort gap in order to maximise their separation from the preceding colliding pairs.
In this position the afterglow, i.e. long-lived debris from the collisions\,\cite{DAPR-2014-01}, is minimised.
 
Five pressure bump tests were performed in LHC fills 4905, 5005, 5331 (all in 2016), 6239 (in 2017) and 7212 (in 2018).
The principal beam parameters of these five fills are summarised in Table\,\ref{tab:beamcond}.
In order to reduce the risk of a beam dump due to the increased local loss rate, all tests were done during the
intensity ramp-up following regular technical stops of the accelerator.
Thus all five fills were characterised by a smaller than nominal number of bunches and, therefore, lower total beam intensity
than regular physics fills.

\begin{table}
\centering
\caption{Beam conditions of the five fills during which pressure bump tests were performed.
$\beta^*$ is the value of the $\beta$-function at the IP and is related to the transverse size of the luminous region\,\protect\cite{weidermann}.
The $\beta^*$ and crossing angle values refer to the ATLAS IP.}
\label{tab:beamcond}
\begin{tabular}{l|c|c|c|c|c}
& \multicolumn{5}{c}{LHC fill} \\
&  4905 & 5005  & 5331 & 6239 & 7212 \\ \hline
Number of bunches per beam             &  241  & 602   & 169  & 1358 &  603 \\
Number of unpaired bunches             &  12   & 12    & 12   &  12  &   13 \\
Maximum bunch train length             &  72   & 72    & 48   &   8  &   48 \\
$\beta^*$ [cm]                         &  40   & 40    & 40   &  30  &   30 \\
Crossing half-angle [$\mathrm{\mu}$rad]&  185  & 185   & 140  & 150  &  155 \\
Single beam intensity [$10^{13}$\,protons]  &  3.2  & 7.1   & 1.8  &  12  &    6 \\ \hline
\end{tabular}
\end{table}
 
The beams had typical per-bunch luminosities of $\sim\mexpfor{5}{30}$\,cm$^{-2}$s$^{-1}$ for physics production, except for
fills 5331 and 7212, during which the beams were separated about half way through the fill in order to provide low collision rates
for dedicated forward physics data-taking. The beam-gas backgrounds are insensitive to this beam separation.
 
\subsection{LHC beam vacuum description and gas injection procedure}
\label{sect:injection}
 
The ultra-high vacuum conditions inside the LHC beam pipe are achieved by distributed pumping, provided by the
NEG coating in the room-temperature sections and cryopumping in the cold sections.
 
A central parameter for this study is the sticking factor (\effneg) of hydrogen molecules on the NEG coated surface in the room-temperature sections.
This factor represents the probability for a gas molecule, hitting the surface, to stick on it. Thus it corresponds to the pumping efficiency of the NEG coating.
It is shown in Section\,\ref{sect:profiles} that the amount of gas in the introduced pressure bump depends strongly on this parameter.
In validation tests\,\cite{Lanza:IPAC2014-WEPME044}, prior to installation, \effneg was measured to be larger than $10^{-3}$ with $\meffneg=\mexpfor{5}{-3}$ being a
commonly expected value for the installed beam pipe.
 
Table\,\ref{tab:gasXsec} lists the inelastic proton cross sections for the typical molecules found in the beam vacuum.
The gases relevant for this study are hydrogen and neon. Although the latter is not a common gas in ultra-high vacuum systems it
could have influenced these tests significantly, as is discussed in Section\,\ref{sect:neon}.
 
In cryogenic parts of the LHC no NEG coating is needed since the cold bore of a magnet operated at 1.9\,K represents a 100\% efficient pump for
all gas species except helium\,\cite{baglin2020}. Since, however, the perforations cover only about 3\% of the BS surface, the pumping efficiency,
or sticking factor, of the cryogenic sections is estimated at \expfor{3}{-2}. Some magnets in the LSS, in particular the D2 dipole, are kept at an elevated
temperature of 4.5\,K at which the sticking factor, especially for hydrogen, is significantly reduced. In order to ensure a good vacuum in
these warmer magnets, dedicated cryoabsorber strips of woven carbon fibre are incorporated into the BS.

\begin{table}
\centering
\caption{Densities ($\rho$) at STP (0\degree\,C and 1.013\,bar) and inelastic scattering lengths ($\lambda$) of LHC beam protons for the
common residual gas molecules. The densities and scattering lengths are extracted from the \fluka code.
The cross sections derived from $\lambda$ and $\rho$ are given for reference only. The uncertainty on the density is negligible and that on the
scattering length a few percent (Section\,\protect\ref{sect:syscommon}).
The last column gives the relative response of the vacuum gauge for each gas species with respect
to N$_2$\,\protect\cite{Jousten:1046855}. The uncertainties on these sensitivities are estimated at about 10\%, but are included in the gauge accuracies, which are discussed in
Section\,\protect\ref{sect:syscommon}.}
\begin{tabular}{l|c|c|c|c}
& Density ($\rho$)                      & Inelastic scattering length ($\lambda$)    & Inelastic cross section            & Gauge response     \\
Gas                                  & [g/m$^3$] at STP                      & [m] at STP at 6.5\,\TeV                    & [mb] at 6.5\,\TeV                  & [N$_2$ equivalent] \\ \hline
\makebox[10mm][r]{H$_2$}             & \makebox[10mm][r]{89.9}               & \makebox[10mm][r]{4900}                    & \makebox[10mm][r]{\phantom{4}76}   & 0.42               \\
\makebox[10mm][r]{Ne}                & \makebox[10mm][r]{900\phantom{.9}}    & \makebox[10mm][r]{980}                     & \makebox[10mm][r]{380}             & 0.22               \\
\makebox[10mm][r]{CH$_4$}            & \makebox[10mm][r]{716\phantom{.9}}    & \makebox[10mm][r]{890}                     & \makebox[10mm][r]{420}             & 1.4\phantom{0}     \\
\makebox[10mm][r]{CO\phantom{$_2$}}  & \makebox[10mm][r]{1250\phantom{.9}}   & \makebox[10mm][r]{630}                     & \makebox[10mm][r]{590}             & 1.05               \\
\makebox[10mm][r]{CO$_2$}            & \makebox[10mm][r]{1970\phantom{.9}}   & \makebox[10mm][r]{410}                     & \makebox[10mm][r]{920}             & 1.4\phantom{0}     \\ \hline
\end{tabular}
\label{tab:gasXsec}
\end{table}
 
In the presence of an intense proton beam, gas is desorbed from the beam-pipe and BS walls by
synchrotron radiation, particles hitting the surface and radio-frequency heating induced by the passing bunches.
The rate of gas release depends on the surface condition and is reduced by the NEG coating
and beam scrubbing in the warm and cryogenic sections, respectively. The latter procedure consists of keeping increasingly
intense low-energy beams in the LHC in order to release the gas molecules from the surface prior to
actual physics operation.
The degree of conditioning cannot be directly measured and the actual state of the surface introduces
a significant uncertainty into the pressure around the ring, especially in cold sectors where no vacuum gauges
are located. Since, however, the pressure bumps are introduced by controlled gas injection and the steady-state
background is subtracted, as is explained in Section\,\ref{sect:BkgdVsP}, the uncertainty of the overall
pressure distribution is only of minor importance for this study.
 
The typical densities of the residual gases are about 12 orders of magnitude below those at standard temperature and pressure (STP),
which means that the mean free path of a beam proton is of the order of $10^{12}$\,km. This implies that the average lifetime of an
individual 6.5\,\TeV proton, before it undergoes a beam-gas interaction, is about 1000\,h.
 
If the gas densities are known, \bgr can be derived by using the data of Table\,\ref{tab:gasXsec}. The mean free
path $\lambda(z)$ is given by
 
\begin{equation}
\frac{1}{\lambda(z)} = \sum_j \frac{N_A \rho_j(z) \sigma_j}{A_j},
\end{equation}
 
where $N_A$ is the Avogadro number and $\sigma_j$ and $A_j$ are the interaction cross section and atomic mass of residual
gas species $j$, respectively. When the proton beam traverses the gas, it sees a $z$-dependent density $\rho_j(z)$.
In the ultra-high vacuum $\lambda$ is very large and the total interaction probability for a single proton traversing a
gas density distribution between $z_1$ and $z_2$ is obtained from a linear approximation
 
\begin{equation}
P_\mathrm{BG} = \int_{z_1}^{z_2} \frac{{\mathrm d}z}{\lambda(z)} = \sum_j\frac{N_A\sigma_j}{A_j}\int_{z_1}^{z_2}\rho_j(z)dz.
\label{eq:intrate}
\end{equation}
 
When $P_\mathrm{BG}$ is multiplied by the number of protons per bunch ($n_b$) -- by convention expressed in units of $10^{11}$\,protons -- and
the revolution frequency $\nu$ of the LHC beam, \bgr is obtained in units of \HzPerB for the $z$-range considered. Thus the factor $f$ in Eq.\,(\ref{eq:epsdef}),
for a single gas, is given by
\begin{equation}
f=n_b \nu \frac{N_A \sigma}{A},
\end{equation}
i.e. it comprises only constants and measured quantities.
 
The \bgr arising from the pressure distribution during normal operation was studied thoroughly
with simulations for LHC \run{1}\,\cite{DAPR-2017-01} and good agreement between simulated and measured BIB rates was obtained.
Although the bunch spacing was reduced from 50\,ns to 25\,ns and the beam
energy was increased for \run{2}, the BIB levels recorded by the BCM in the pressure bump fills, prior to the
gas injection, were very similar to those seen in \run{1}. This suggests that the \bgr distribution
as a function of $z$, presented in Ref.\,\cite{DAPR-2017-01}, provides a good estimate for the
pedestal \bgr in the pressure-bump fills.
According to that study the \bgr is about 0.01\,\HzPerBm for $|z|\lesssim 20$\,m and varies between
0.05\,\HzPerBm and 0.5\,\HzPerBm for larger $|z|$.
These \bgr values were used to define the target pressure for the bump tests, i.e. to determine  how  much
additional gas was needed in order to induce about an order of magnitude increase in the BIB rate seen by the BCM.
For the most distant bump location, at 151\,m, only a factor of about five increase of the BIB rate was reached at a
pressure of $\sim$1\,nbar, which was considered the safe limit.
 
Local vacuum stations, typically comprising ion pumps, NEG cartridges and vacuum gauges, are installed in the warm sections.
The NEG cartridges, Capacitorr D-400 of SAES getters\,\cite{capacitorr-d400}, are composed of
sintered discs of non-evaporable getter material, contained in a protective stainless steel grid. They are equipped with a heater element for
activation. In these tests the heaters were used to provoke an emission of gas from the cartridges
at the locations indicated in Figure\,\ref{fig:lss1-schema}.
The rate of gas release was controlled by the current used to heat the NEG and the pressure was monitored with the nearby gauges.
Upon heating the NEG is assumed to release predominantly hydrogen, which is least strongly bound into the NEG material. Thus
the gas composition of an introduced pressure bump differs significantly from the average in that region during normal operation.
Given that the pressure in the bumps was up to four orders of magnitude above normal, the additional beam-gas interactions were
initially assumed to all be $p$-H collisions, although this assumption had to be revised in the course of the analysis.
 
Since only BIB events from unpaired bunches were used, the observed background could be unambiguously associated with one of
the two incoming beams.
Therefore it was possible to heat two NEG cartridges simultaneously, one on either side of ATLAS. It took typically 5--10\,minutes to
reach the target level, where the pressure was stabilised for
about 20\,minutes in order to collect a sufficient number of events. Between consecutive pressure bumps on the same beam about 20\,minutes
were left for the BIB to fall close enough to the initial level.
 
\subsection{Pressure measurement}
\label{sect:press-meas}
 
The warm sections of the LHC beam pipe are equipped with vacuum gauges in order to monitor the pressure constantly.
In the regions where the pressure bumps are introduced, two different gauge technologies are utilised:
 
\vspace{3mm}
\hspace{10mm}
\begin{minipage}{130mm}
\begin{itemize}
\item[{\makebox[2cm][r]{VGPB:}}] cold cathode gauges\,\cite{Jousten:1046855}, Pfeiffer vacuum IKR 070, triaxial and
\item[{\makebox[2cm][r]{VGI:}}]  hot cathode gauges\,\cite{hot-cathode-gauges}, custom CERN made.
\end{itemize}
\end{minipage}
\vspace{3mm}
 
The measurement ranges of both gauge types cover the full pressure variation during the bump tests.
 
During the tests two or three gauges provided data in the vicinity of each pump but the positioning of the gauges with respect
to the beam line and the NEG cartridge is different at each pressure bump location.
Some gauges exhibited erratic behaviour in one or more fills. Therefore only the data
from the most appropriate devices, in terms of vicinity to the activated NEG cartridge and consistency of the data in all fills, were used.
In most cases, selecting an alternative gauge would have had only a very small influence on the results. This is discussed as
a systematic uncertainty in Section\,\ref{sect:syscommon}.
 
\subsection{Estimation of pressure bump profiles}
\label{sect:profiles}
 
Because a vacuum gauge does not provide any information about the gas density profile along the beam line,
pressure calculations have to be invoked to estimate the profiles, $\rho(z)$, which are needed to calculate \bgr
from Eq.\,(\ref{eq:intrate}).
 
For the calculation of pressure profiles at the various bump locations, two conceptually different vacuum calculation/simulation programs were used:
\begin{enumerate}
\item \vasco \,\cite{Rossi:728512} uses a simple one-dimensional model to produce a pressure profile along the beam line, based on the
requirement that various sources (outgassing mechanisms), drains (local or distributed pumps) and the diffusion of gas between sources and drains,
are in equilibrium.
\item \molflow\,\cite{Kersevan:2694236} is a detailed three-dimensional Monte Carlo simulation which transports individual gas molecules
until they are removed by a pump or absorbed on a surface. In order to benefit fully from the power of \molflow, a detailed three-dimensional
description of the vacuum system is needed.
\end{enumerate}
 
Initially, only \vasco calculations, based on the best knowledge of the pumping efficiency in the room temperature and cryogenic parts of the beam
pipe, were produced for all four pressure bump locations. These, however, resulted in unphysically low values of \bgr for the
nearby pressure bumps at $|z|$<23\,m, as is seen in Section\,\ref{sect:bibeff}. Therefore more detailed \molflow simulations were set up
for the 19\,m and 22\,m pressure bumps, exploring various alternative pumping efficiencies of the NEG coating. No sufficiently detailed \molflow geometry models are available for
the 58\,m and 151\,m locations, so the original \vasco results are used for these distant pressure bumps. The uncertainties associated with
these calculations are discussed in Section\,\ref{sect:sysmeas}.
 
The NEG cartridges and vacuum gauges are not directly on the beam pipe, but recessed on the same or different chimneys. When
gas is emitted from the NEG cartridge, the pressure at the gauge is typically different than at the chimney exit.
Being one-dimensional, the \vasco calculations do not consider the various chimneys in detail, so they do not provide estimates for the
pressures at the gauge locations. The \molflow simulations, on the other hand, can provide an estimate of the gas density profiles along
the beam line where the protons pass, but also at the location of the gauge where the pressure is measured.
 
\begin{figure}
\centering
\mbox{
\begin{subfigure}{0.49\textwidth}
\centering\captionsetup{width=0.9\textwidth}
\includegraphics[width=0.95\textwidth]{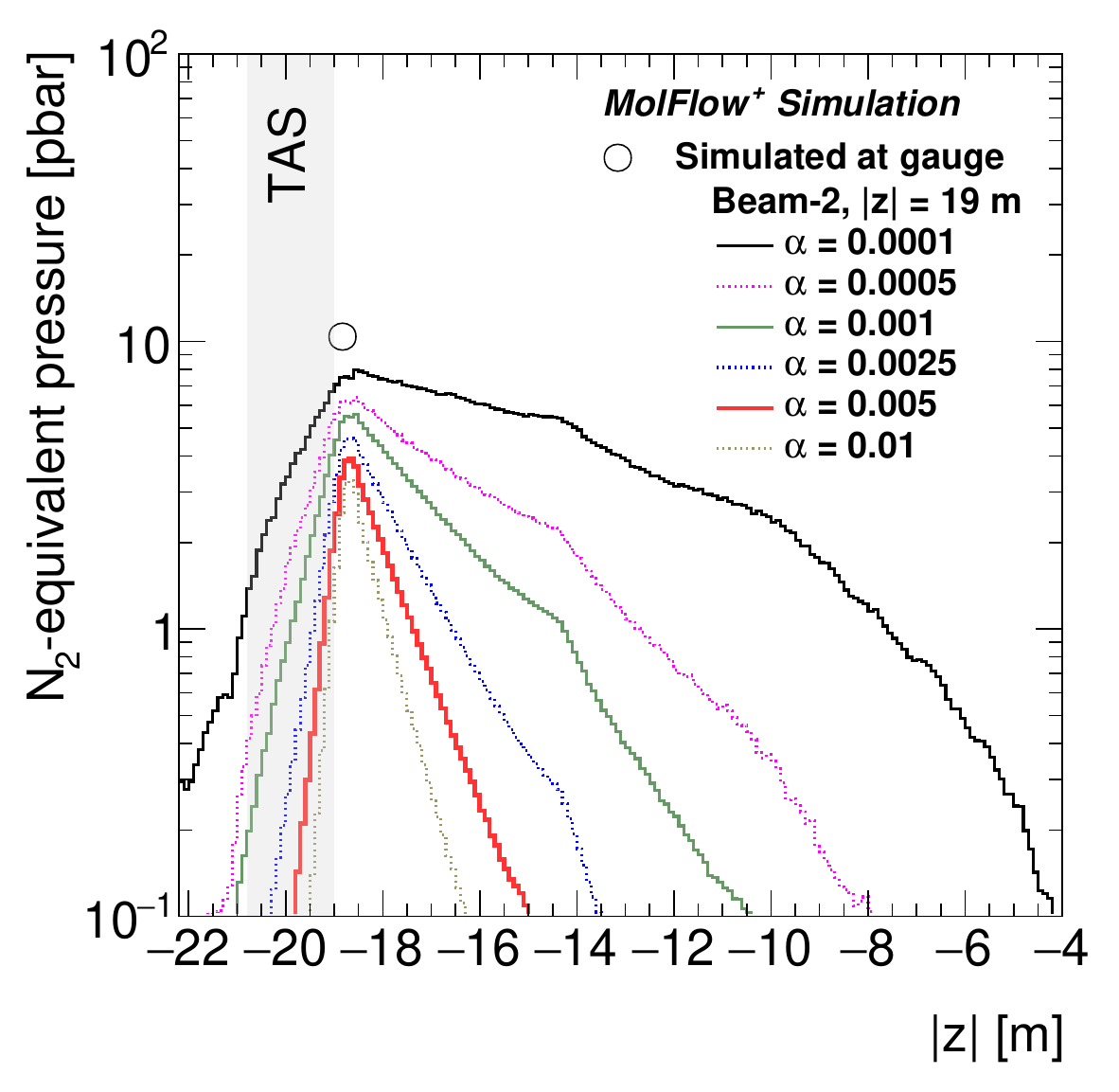}
\caption{}
\label{fig:bumpNeqNewProfiles19m}
\end{subfigure}
\begin{subfigure}{0.49\textwidth}
\centering\captionsetup{width=0.9\textwidth}
\includegraphics[width=0.95\textwidth]{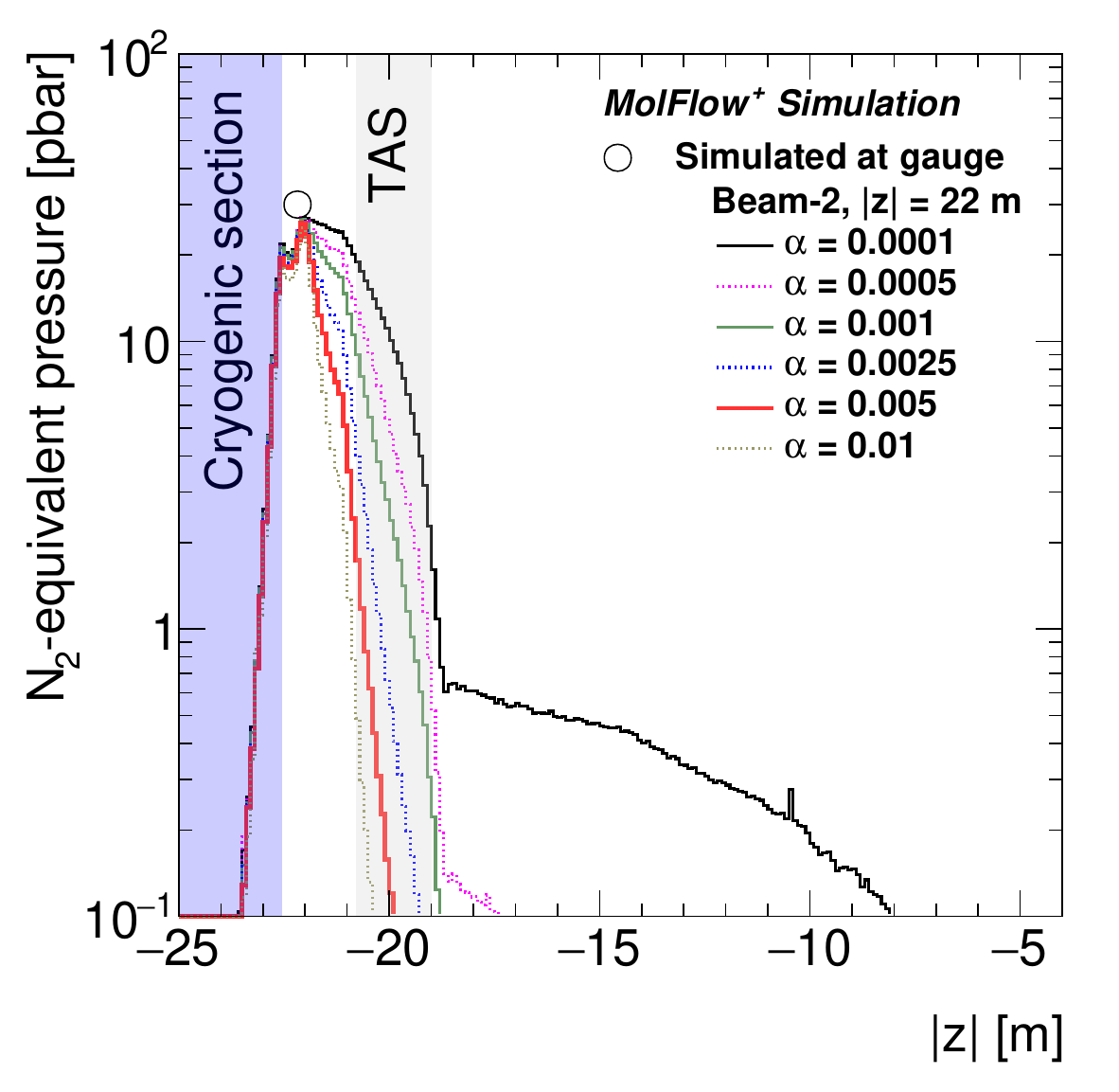}
\caption{}
\label{fig:bumpNeqNewProfiles22m}
\end{subfigure}
}
\caption{Pressure profiles at \subref{fig:bumpNeqNewProfiles19m} 19\,m and \subref{fig:bumpNeqNewProfiles22m} 22\,m as
simulated with \molflow for different assumptions of \effneg. The profile for the
expected value, $\meffneg=\mexpfor{5}{-3}$, is indicated by the thicker line.
The \molflow simulations
also provide an estimate of the pressure at the relevant vacuum gauge, which can be used to scale the simulation results to the pressure
actually measured in any particular fill, as is detailed in Section\,\protect\ref{sect:pressure-scaling}.
The blue shaded area in \subref{fig:bumpNeqNewProfiles22m} indicates the cryogenic region of the inner triplet.}
\label{fig:bumpNeqNewProfiles}
\end{figure}
 
Figure\,\ref{fig:bumpNeqNewProfiles} shows the \molflow results for pressure bumps at 19\,m and 22\,m, for different values of the NEG sticking factor \effneg.
The value of \effneg is found to have a particularly large effect on the pressure profile of the 19\,m bump. The reason is that between 19\,m and 20.8\,m
the TAS, with its small aperture, reduces the gas flow while the beam pipe towards the IP is wide, and for low \effneg the pressure bump extends
almost to the IP. It is also seen that with increasing \effneg the difference between the value at the gauge
and the maximum at the beam line increases.
 
The situation is quite different at 22\,m where the pressure bump is constrained between the TAS on the IP side and
the $T=1.9$\,K region of the inner triplet, starting at $|z|\approx 23.5$\,m. The cold bore of the quadrupole magnet is considered a
100\% efficient pump for all gas species. So any gas molecule traversing the pumping holes, which cover about 3\% of the BS,
is considered to be absorbed. Since this cold part has no NEG coating, the effect of \effneg is
restricted to the region between the inner triplet and the TAS and, for the small amount of gas passing through it, the beam pipe
on the IP side of the TAS. This restriction from both sides is very evident when comparing Figures\,\ref{fig:bumpNeqNewProfiles22m} and
\ref{fig:bumpNeqNewProfiles19m}: for $\meffneg=0.005$ the pressure drops by two orders of magnitude when the gas is injected at 22\,m and
has to traverse the TAS. If the injection point is at 19\,m the profile towards the IP is much wider.
 
\section{Beam background simulations with \fluka}
\label{sect:fluka}
 
Most LHC beam background simulations\,\cite{BRUCE2013825} are performed with the \fluka Monte Carlo simulation code.
Recently these have been extended\,\cite{DAPR-2017-01} to include hit simulation and trigger
emulation in the ATLAS sub-detectors that are relevant for studies of BIB. Thus the \fluka
simulations can be used to predict the efficiency of ATLAS to detect a beam-gas event as a function of its $z$-position.
 
The \fluka simulations are performed as outlined in Ref.\,\cite{DAPR-2017-01}.
Beam-gas events are modelled as inelastic $p$-H scattering and the $z$-coordinates are sampled from the simulated
pressure profiles, which are described in Section\,\ref{sect:profiles}. For the nearby pressure bumps, i.e. $|z|<23$\,m, the position of the beam-gas interaction
in the $xy$-plane is determined from the offset due to the vertical crossing angle and smeared according to the divergence caused by
the $\beta$-squeeze and transverse emittance of the beams.  In all simulations 185\,${\mathrm \mu}$rad and 40\,cm are assumed for the crossing half-angle and $\beta^*$, respectively.
These do not match the values given in Table\,\ref{tab:beamcond} for all fills but the results depend only very weakly on the
exact values of these parameters.
In the case of the two distant pressure bumps, the transverse position of the event is placed on the ideal beam orbit and the shower is transported with \fluka
to a virtual interface plane at $z=22.6$\,m where particles are picked up by the ATLAS simulation, as described in Ref.\,\cite{DAPR-2017-01}.
In these computing-intensive LHC simulations, where particle showers are tracked in the accelerator structure and LHC tunnel over more than 100\,metres,
the primary $p$-H collisions have a uniform $z$-distribution of $\pm$5\,m around the centre of each pressure bump. For the ATLAS simulations the events are selected
according to the calculated pressure bump profiles along $z$. This allowed different pressure profiles to be studied, using the same set of events from the
LHC simulations.
 
Fake-jet events are rare, and the number of events available from the LHC simulations determines the statistical uncertainty:
9.5\,million and 22.5\,million events were selected for the 58\,m and 151\,m locations, respectively.
As practically all fake jets are generated by radiative energy losses of muons, only these were picked up at the interface
plane for further transport through ATLAS.
 
Obtaining a sufficient number of BIB events in the BCM background simulations is not an issue for any of the pressure bump locations.
Thus only a small fraction of the events produced in the LHC simulations
was used: \expfor{2}{4} for the pressure bumps at 58\,m and \expfor{2}{5} for the 151\,m location. For the nearby bumps, where the
efficiency is very high, 8000 events, sampled from the pressure profiles, sufficed to obtain a negligible statistical error.
All particles were transported until their energy fell below 100\,keV, which is lower than the hit threshold of the BCM modules.
In order to account for the dead-time of a BCM module, only the first
hit within the $\pm$12.5\,ns window, centred around the bunch passage at the IP, was considered for any individual module.

\section{Background versus measured pressure}
\label{sect:BkgdVsP}

\subsection{Determination of pedestal and bump pressure}
 
During normal operation the BIB is dominated\,\cite{PhysRevAccelBeams.22.021004} by beam-gas interactions, distributed
along the LSS and adjacent arc\,\cite{DAPR-2017-01}.
The gas injection causes localised pressure bumps into this distributed source. The observed background ($B$) is therefore parameterised
as a function of the change of local pressures ($\Delta P(z_i)$) reported by vacuum gauges at locations $z_i$:
 
\begin{equation}
B = C + \sum_i \omega(z_i) \Delta P(z_i),
\label{eq:pfit}
\end{equation}
 
where $z_i$, with $i=1-4$, refers to the four pressure bump locations and the pressure differences are measured with respect to the
value prior to the gas injection.
The parameter \omegf{z_i} relates $\Delta P(z_i)$ and the change of detected background while $C$ describes
the pedestal BIB level, arising from beam-halo losses at the TCT collimators and from beam-gas
interactions in the absence of the pressure bump. Another minor contribution to $C$, for fake jets, comes from energy
depositions by cosmic rays.
 
The accuracy with which $\Delta P(z_i)$, as measured by a vacuum gauge, reflects the pressure at the beam line depends on the
local configuration of the vacuum equipment.
In addition, the local pressure measurements provide no information on how
far the injected gas spreads in the beam pipe, i.e. what is the gas density profile traversed by the beam.
The dependence between $\Delta P(z_i)$ and the gas density, integrated along the beam trajectory,
is reflected in the value of \omegf{z_i}, which therefore differs from the absolute BIB detection
efficiency that is discussed in Section\,\ref{sect:abseff}.
Since BIB is generated only by the incoming beam, Eq.\,(\ref{eq:pfit}) is evaluated separately for positive and
negative $z$.

\begin{figure}
\centering
\includegraphics[width=0.95\textwidth]{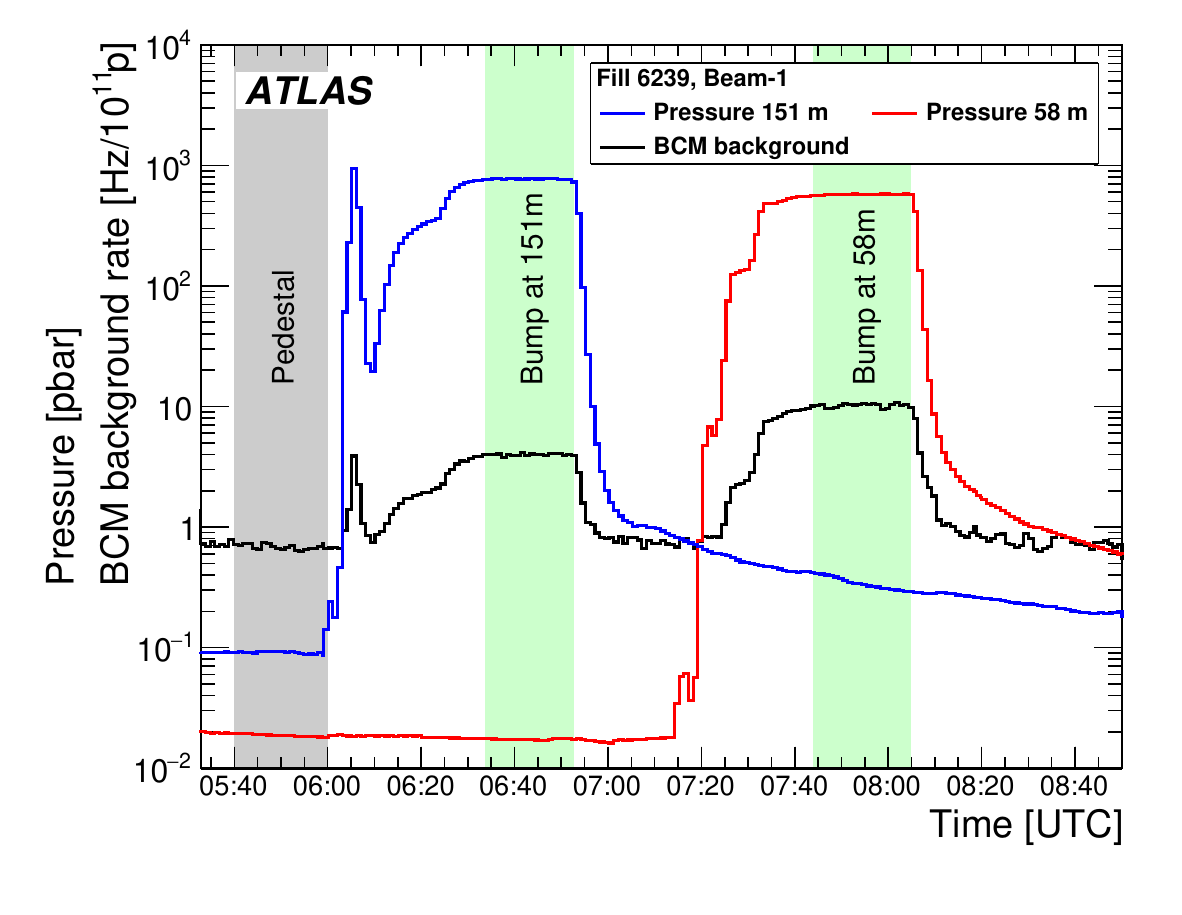}
\caption{BCM background observed on \beam{1} in fill 6239 and pressures measured by the vacuum gauges at $z=58$\,m and $z=151$\,m.
The grey-shaded area before the first bump indicates the time periods over which the pedestal pressure and background are averaged.
The green shaded areas correspond to the averaging over the stable pressure periods of each of the two bumps.
}
\label{fig:BCMandP}
\end{figure}
 
Figure\,\ref{fig:BCMandP} shows, as an example,  the BCM background rates and pressures at 58\,m and 151\,m for \beam{1} in fill 6239.
The pedestal background, $C$, is determined, both for BCM and fake-jet rates, by averaging over the grey-shaded area indicated in
Figure\,\ref{fig:BCMandP}.
The pressures reported by the selected gauges are averaged over the same period in order to determine
the corresponding pedestal pressures.  Figure\,\ref{fig:BCMandP} reveals that after the gas injection is terminated the pressure
does not promptly return to its initial value, i.e. some pressure, additional to the pedestal, remains when
further bumps are introduced at other locations. This residual, however, is at the percent level or below and therefore has no significant
impact on the value of $\Delta P(z_i)$.
 
The background rates and pressures due to the gas injection are taken from the green-shaded time intervals, which
correspond to the stable pressure periods at the top of each bump.
For each fill and pressure bump these shaded areas are determined by eye, requiring that the quality of the
pressure and background measurements is as good and stable as possible throughout each period. In particular,
periods obviously affected by instrumental noise or beam instabilities, not related to the test itself, are excluded.
 
\subsection{Calculation of the \omegf{z_i} parameters}
 
The rate of BIB, especially the beam-gas component, depends on the bunch intensity and residual pressure around the accelerator ring. The latter
varies with the total beam  intensity. This variation, however, is a function of the $z$-location, and the information available is
insufficient to take all local pressure changes into account when propagating $C$. Therefore only the intensity decrease of the unpaired
bunches is considered when propagating $C$ to later times.
 
The pedestal $C$ is subtracted from the background rate $B$, measured at maximum bump pressure, i.e. the periods indicated by the
green shades in Figure\,\ref{fig:BCMandP}. The $\Delta P(z_i)$ is determined as the difference between the pressure in the green- and
grey-shaded periods.
 
Since only one pressure bump at a time is introduced per beam, \omegf{z_i} can be solved from Eq.\,(\ref{eq:pfit}):
 
\begin{equation}
\omega(z_i) = \frac{B - C}{\Delta P(z_i)} = \frac{\Delta B}{\Delta P(z_i)},
\label{eq:omega}
\end{equation}
 
which is the same as Eq.\,(\ref{eq:omegadef}).
 
Both, $B-C$ and $\Delta P(z_i)$ are evaluated separately for each LB in the green shaded area.
 
The final \omegf{z_i} for each $z$-location is obtained by averaging over all LBs in the green-shaded area of the corresponding
pressure bump. The uncertainties on the mean values are estimated from the LB to LB variation. Upon insertion of the constant $C$
and the obtained \omegf{z_i}-values, Eq.\,(\ref{eq:pfit}) represents a global parameterisation of the BIB as a function of local
pressure changes with time.
 
\subsection{Parameterisation results}
 
\begin{figure}
\centering
\mbox{
\begin{subfigure}{0.49\textwidth}
\centering\captionsetup{width=0.9\textwidth}
\includegraphics[width=0.95\textwidth]{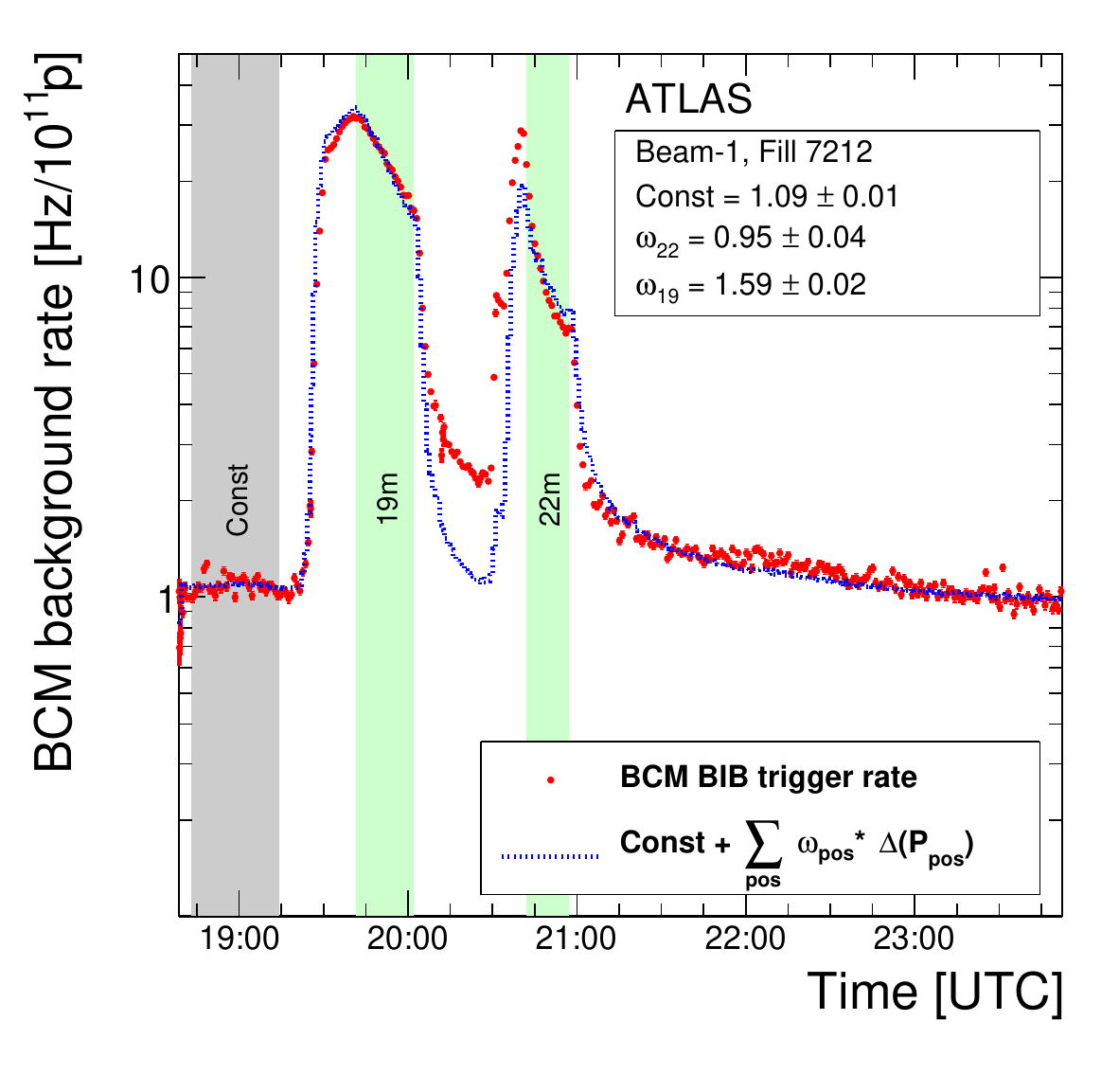}
\caption{}
\label{fig:bcmb1fit7212}
\end{subfigure}
\begin{subfigure}{0.49\textwidth}
\centering\captionsetup{width=0.9\textwidth}
\includegraphics[width=0.95\textwidth]{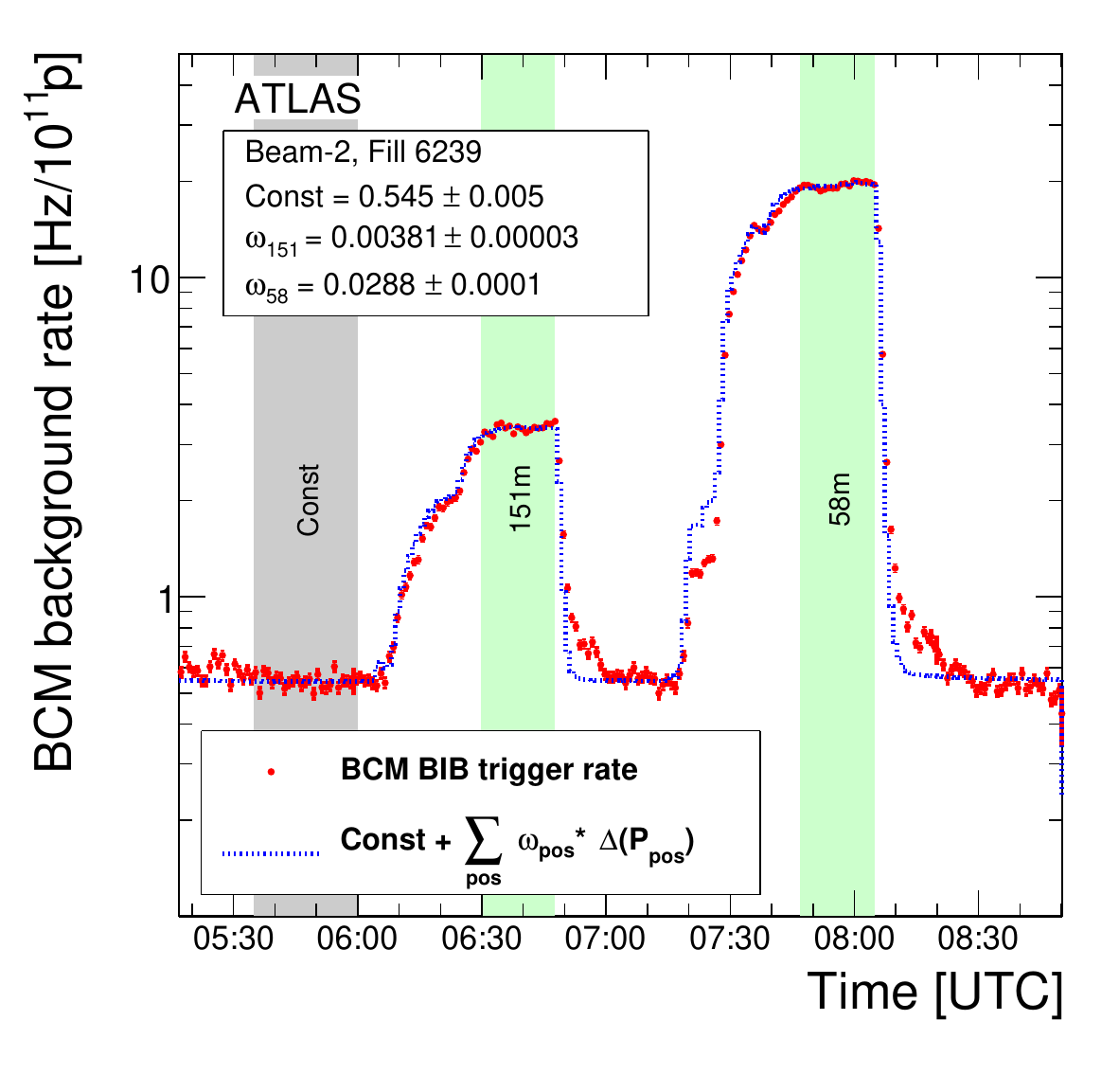}
\caption{}
\label{fig:bcmb2fit6239}
\end{subfigure}
}
\\
\mbox{
\begin{subfigure}{0.49\textwidth}
\centering\captionsetup{width=0.9\textwidth}
\includegraphics[width=0.95\textwidth]{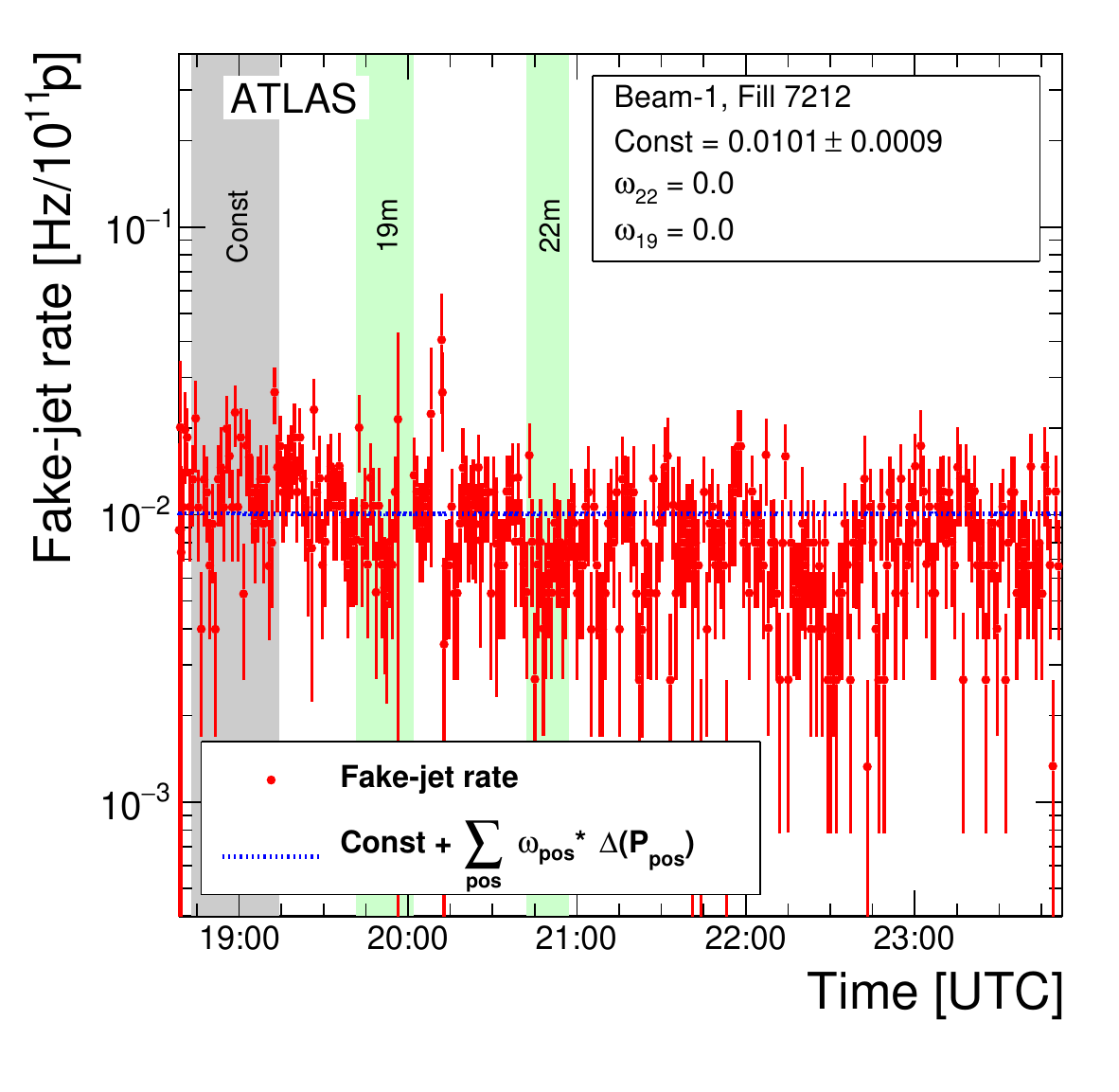}
\caption{}
\label{fig:jetb1fit7212}
\end{subfigure}
\begin{subfigure}{0.49\textwidth}
\centering\captionsetup{width=0.9\textwidth}
\includegraphics[width=0.95\textwidth]{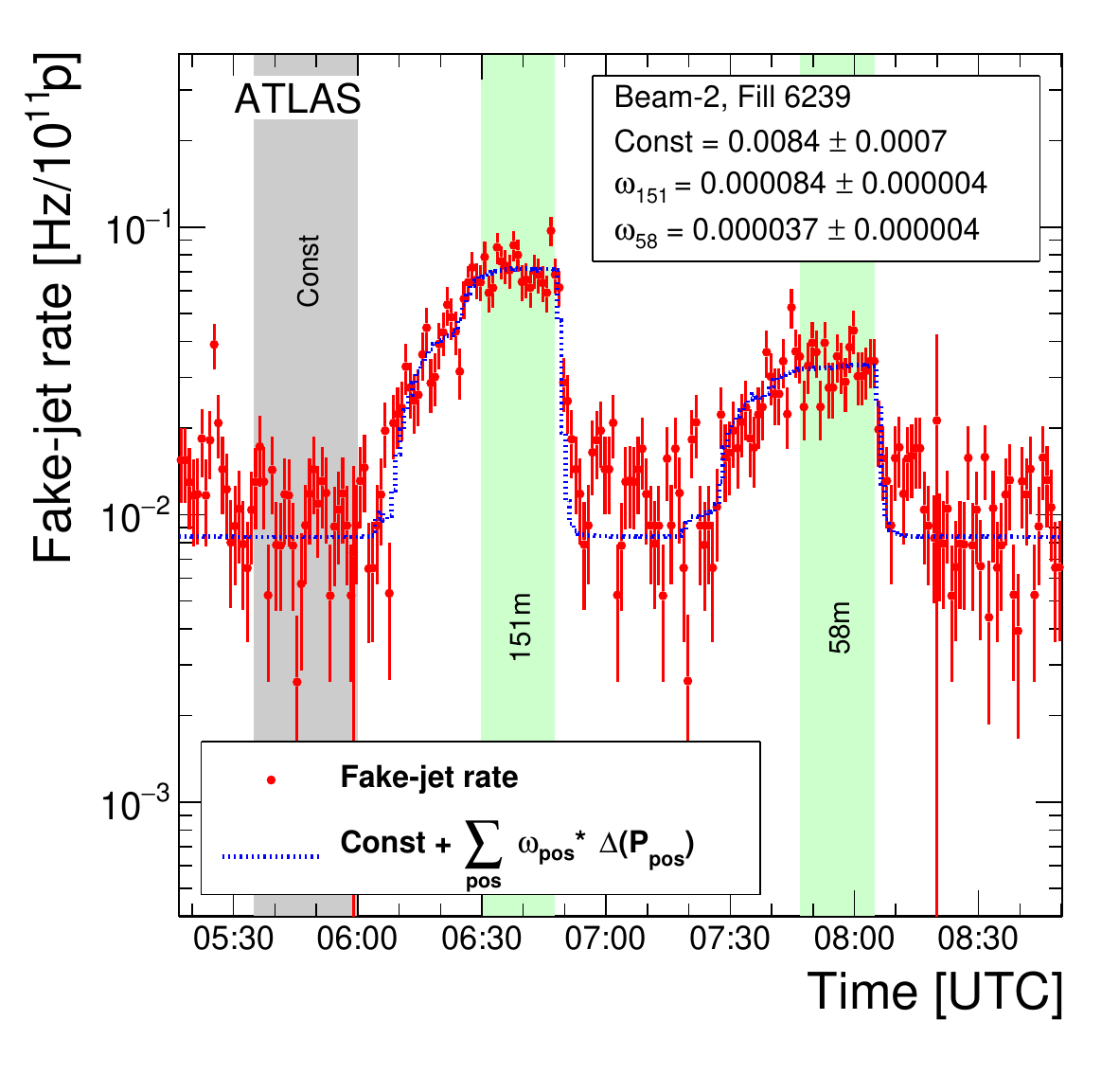}
\caption{}
\label{fig:jetb2fit6239}
\end{subfigure}
}
\caption{Comparison of BIB predictions using Eq.\,(\protect\ref{eq:pfit}), with the observed \subref{fig:bcmb1fit7212} BCM and
\subref{fig:jetb1fit7212} fake-jet backgrounds from \beam{1} pressure bumps at 19\,m and 22\,m in fill 7212. Figures\,\subref{fig:bcmb2fit6239}
and \subref{fig:jetb2fit6239} show the corresponding results at 58\,m and 151\,m for BCM and fake jets, respectively, for \beam{2} in fill 6239.
Only the data in the shaded regions are used to determine the parameters $C$ and \omegf{z_i}, which are needed in Eq.\,(\protect\ref{eq:pfit})
to produce the dotted blue histogram. The time-dependence arises from the time variation of the local pressures $P(z_i)$.
}
\label{fig:fits}
\end{figure}

Figure\,\ref{fig:fits} compares some representative BIB predictions, using the pressure parameterisation defined in Eq.\,(\ref{eq:pfit}),
with the observed background. The data-points outside of the shaded areas, shown in the plots, are not included when determining
the parameters \omegf{z_i}. Thus any background anomalies that are not explained by the pressures at the considered
gauges appear as differences instead of biasing the \omegf{z_i}-values.
Equation\,(\ref{eq:pfit}) agrees with the data also at the right edge of each plot. This confirms that
the variation of $C$ is negligible over the short duration of the tests, even though a correction
with the overall decrease of dynamic pressure has not been applied.

Ideally \omegf{z_i} should be determined from a stable period around the maximum bump pressure.
For several bumps, especially those at 19\,m and 22\,m, the pressure, however, could not be stabilised at a plateau value.
In these cases a range with as stable as possible a pressure was selected, while taking care to maintain a
reasonable number of events. In any case, the effect of a pressure variation should largely cancel since \omegf{z_i} is
the ratio between the observed background and measured pressure and is evaluated for each LB separately.
Thus \omegf{z_i} is constant, as long as the background change is proportional to the change of the pressure.
Should, however, this proportionality be violated by a different time dependence of the two quantities, then this will
be reflected in the uncertainty of \omegf{z_i} such that it includes in addition to statistical fluctuations
also an unresolved systematic component. Such a non-linearity might appear, especially, at the start and end of the gas injection
when the pressure bump profile is not yet, or not any more, stabilised.
 
In fill 7212 the tests started with a pressure bump at 19\,m. Figure\,\ref{fig:bcmb1fit7212} shows
that initially the pressure fit follows the rise of the background very closely. The pressure could not be
perfectly stabilised at the target level but was gradually decreasing with time, as was the background rate.
When the heating of the NEG cartridge was terminated, at about 20:00 UTC, the pressure dropped back to
the initial value within about 20\,minutes but the background remained at a higher level.
Thus the 22\,m pressure bump started with a factor of about two offset with respect to the initial pedestal background. The level,
however, still was only about 10\% of
that at the top of the bump. An interesting feature of the  leading edge of the 22\,m pressure bump is that the background rises
more rapidly than the pressure and overshoots the parametrisation in the first few minutes after the maximum pressure
is reached. A possible reason for this is discussed in Section\,\ref{sect:neon}.
No stable pressure could be established for the 22\,m bump; both pressure and background dropped by about a factor of three over
the shaded region which was considered for the averaging. When the gas injection was terminated it took about 2\,hours
for the background and pressure to return to their initial levels.
 
Figure\,\ref{fig:jetb1fit7212} shows that no additional fake jets can be resolved from the pedestal for the
19\,m and 22\,m pressure bumps. This is the case also for all other fills where gas was injected at these nearby locations.
On the right side of Figure\,\ref{fig:jetb1fit7212} the parameterisation overestimates the data. This is due to a separation of
the beams at about 20:30\,UTC, which causes a drop of the jet background from ghost collisions. Despite the
applied vertex veto some contribution of these remains in the pedestal that is derived from the grey-shaded period,
which corresponds to a time when the beams were still colliding head-on.
 
Contrary to the nearby bumps, pressures were stable for the 58\,m and 151\,m bumps in
fill 6239, shown in Figures\,\ref{fig:bcmb2fit6239} and \ref{fig:jetb2fit6239} for BCM and fake-jet rates,
respectively. The pressure parameterisations follow very closely the BCM backgrounds, except for a minor overshoot
on the rising edge of the 58\,m bump and a slightly faster return to the initial level.
Excess jet rates are clearly resolved for these distant pressure  bumps, even though the pedestal level is rather noisy.
 
A common feature of several pressure bumps is that when the outgassing is stopped, the background decays slower than the measured
pressure. This effect is particularly clear at 19\,m and less pronounced or totally absent at the three other locations.
A possible explanation is that some gas that has diffused far into the beam pipe is pumped out slower, i.e. when the
outgassing from the NEG is stopped the low pressure is quickly re-established in the vicinity of the NEG cartridge where the
gauges and additional ion pumps are located.
Further away along the beam pipe a small fraction of the injected gas seems to reside for longer, maintaining a background excess.
This is discussed in more detail in Section\,\ref{sect:abseff}.

\renewcommand{\arraystretch}{1.25}
\begin{table}
\centering
\caption{Values of the pedestal background ($C$) and \omegf{z_i} averaged over all fills where the  corresponding pressure bump
was introduced. The uncertainties include the statistical variation within any given fill but are dominated by the fill-to-fill variation.
As explained in the text, they might also include a component due to a non-linearity between measured pressure and observed background.}
{\small
\begin{tabular}{r|c||c|c|c|c}
\multicolumn{2}{c||}{}   & \multicolumn{2}{c|}{BCM background rate}                & \multicolumn{2}{c}{Fake-jet background rate}            \\
& Unit        &\Beam{1}                    &  \Beam{2}                  &  \Beam{1}                  & \Beam{2}                   \\ \hline
$C$                      & \HzPerB     & $0.80 \pm 0.09$            & $0.60 \pm 0.07$            & $0.012 \pm 0.002$          &  $0.012 \pm 0.002$         \\
\omegf{151\,{\mathrm m}} & \!\HzPerBpbar\! &\!\expfor{(4.2 \pm 0.2)}{-3}\!& \!\expfor{(3.4 \pm 0.5)}{-3}\! & \!\expfor{(8.8 \pm 0.5)}{-5}\! & \!\expfor{(8.1 \pm 0.8)}{-5}\! \\
\omegf{58\,{\mathrm m}}  & \!\HzPerBpbar\! &\!\expfor{(1.8 \pm 0.3)}{-2}\!& \!\expfor{(2.9 \pm 0.2)}{-2}\! & \!\expfor{(1.9 \pm 0.3)}{-5}\! & \!\expfor{(4.0 \pm 0.4)}{-5}\! \\
\omegf{22\,{\mathrm m}}  & \!\HzPerBpbar\! & $0.82 \pm 0.03$            & $0.610 \pm 0.006$          &   ---                      & ---                        \\
\omegf{19\,{\mathrm m}}  & \!\HzPerBpbar\! & $1.53 \pm 0.05$            & $0.80 \pm 0.04$            &   ---                      & ---                        \\ \hline
\end{tabular}
}
\label{tab:averageomegas}
\end{table}
 
The \omegf{z_i} values obtained from Eq.\,(\ref{eq:omega}) and averaged over all relevant fills are listed in Table\,\ref{tab:averageomegas}.
They have units of \HzPerB per pbar because they express the change of the
observed background rate with respect to the measured N$_2$-equivalent pressure change. Since the \omegf{z_i}-values are based on the
pressure measurement and not on the actual amount of additional gas traversed by the beam, they are not exactly comparable for locations
with different vacuum gauge configurations. Nevertheless, they still
give some indications of the relative importance of background sources as a function of $z$.
The general trend is that the efficiency of the BCM to detect beam-gas events decreases as the distance between the pressure bump and the IP increases.
This is expected from the fact that the BCM probes the BIB at very small radii; even from beam-gas collisions only a few metres from the IP
some secondaries can reach the radial range of the BCM sensors. This is consistent with simulations\,\cite{DAPR-2017-01}, where the efficiency is found
to increase between $z\approx5$\,m and $z\approx10$\,m and then stay roughly constant up to $z\approx25$\,m.  Beyond this the attenuation in accelerator
components, notably the TAS, and the lateral spread of the shower start to reduce the probability of a BCM module to be hit.
 
One striking observation that can be made from Table\,\ref{tab:averageomegas} is that for the BCM background, \omegf{19\,{\mathrm m}} and \omegf{22\,{\mathrm m}}
are significantly higher for \beam{1} than for \beam{2} although the beam line, the instrumentation and the BCM detector are
symmetric on both sides. This finding, which here is based purely on data, has implications for the discussion in Section\,\ref{sect:abseff}.
 
Contrary to the BCM, the probability to observe fake jets increases with distance from the IP. This is explained by the fact that
fake jets are produced by energetic muons that reach the calorimeter radii only if they are produced sufficiently far away.
In particular, if they originate from beyond the D1 magnet they are deflected by its field, which increases their spread
in the horizontal plane.
 
\begin{table}
\centering
\caption{Ratio (BCM/Jet) of the average values given in Table\,\protect\ref{tab:averageomegas}.
The simulated ratios are derived from Tables\,\ref{tab:bcmSimEfficiency} and \ref{tab:jetSimEfficiency} in Section\,\ref{sect:abseff}, i.e.
their uncertainties include only the statistical errors given there.
No simulations exist for the value of $C$.}
\begin{tabular}{r|c|c|c}
Position & \Beam{1}              & \Beam{2}              & Simulation    \\ \hline
$C$      & 64$\pm$11             & 49$\pm$8              &    N.a.       \\ \hline
58\,m    & 950$\pm$230           & 740$\pm$80            & $790 \pm 45$  \\
151\,m   & 48$\pm$3              & 41$\pm$7              & $52 \pm 2$    \\ \hline
\end{tabular}
\label{tab:bcmPerJet}
\end{table}
 
The rates reported in Table\,\ref{tab:averageomegas} provide a purely experimental determination of the relative sensitivity of ATLAS to
detect BIB through BCM trigger and fake-jet rates: since for any given position the backgrounds originate from the same pressure distribution,
all unknown factors, especially the pressure profile, cancel from the ratios of the BCM and fake-jet rates.
These ratios, presented in Table\,\ref{tab:bcmPerJet}, reveal several interesting features:
\begin{itemize}
\item The ratio of the constant terms is slightly higher for \beam{1}. Table\,\ref{tab:averageomegas} shows that this difference is
entirely due to an asymmetric BCM background. The BCM sensitivity is highest for nearby beam-gas events, which do not produce fake jets.
Thus a \beam{1}/\beam{2} ratio larger than one suggests that there are more nearby beam gas events on the incoming \beam{1} side, i.e. for an unknown
reason the vacuum in the vicinity of ATLAS seems worse on that side.
\item  For both distant pressure bump locations the BCM/Jet ratios are consistent between the two beams, as would be expected from the symmetry
of the beam line and the ATLAS detector.
\item The BCM/fake-jet ratio is almost 20 times larger for beam-gas events at 58\,m than at 151\,m.
\item The agreement with simulations is nearly perfect. Since the simulations are done with the same methods as in Ref.\,\cite{DAPR-2017-01}, these
measurements provide a valuable benchmark for that earlier work.
\item The most interesting observation arises from a comparison with Ref.\,\cite{PhysRevAccelBeams.22.021004}, where the ATLAS beam backgrounds resulting from
proton losses on the TCTs, at $|z|\approx$145--148\,m from the IP, are reported. In those data, at $\beta^*=40$\,cm, the
BCM/fake-jet ratio was found to be \mbox{5.4\,$\pm$\,0.3} for \beam{1} and \mbox{6.1\,$\pm$\,0.3} for \beam{2}.
The corresponding ratios in Table\,\ref{tab:bcmPerJet} for the 151\,m pressure bump are higher by a factor of about seven.
The explanation of this substantial difference is deferred to the end of Section\,\ref{sect:bibdeteff}.
\end{itemize}

\section{Absolute beam-gas detection efficiency}
\label{sect:abseff}
 
\subsection{Definition}
 
The values of \omegf{z_i}, derived from Eq.\,(\ref{eq:omega}), are based entirely on data and show that
the observed background rates are strongly correlated with the magnitude of the locally introduced pressure bumps.
Although they provide an experimental determination of the BCM/fake-jet background ratio as a function of the pressure bump location,
they do not allow for determining the real efficiency to detect beam-gas events by the BCM
(\effbcm{z}) or as fake jets (\effjet{z}) from different $z$-locations. Furthermore, since the \omegf{z_i} are specific
to the local configuration of vacuum gauges and NEG cartridges used to inject the gas, they are not even suited to accurately determine
the relative efficiency between different $z$-locations.
 
The generally good agreement of the time-dependence between the observed background and the measured pressure, seen in Figure\,\ref{fig:fits},
confirms that the introduced pressure bumps dominate the background rates. Other contributions are negligible, especially for the BCM.
 
The ultimate goal of these pressure bump tests was to determine the absolute efficiency of ATLAS to detect beam-gas events as a
function of their $z$-location. This, however, requires that \bgr is known for each bump.
Once \bgr is determined, \effbcm{z} and \effjet{z} are obtained from Eq.\,(\ref{eq:epsdef}), i.e.
as the ratio between the observed background rate increase and the additional \bgr due to the introduction of the pressure bump.
 
\subsection{Absolute scale of the pressure profiles}
\label{sect:pressure-scaling}
 
\begin{figure}
\centering
\mbox{
\begin{subfigure}{0.49\textwidth}
\centering\captionsetup{width=0.9\textwidth}
\includegraphics[width=0.95\textwidth]{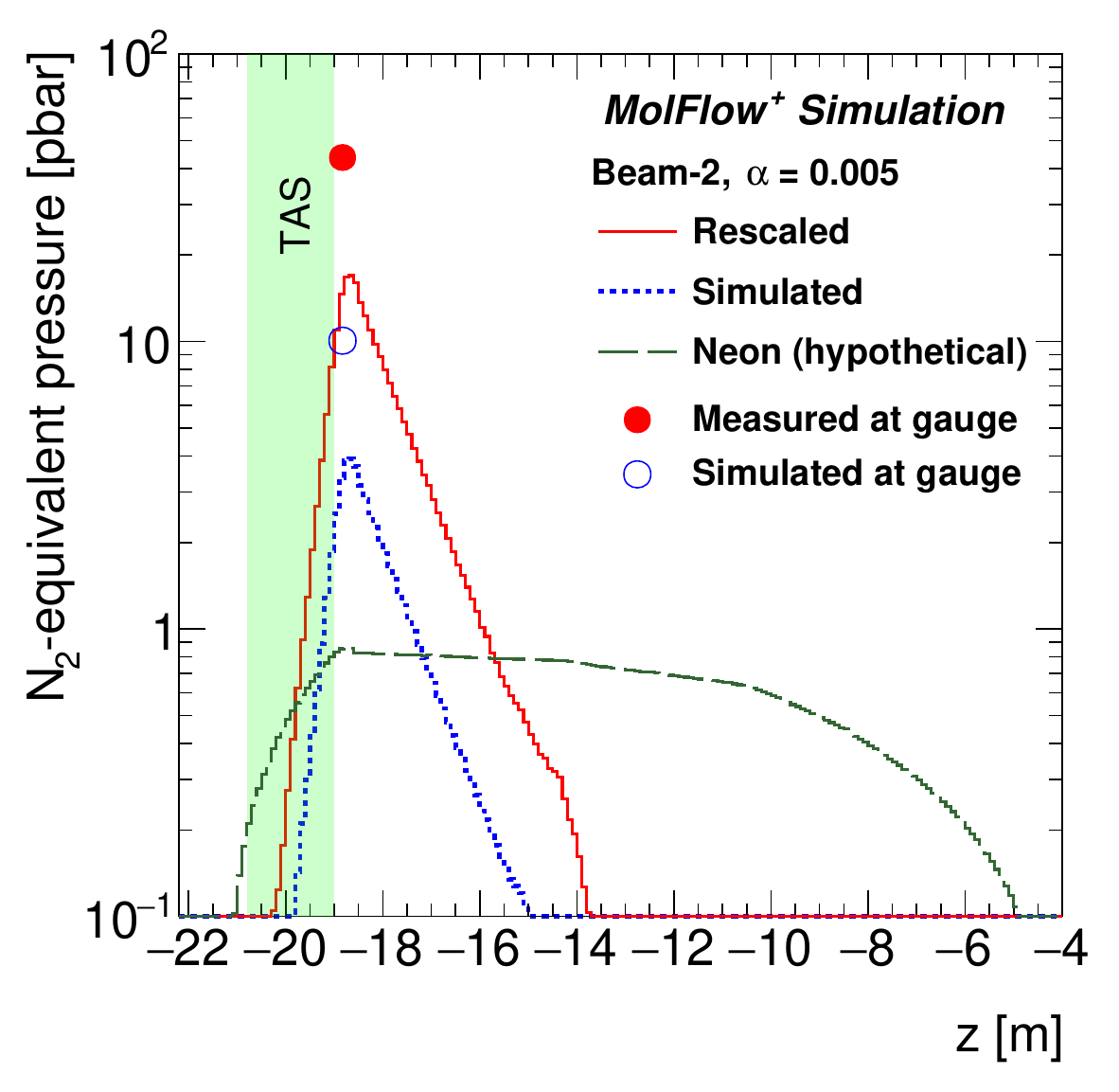}
\caption{}
\label{fig:bumpNeqNewProfilesRescaled19m}
\end{subfigure}
\begin{subfigure}{0.49\textwidth}
\centering\captionsetup{width=0.9\textwidth}
\includegraphics[width=0.95\textwidth]{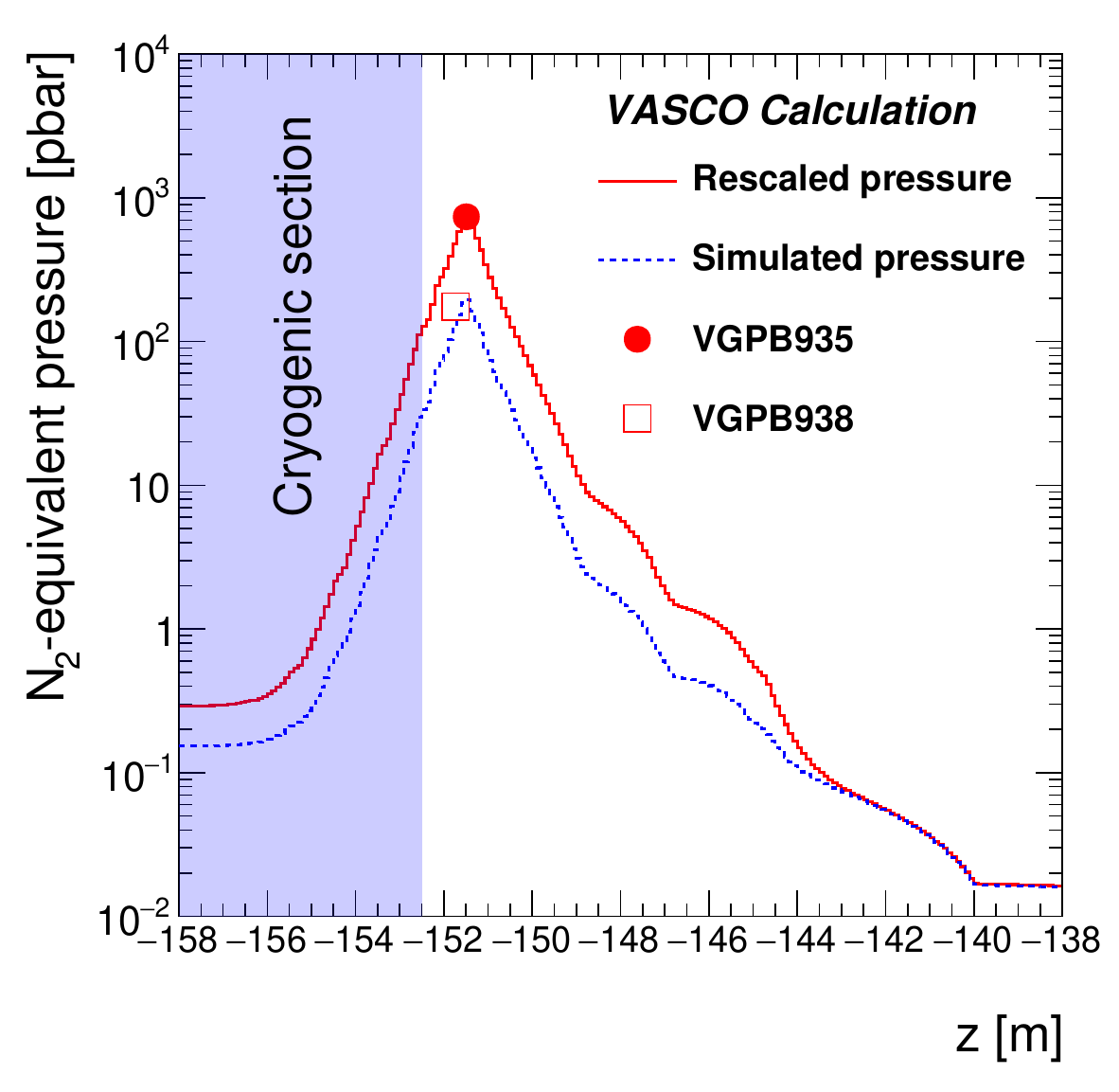}
\caption{}
\label{fig:bumpNeqProfiles151}
\end{subfigure}
}
\caption{Illustration of the profile rescaling to the measured pressures in the case of \beam{2}.
\subref{fig:bumpNeqNewProfilesRescaled19m} \molflow-simulated
pressure profiles at 19\,m assuming $\meffneg=\mexpfor{5}{-3}$
before and after rescaling the calculations to the pressure reported by the vacuum gauge in fill 7212. The neon profile,
shown in the figure, is normalised to a total of \expfor{6}{12} atoms, as is discussed in Section\,\protect\ref{sect:neon}.
\subref{fig:bumpNeqProfiles151} \vasco-calculated
pressure profiles at 151\,m before and after rescaling the value at the
profile maximum to the value reported by the VGPB935 gauge in fill 6239.}
\end{figure}

The pressure profile calculations, described in Section\,\ref{sect:profiles}, are performed for a single outgassing rate of the NEG cartridge.
In order to adjust the calculations to the conditions of a particular fill
the pressure measurement provided by the vacuum gauges is invoked: the calculated density profiles ($\rho_{\mathrm{calculated}}(z)$)
are rescaled, as illustrated in Figure\,\ref{fig:bumpNeqNewProfilesRescaled19m},
such that the calculated $P$ at the gauge position matches the measured $P$, averaged over the period considered for the analysis.
Formally the rescaling can be described as
\begin{equation}
\rho_{\mathrm{rescaled}}(z) = \frac{P_{\mathrm{measured}}({\mathrm{gauge}})}{P_{\mathrm{calculated}}({\mathrm{gauge}})} \rho_{\mathrm{calculated}}(z).
\label{eq:rescaling}
\end{equation}
The right side of Eq.\,(\ref{eq:rescaling}) is equivalent to the pressure ratio and density integral in the denominator of Eq.\,(\ref{eq:epsdef}).
The rescaling brings two advantages:
\begin{enumerate}
\item it accounts for the measured pressure data, and
\item it renders the single pressure profile calculation universal, i.e. the same profile, at a given location, is equally applicable to all fills.
\end{enumerate}
It is important to emphasise that this rescaling procedure is in no sense a correction introduced to compensate for deficiencies
of the calculation. It is purely a factor applied to adjust the assumptions made in the calculations to the observed conditions
of a particular fill.
 
The gauges report the nitrogen equivalent pressure (\pneq). In order to compare this with the calculated density of
residual gas molecules $n_i$, the latter is converted into \pneq at room temperature using the formula
 
\begin{equation}
P_{\mathrm{N_{eq}}} = k_B T \sum_i r_i n_i,
\end{equation}
 
where $k_B$ is the Boltzmann constant and $r_i$ the gas-dependent gauge response taken from Table\,\ref{tab:gasXsec}.
The sum runs over all gas species.
The ratio of measured and calculated \pneq is used in Eq.\,(\ref{eq:rescaling}) to perform the
rescaling of the calculated gas density. The \molflow simulations provide the pressure at the gauge location
and consider only the injected hydrogen, ignoring the other gases, which contribute only to the pedestal pressure.
In this case only one single gas contributes to the pressure change, and the rescaling procedure is straightforward.
The procedure is slightly more complicated at 58\,m and 151\,m where \vasco is used to calculate the pressure profiles.
\vasco takes into account also the other gases, which contribute to the pressure in the absence of the gas injection, so the
breakdown of \pneq into individual gas densities is ambiguous. The injected gas, however, dominates the pressure
in the bump by orders of magnitude. Since it is assumed to be hydrogen, only the partial pressure of hydrogen is
changed in the rescaling procedure.
Contrary to \molflow, \vasco calculations do not consider the gauge explicitly. Thus the beam-line pressure at the $z$-location of the
gauge is compared with the measured value, i.e. $P_{\mathrm{calculated}}({\mathrm{gauge}})$ is replaced
by $P_{\mathrm{calculated}}(z_{\mathrm{gauge}})$ in Eq.\,(\ref{eq:rescaling}).
The pressure profiles for the bump at $z=151$\,m are shown in Figure\,\ref{fig:bumpNeqProfiles151}.
At this location two vacuum gauges, separated by only 25\,cm, provide pressure measurements. A significant difference
between the values from the two gauges is seen. The gauge VGPB935 is located next to the NEG cartridge used to emit
the gas and is therefore considered to provide the more relevant measurement. The mismatch, seen in Figure\,\ref{fig:bumpNeqProfiles151},
between the rescaled profile and the pressure reported by the VGPB938 gauge is discussed in Section\,\ref{sect:sysfar}.

\begin{figure}
\centering
\mbox{
\begin{subfigure}{0.49\textwidth}
\centering\captionsetup{width=0.9\textwidth}
\includegraphics[width=0.95\textwidth]{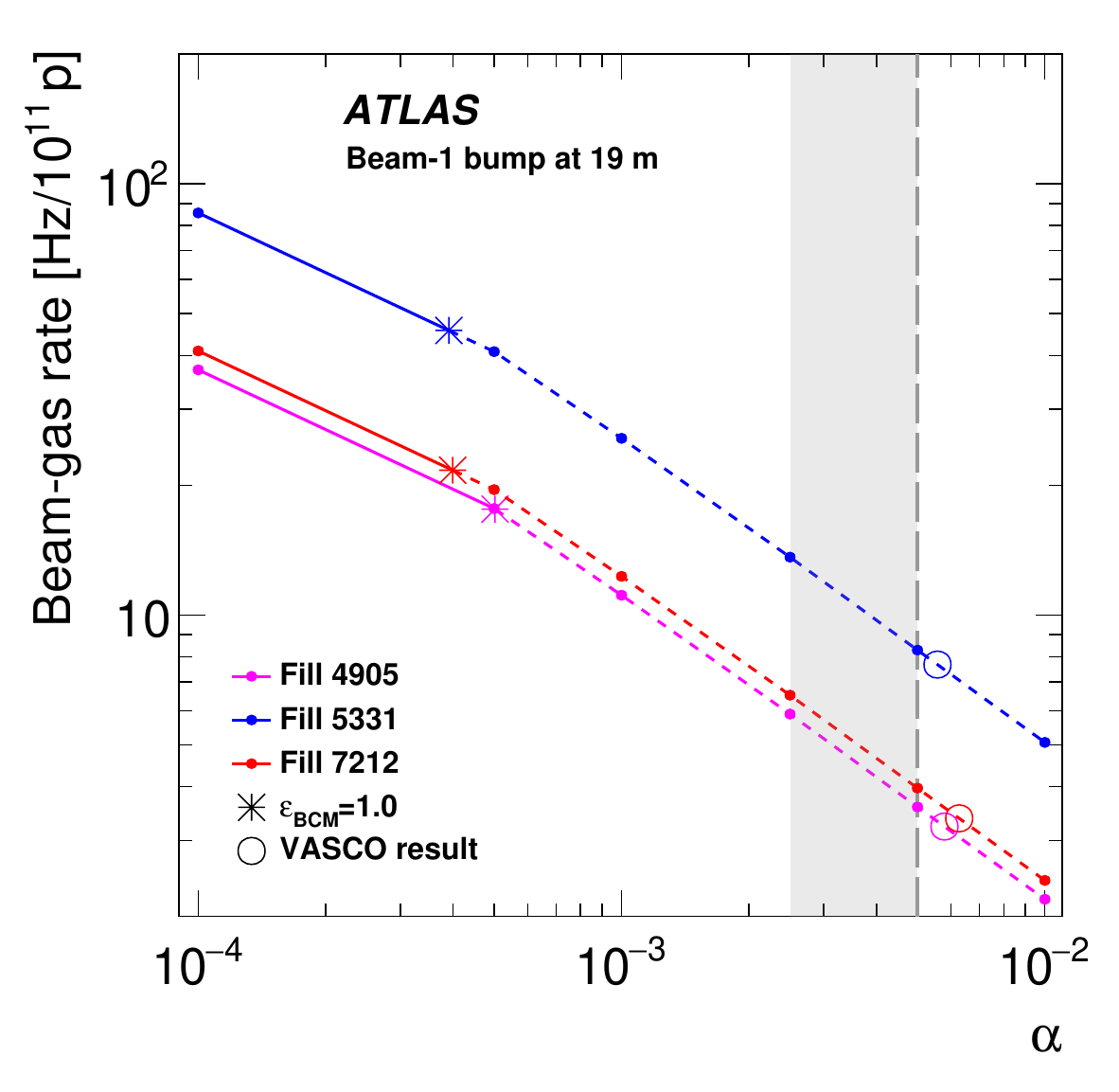}
\caption{}
\label{fig:beamGasRate19B1}
\end{subfigure}
\begin{subfigure}{0.49\textwidth}
\centering\captionsetup{width=0.9\textwidth}
\includegraphics[width=0.95\textwidth]{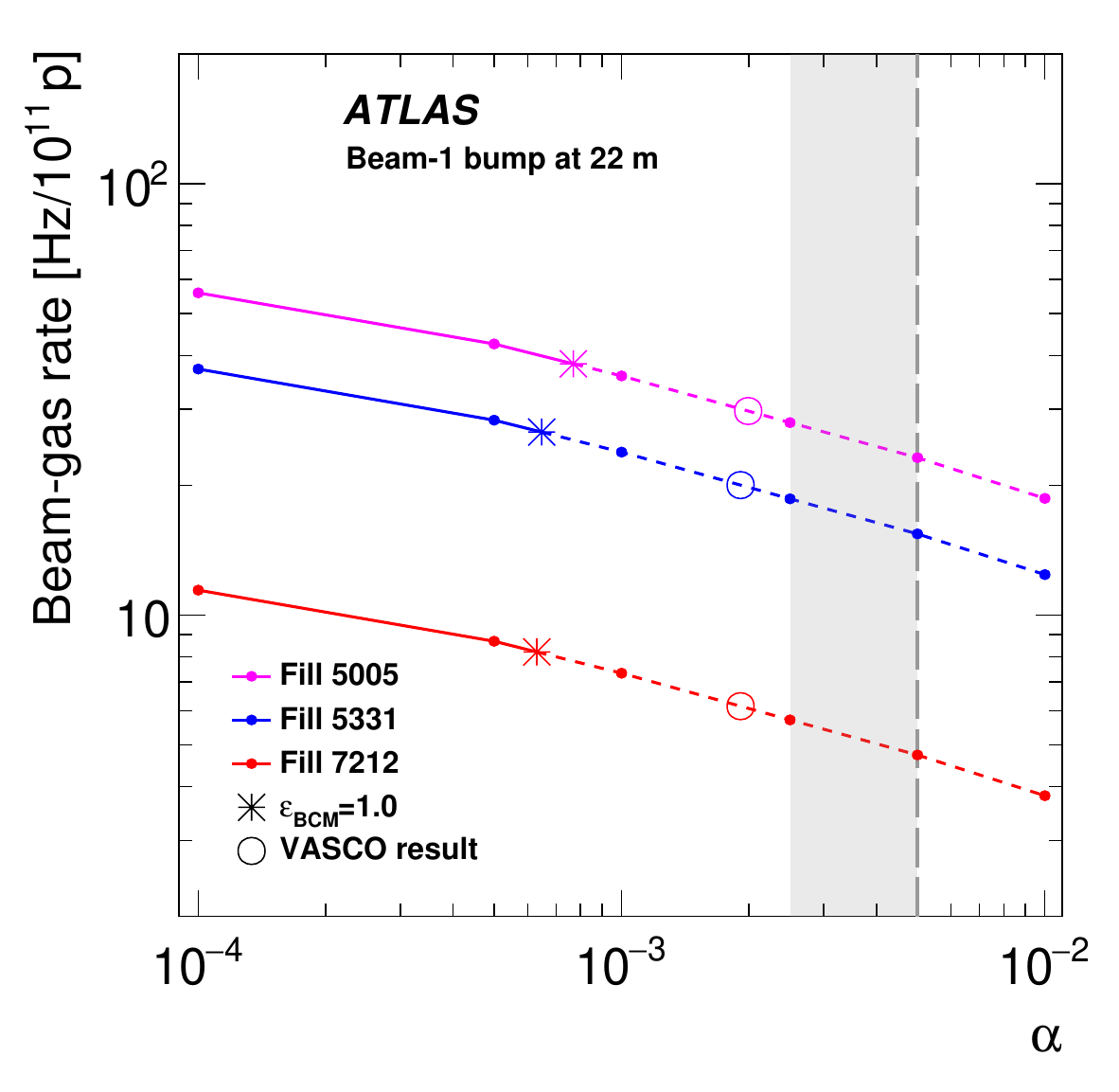}
\caption{}
\end{subfigure}
} \\
\mbox{
\begin{subfigure}{0.49\textwidth}
\centering\captionsetup{width=0.9\textwidth}
\includegraphics[width=0.95\textwidth]{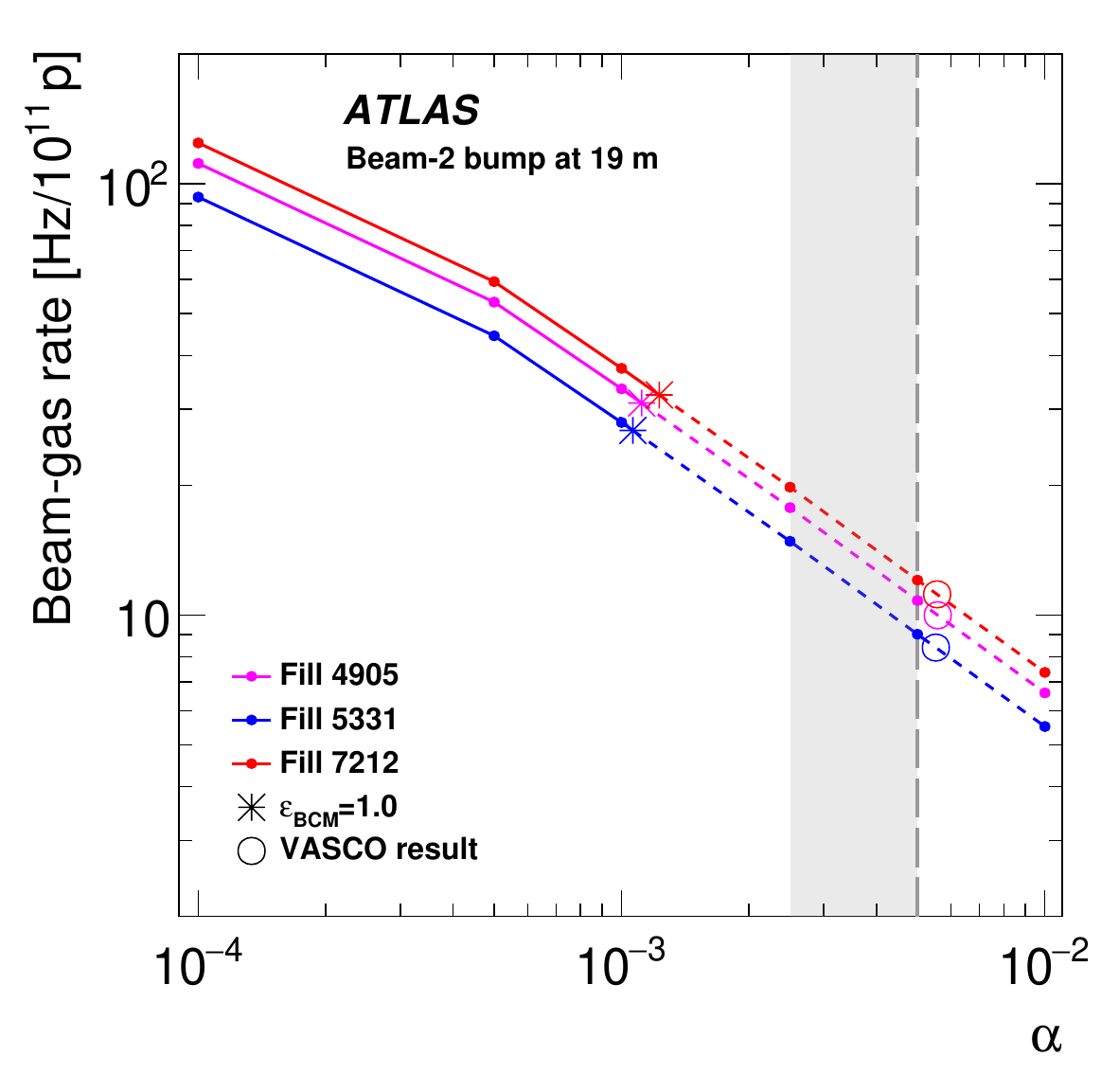}
\caption{}
\end{subfigure}
\begin{subfigure}{0.49\textwidth}
\centering\captionsetup{width=0.9\textwidth}
\includegraphics[width=0.95\textwidth]{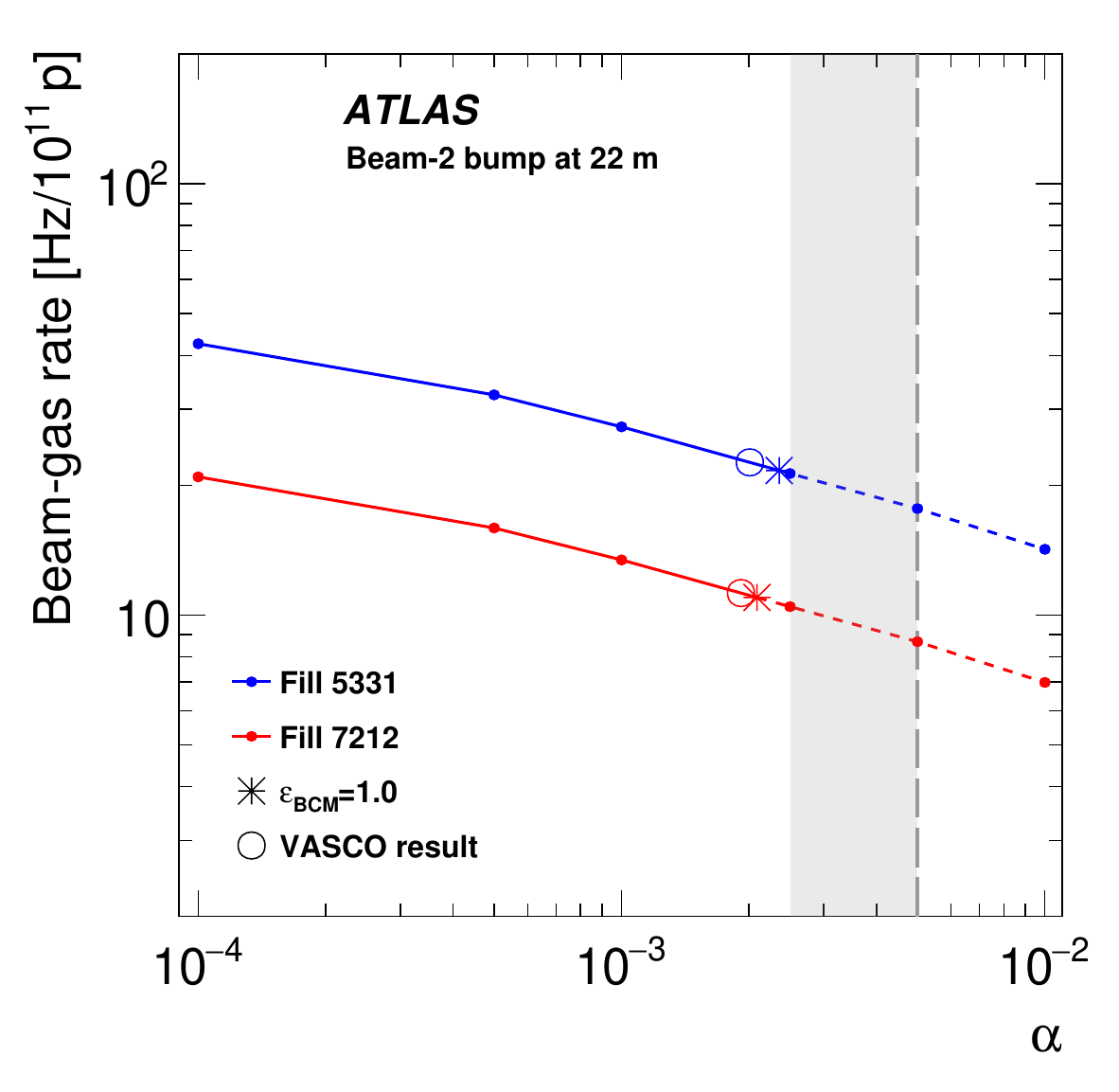}
\caption{}
\end{subfigure}
}
\caption{Beam gas rates in the 19\,m and 22\,m bumps, calculated from the pressure profiles obtained with \molflow when assuming
different values of \effneg. The dashed vertical line indicates the standard assumption for \effneg and the adjacent shade
shows the reasonable range of downward variation. The asterisks indicate the point where \bgr equals the observed
BCM background rate increase in the given fill, i.e. $\meffbcm{z}=1.0$. The dashed part of the lines represents the unphysical region $\meffbcm{z} > 1.0$.
The open circles show, for each individual fill, the \effneg values for which \molflow gives the same \bgr as \vasco.}
\label{fig:beamGasRates}
\end{figure}
 
The calculated pressure profiles are rescaled such that they correspond to the conditions of a particular fill.
Then the integral over the rescaled profiles is used in Eq.\,(\ref{eq:intrate}) to determine the increase of \bgr due to the introduction of the bump.
The results, as a function of \effneg, are shown in Figure\,\ref{fig:beamGasRates} for all nearby, i.e. 19\,m and 22\,m, bumps.
For each fill and pressure bump, the
\bgr that equals the observed BCM background rate increase is indicated on the curves by an asterisk.
For all fills these points agree rather well on a unique \effneg value for any given bump location.
An \bgr lower than the indicated value is unphysical, as it would imply that more beam-gas events are detected than produced.
The \effneg values for which \molflow gives the same \bgr as \vasco are indicated by the open circles.
With the exception of the 22\,m pressure bump in \beam{2}, they all fall into the unphysical region.
 
\subsection{Determination of BIB detection efficiencies}
\label{sect:bibeff}

Tables\,\ref{tab:bcmBkgdEff} and \ref{tab:jetBkgdEff} give, for each individual fill, the BIB detection efficiencies that are obtained as the
ratio between observed additional background events and the estimated \bgr at the pressure bump location.
It is expected that \effneg is around \expfor{5}{-3} for gases, including H$_2$, which are pumped by the NEG coating of the beam-pipe wall.
Therefore the \molflow profiles corresponding to $\meffneg=\mexpfor{5}{-3}$ are used for the nearby pressure bumps.
The pedestals of \bgr and the background event rate, in the absence of a pressure bump, are determined
from the grey-shaded areas in Figure\,\ref{fig:fits} (and equivalent figures for other fills) and are subtracted before
forming the ratio. Despite varying beam conditions, the efficiencies in different fills are consistent to better
than 30\%. The only exception is the \effbcm{58\,\mathrm{m}} for \beam{2}, which is about 60\% higher in fill 6239 than in 4905.

\begin{table}
\centering
\caption{Ratio between the BCM background rate and the estimated \bgr due to the pressure bump.
A '---' means that no bump was introduced in that position in the given fill.
The uncertainties are based on the LB-to-LB variation, i.e. are mostly statistical but reflect also a possible non-linearity
between the pressure and background measurements.
The values for the 19\,m and 22\,m bumps correspond to $\meffneg=\mexpfor{5}{-3}$. As discussed in the text, as efficiencies they are
unphysical, but they serve here only to indicate the consistency of data obtained in different fills.}
\begin{tabular}{l||c|c|c|c}
& \multicolumn{4}{c}{Position, \beam{1}} \\
LHC Fill  & 19\,m                   & 22\,m                      & 58\,m                       & 151\,m                       \\ \hline
4905      & $4.9\pm 0.4$            & ---                        & \expfor{(2.84\pm 0.03)}{-2} & ---                          \\
5005      & ---                     & $1.65\pm 0.02$             & \expfor{(2.99\pm 0.02)}{-2} & ---                          \\
5331      & $5.5\pm 0.2$            & $1.7\pm 0.2$               & \expfor{(2.63\pm 0.03)}{-2} & \expfor{(6.34\pm 0.08)}{-3}  \\
6239      & ---                     & ---                        & \expfor{(3.13\pm 0.03)}{-2} & \expfor{(7.29\pm 0.04)}{-3}  \\
7212      & $5.5\pm 0.3$            & $1.7\pm 0.2$               & ---                         & ---                          \\ \hline
\multicolumn{5}{c}{} \\
& \multicolumn{4}{c}{Position, \beam{2}} \\
LHC Fill  & 19\,m           & 22\,m           & 58\,m                       & 151\,m                      \\ \hline
4905      & $2.9\pm 0.3$    & ---             & \expfor{(3.29\pm 0.03)}{-2} & ---                         \\
5005      & ---             & ---             & \expfor{(4.07\pm 0.03)}{-2} & ---                         \\
5331      & $3.0\pm 0.4$    & $1.23\pm 0.07$  & \expfor{(4.7\pm 0.1)}{-2}   & \expfor{(4.88\pm 0.06)}{-3} \\
6239      & ---             & ---             & \expfor{(5.46\pm 0.03)}{-2} & \expfor{(6.47\pm 0.07)}{-3} \\
7212      & $2.69\pm 0.08$  & $1.27\pm 0.01$  & ---                         & ---                         \\ \hline
\end{tabular}
\label{tab:bcmBkgdEff}
\end{table}

\begin{table}
\centering
\caption{Ratio between the fake-jet background rate and the estimated \bgr due to the pressure bump.
A '---' means that no bump was introduced in that position in the given fill.
The uncertainties are based on the LB-to-LB variation, i.e. are mostly statistical but reflect also a possible non-linearity
between the pressure and background measurements.
No ratios are given for the 19\,m and 22\,m bumps, where no excess of fake jets could be resolved from the pedestal background.}
\begin{tabular}{l||c|c||c|c}
&\multicolumn{2}{c||}{Position, \beam{1}}&\multicolumn{2}{c}{Position, \beam{2}} \\
LHC Fill & 58\,m                    & 151\,m                      &  58\,m                       & 151\,m                      \\ \hline
4905     &\expfor{(2.6\pm 0.6)}{-5} & ---                         &  \expfor{(5.3\pm 0.8)}{-5}   & ---                         \\
5005     &\expfor{(2.7\pm 0.4)}{-5} & ---                         &  \expfor{(6.3\pm 0.5)}{-5}   & ---                         \\
5331     &\expfor{(1.5\pm 0.8)}{-5} & \expfor{(1.4\pm 0.1)}{-4}   &  \expfor{(6.7\pm 0.7)}{-5}   & \expfor{(1.20\pm 0.08)}{-4} \\
6239     &\expfor{(2.6\pm 0.5)}{-5} & \expfor{(1.49\pm 0.07)}{-4} &  \expfor{(7.0\pm 0.5)}{-5}   & \expfor{(1.46\pm 0.06)}{-4} \\ \hline
\end{tabular}
\label{tab:jetBkgdEff}
\end{table}
 
\begin{figure}
\centering
\includegraphics[width=0.95\textwidth]{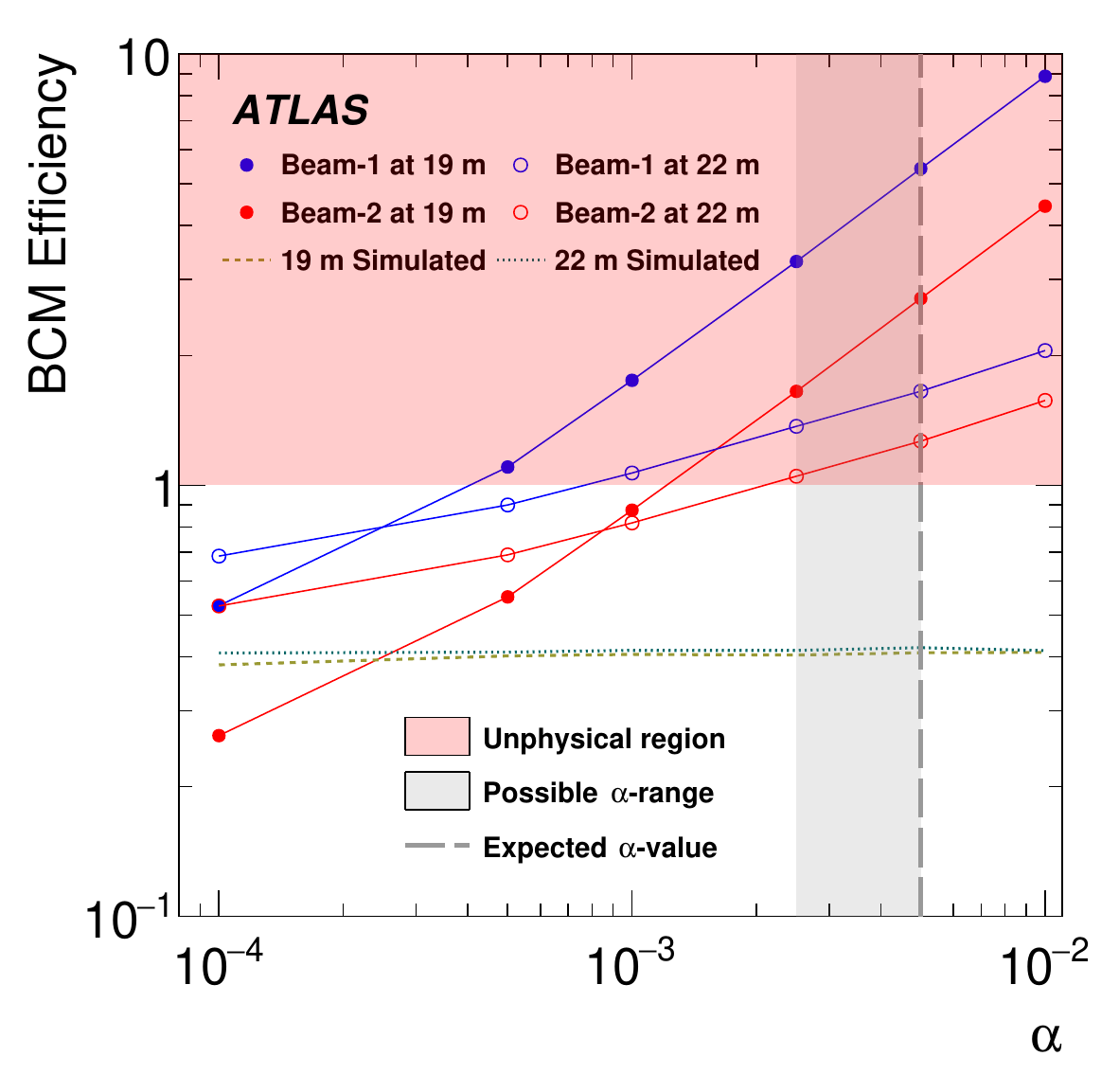}
\caption{Beam-gas induced background detection efficiency of the BCM as a function of \effneg
assumed in the \molflow simulations. The dashed vertical line indicates the standard assumption for \effneg, and the adjacent grey shade
shows the reasonable range of downward variation. The values are averaged over all fills in which a pressure bump at the indicated location was
introduced, i.e. those shown in Figure\,\protect\ref{fig:beamGasRates}. In all cases the error bars due to the fill-to-fill variation are smaller than the symbols.
The red-shaded area indicates the unphysical region where the number of detected background events would exceed the number of
produced beam-gas collisions. The simulated \effbcm{z} should not depend on \effneg, but a weak dependence arises because the $z$-locations of the
events are sampled from the pressure profiles shown in Figure\,\protect\ref{fig:bumpNeqNewProfiles}.
Especially at 19\,m the average of this distribution moves away from the IP when \effneg increases.
}
\label{fig:effGraphs}
\end{figure}

The \effbcm{19\,\mathrm{m}} and \effbcm{22\,\mathrm{m}} values given in Table\,\ref{tab:bcmBkgdEff} exceed unity by a significant amount.
This is obviously unphysical and therefore suggests that the density of gas traversed by the beam is severely underestimated.
Figure\,\ref{fig:beamGasRates} indicates that \effneg should
be much smaller than assumed in order to increase \bgr to the same level as the observed background.
This is even better illustrated by Figure\,\ref{fig:effGraphs} where \effbcm{z}, averaged over all relevant fills, is plotted as a function of \effneg.

Several interesting observations can be made from Figure\,\ref{fig:effGraphs}:
\begin{itemize}
\item for both locations, \effbcm{z} for \beam{1} is systematically higher than for \beam{2},
\item the dependence of \effbcm{z} on \effneg is much stronger for the 19\,m than for the 22\,m pressure bumps,
\item the \fluka simulations predict almost equal \effbcm{z} for beam-gas events at the 19\,m and 22\,m locations.
\end{itemize}
 
The last observation is particularly interesting: while the absolute rates obtained from the simulations have substantial uncertainties, many of these
cancel from the ratio $\meffbcm{19\,{\mathrm m}}/\meffbcm{22\,{\mathrm m}}$, notably those associated with the BCM hit and trigger simulation,
beam-gas event generation and accuracy of the beam-line geometry description at $z<19$\,m. Therefore this ratio is considered to be a much more
reliable prediction than the absolute value of \effbcm{z}, obtained from the simulations. This means that
$\meffbcm{19\,{\mathrm m}} = \meffbcm{22\,{\mathrm m}}$ can be considered a rather strong constraint, although it arises from the \fluka simulations.
 
Figure\,\ref{fig:effGraphs} is based on the \molflow simulations, assuming that only hydrogen is released from the NEG cartridge.
The points of intersection of the curves suggest that \effneg must be slightly below $10^{-3}$ for \beam{2} and about three
times lower for \beam{1} in order to obtain $\meffbcm{19\,{\mathrm m}} = \meffbcm{22\,{\mathrm m}}$ for a given beam.
 
Table\,\ref{tab:averageomegas} reveals that the ratio between BCM background and measured pressure is highest for the 19\,m bump
on the \beam{1} side, lowest for the 22\,m bump on the \beam{2} side and intermediate for the two other nearby locations.
This is qualitatively consistent with the order of the curves in Figure\,\ref{fig:effGraphs} at $\meffneg=\mexpfor{5}{-3}$, i.e. the purely
data-driven results also indicate a significant difference for the two sides of ATLAS.
 
An independent verification of the calculated pressure profiles is obtained by monitoring the pressure increase at 19\,m when gas is injected at 22\,m
and vice versa. Since \molflow provides the pressure at the gauges, this ratio can also be derived from the pressure simulations. A comparison
of these measured and simulated ratios reveals that the best agreement is obtained for \effneg$<10^{-3}$. This value,
as well as the fact that it is found to be lower for \beam{1} than \beam{2}, is in good agreement with Figure\,\ref{fig:effGraphs}.
 
Such low values of \effneg, and a large difference on the two sides of the IP, do not seem
plausible\,\cite{Lanza:IPAC2014-WEPME044}. One possible explanation for the apparently low and non-uniform \effneg
is that in addition to hydrogen, another gas, not considered in the \molflow simulations, was emitted from the NEG cartridge.
 
\subsection{Possible neon contribution}
\label{sect:neon}
 
During each annual technical stop, the experimental beam pipe within $|z|<22$\,m is filled with ultra-pure neon to atmospheric
pressure\,\cite{Baglin:IPAC11-TUPS017} in order to protect the pipe walls from accidental collapse during maintenance of the ATLAS experiment.
Although noble gases are not pumped by NEGs, it is possible that some
neon atoms reside on the large surface of very porous material of the NEG cartridge\,\cite{Bregliozzi:2010kva}. Upon heating,
this neon would be released alongside the hydrogen.
 
Such a two-gas hypothesis could provide a plausible qualitative explanation for the apparently low \effneg: \effneg could well be
close to the expected value of \expfor{5}{-3} for H$_2$, but since it is zero for neon, a sizeable amount of emitted neon could result
in an apparent\footnote{Apparent \effneg refers to Figure\,\protect\ref{fig:effGraphs}, i.e. it is the \effneg that
produces the observed \effbcm{z}.} \effneg around, or even below, $10^{-3}$.
 
There are two reasons why even a small amount of neon can have a significant impact on \effbcm{z}:
\begin{enumerate}
\item Table\,\ref{tab:gasXsec} shows that the inelastic cross section for $p$-Ne scattering is five times larger than for $p$-H$_2$ but for
the same density of molecules the vacuum gauge will report two times higher pressure for hydrogen than for neon. Thus, for the
same measured pressure, \bgr in pure neon is about ten times higher than in pure H$_2$.
In addition, the \fluka simulations indicate that, due to the higher particle multiplicity, the efficiency of the  BCM to detect $p$-Ne events
is about 15\% higher than for $p$-H collisions.
\item Since neon is not removed by the NEG coating of the beam pipe, i.e. \effneg$=0$, its distribution can be expected to be much wider in $z$ than for hydrogen.
This is the case, especially, for the 19\,m location, where the neon flow towards the IP is restricted only by the ion pump at $|z|=3.7$\,m.
As a consequence the ratio between the total amount of gas traversed by the beam and the pressure measured at the gauge is likely to be much larger than for hydrogen.
\end{enumerate}
 
Hydrogen is quickly absorbed by the NEG coating of the beam pipe, but if neon is emitted from the NEG at $z=-19$\,m it
will form a wide bump between $z\approx -8$\,m and $z\approx -19$\,m, as indicated in Figure\,\ref{fig:bumpNeqNewProfilesRescaled19m}. In order
to be removed, this neon has to diffuse to the ion pumps, so it will reside in the beam pipe for a longer time than hydrogen.
The neon profile shown in Figure\,\ref{fig:bumpNeqNewProfilesRescaled19m} is obtained from \molflow, but since
the rate of neon emission is not known, the (possibly time-dependent) mixing ratio of hydrogen and neon cannot be derived by combining
the pressure data with the simulations. Thus only an approximate range of \bgr can be determined for the mixture, as is discussed in Section\,\ref{sect:sysnear}.
 
In the $|z|$-region where the neon is contained, according to the \molflow simulations, \effbcm{z} is roughly constant\,\cite{DAPR-2017-01}.
Thus Eq.\,(\ref{eq:intrate}) can be used to estimate that $\sim\mexpfor{6}{12}$ neon atoms are sufficient to lower \effbcm{19\,\mathrm{m}}
below unity even if \effneg is \expfor{5}{-3} for hydrogen.
In order to verify if neon could be a viable explanation for the observations, dedicated laboratory tests were carried out by the CERN vacuum group\,\cite{neon-labtest}.
In these tests a NEG cartridge was exposed to pure neon at atmospheric pressure for three days before re-establishing the vacuum. Then the heating on
the NEG cartridge was turned on and released gas composition was monitored with a residual gas analyser and the pressure by a hot cathode vacuum gauge.
The tests indicated that about $\sim\expfor{5}{13}$ neon atoms were released from the cartridge over a 55\,minute period of heating. This result has
large uncertainties but it exceeds by almost an order of magnitude the amount needed to explain the present results.
A detailed time dependence of the neon emission, however, is not determined in these tests. It is not clear if the neon is emitted gradually such that
the total amount would scale with the heating time or predominantly soon after the heating of the NEG cartridge starts.
The early rise and overshoot of the background in Figure\,\ref{fig:bcmb1fit7212}, with respect to Eq.\,(\ref{eq:pfit}), suggests that a significant
fraction of the neon is released promptly when the heating of the NEG starts: \omegf{22\,{\mathrm m}} is derived from Eq.\,(\ref{eq:omega}) after this
overshoot. If hydrogen dominates during that time, \omegf{22\,{\mathrm m}} will underestimate the BIB from the earlier gas composition if that is
richer in neon.
 
No such overshoot is seen for the 19\,m bump in Figure\,\ref{fig:bcmb1fit7212}. A possible explanation is provided by the
longer residence time, discussed above: compared to the 22\,m bump, the total amount of neon varies less over time for the 19\,m bump, i.e.
neon contributes also during the time-interval used to determine \omegf{19\,{\mathrm m}}.
This argument is also consistent with the discrepancy between the background rate and the pressure parameterisation seen in Figure\,\ref{fig:bcmb1fit7212}
after the 19\,m bump: when the NEG heating is terminated, neon probably still resides far from the ion pumps, e.g. around $|z| \approx 10 - 15$\,m,
and causes an excess of beam-gas events.
 
Another observation that supports the hypothesis of neon emission is that a small pressure increase is observed on the
other side of the TAS if the gas is injected at either 19\,m or 22\,m.
The \molflow profiles shown in Figure\,\ref{fig:bumpNeqNewProfilesRescaled19m} indicate that
when the NEG at $z=-19$\,m is heated, practically none of the emitted hydrogen can traverse the NEG-coated TAS if $\meffneg=\mexpfor{5}{-3}$
is assumed.
However, the neon pressure on the opposite side, at $z=-22$\,m, is non-zero and sufficient to cause the pressure gradient
through the TAS to appear equivalent to \mbox{\effneg$\lesssim 10^{-3}$}.
Without the annular pump at 3.7\,m, the neon from an injection at $|z|=19$\,m would spread to the other side
of the IP, cause some increase of BIB in the opposite beam, and thereby provide firm evidence for the two-gas hypothesis. Unfortunately the pump
prevents this, as evidenced by the drop of the \molflow-simulated neon pressure for $|z|\lesssim 10$\,m in Figure\,\ref{fig:bumpNeqNewProfilesRescaled19m}.
 
Although these arguments are consistent with the neon emission hypothesis, and the laboratory measurements indicate the
right order of magnitude of released neon atoms, they cannot be taken as a firm proof for the neon hypothesis but merely as an indication that it is
a likely explanation for the unphysically high \effbcm{19\,{\mathrm m}} and \effbcm{22\,{\mathrm m}} values in Table\,\ref{tab:bcmBkgdEff}.
A firm verification would require further tests with a detailed analysis of the residual gas composition.
 
One feature, however, remains unexplained: the injection of neon takes place on the incoming \beam{1} side,
close to the 22\,m NEG cartridge. Although this creates an asymmetry between the sides, it is unclear why more neon seems to be released from
the NEG cartridge on the \beam{1} side, although the entire ATLAS beam pipe is filled with neon at atmospheric pressure for several
months during the annual technical stop.
 
\section{Discussion of the level of systematic uncertainties}
\label{sect:syserrors}
 
\subsection{Uncertainties affecting the measured values}
\label{sect:sysmeas}
 
\subsubsection{Common effects}
\label{sect:syscommon}
 
The statistical errors, reported, e.g., in Tables\,\ref{tab:bcmBkgdEff} and \ref{tab:jetBkgdEff}, are in general negligible
compared to systematic uncertainties, which comprise several contributions. Given the nature of the problem and,
especially, the incomplete knowledge of the injected gas composition, only a qualitative discussion of the
magnitude of the systematic uncertainties is provided.
 
In some cases, when the observed BIB is not well correlated with the pressure change, the $\omega$-factor obtained from
Eq.\,(\ref{eq:omega}) depends on the averaging range. Such deviations appear, especially, at the rising and falling edges of
the pressure bumps.
For instance, if the averaging range for the 22\,m pressure bump in Figure\,\ref{fig:bcmb1fit7212} would be moved earlier
a higher value would be obtained.
The effect of small variations of the averaging ranges was investigated and, as long as the rising and falling edges were
excluded, the effect on the results was found to be negligible.
A possible bias caused by erroneously excluding periods that appear suspicious would be revealed when averaging over
several fills. The uncertainties reported in Table\,\ref{tab:averageomegas} are dominated by the
fill-to-fill variation, and they are all of the order of 10\% or less. Thus, any significant selection bias
in the analysed fills can be excluded. As the same averaging periods are used in Section\,\ref{sect:abseff} to
determine the absolute efficiencies, these fill-to-fill uncertainties apply also to \effbcm{z} and \effjet{z} and are included
in their statistical errors.
 
A systematic uncertainty that affects both the $\omega$ factors and absolute efficiencies is the choice and accuracy of the
vacuum gauge used for the pressure measurement.
For the 19\,m and 22\,m bumps the gauges that provided the most consistent output
in all relevant fills and for both beams were selected. Using an alternative gauge
has an average effect of 25\% at the 22\,m location and no effect at all at 19\,m. This is consistent with
the expected gauge uncertainty of $<$10\% for VGI gauges, used at 19\,m and $<$30\% for VGPB gauges, used at 22\,m.
A gauge uncertainty of $<$30\% is assumed also for the VGPB gauges used at the 58\,m and 151\,m locations.
 
A minor contribution to the uncertainty of the absolute efficiencies arises from the cross sections listed in
Table\,\ref{tab:gasXsec}. These affect \bgr through Eq.\,(\ref{eq:intrate}) and thereby the measured values of
\effbcm{z} and \effjet{z}. CH$_4$, CO and CO$_2$ contribute only to the pedestal beam-gas rate, which cancels from
the final result. The uncertainty associated with the possibly present neon is entirely dominated by the
lack of knowledge of the amount of that gas, as discussed in Section\,\ref{sect:sysnear}. Thus only the cross-section uncertainty of
$p$-H$_2$ is relevant. For this the values in Table\,\ref{tab:gasXsec} can be compared with the proton--proton cross section
from Ref.\,\cite{Workman:2022ynf}, which is about 38\,mb at 6.5\,\TeV laboratory momentum. This is consistent with
Table\,\ref{tab:gasXsec}, and the fit in Ref.\,\cite{Workman:2022ynf} suggests that the uncertainty is less than 5\%, which
is negligible compared to other contributions.
 
\subsubsection{Nearby pressure bumps at 19\,m and 22\,m}
\label{sect:sysnear}
 
It is evident from Section\,\ref{sect:neon} that the possible emission of another gas, notably neon, in addition to
hydrogen, constitutes a major uncertainty impacting the measured values of \effbcm{19\,{\mathrm m}} and \effbcm{22\,{\mathrm m}}.
Within this two-gas hypothesis not enough information is available to establish anything other than an estimated range.
Figure\,\ref{fig:effGraphs} indicates that this range is rather limited: clearly \bgr cannot exceed the rate
of detected additional events, i.e. the requirement $\meffbcm{z} \leq 1$ defines the upper limit.
Other possible sources of spurious BCM background triggers, e.g. noise, ghost collisions or afterglow,
are not affected by the introduction of the pressure bump and therefore are removed by the pedestal subtraction and by rejecting
data in periods that are obviously affected by noise bursts or beam instabilities. The good consistency between individual fills,
seen in Table\,\ref{tab:bcmBkgdEff}, indicates that this cleaning is efficient and no significant anomalies remain in the data.
 
A lower bound for \effbcm{z} is more complicated to establish since in the two-gas hypothesis it arises from the maximum amount
of neon that could contribute to the background.
Figure\,\ref{fig:bumpNeqNewProfilesRescaled19m} indicates that neon emitted at 19\,m would create a wide pressure distribution extending from $|z|\approx 8$\,m to
the TAS, which coincides with the region where \effbcm{z} is roughly constant. The volume of the beam pipe in this
$z$-range is $\sim$70\,litres. Thus every $10^{12}$ neon atoms would increase \bgr by about 7\,\HzPerB.
This can be compared with the curves in Figure\,\ref{fig:beamGasRates}: the highest \bgr required for $\meffbcm{19\,{\mathrm m}} = 1$ is
$\sim$50\,\HzPerB as indicated by an asterisk in Figure\,\ref{fig:beamGasRate19B1} for fill 5331.
The laboratory tests indicate that up to \expfor{5}{13} neon atoms could be emitted by the NEG cartridge upon heating.
Taking this as the maximum amount that could contribute would increase \bgr by
about 350\,\HzPerB. This adds to the $\sim$10\,\HzPerB of $p$-H events indicated in Figure\,\ref{fig:beamGasRate19B1}
if \effneg is assumed to be in the expected range $\sim\mexpfor{2.5-5}{-3}$.
Reaching $\meffbcm{19\,{\mathrm m}} = 1$ requires \bgr\,=\,50\,\HzPerB, which means that 360\,\HzPerB corresponds to
$\meffbcm{19\,{\mathrm m}} \gtrsim 0.15$.
In all other cases 350\,Hz of additional rate would have a larger effect.
Since the \fluka prediction of $\meffbcm{19\,{\mathrm m}}=\meffbcm{22\,{\mathrm m}}$ is considered reliable and as there is no reason
to believe that \effbcm{z} would be different for the two beams, the highest lower limit of \effbcm{z} applies to all cases.
This implies that for the other pressure bumps the role of the neon must be smaller, which is consistent with Figure\,\ref{fig:effGraphs}.
 
\subsubsection{Distant pressure bumps at 58\,m and 151\,m}
\label{sect:sysfar}
 
For both the 58\,m and the 151\,m pressure bumps, \effbcm{z} is far below unity, i.e. this hard limit does not provide any constraint.
On the other hand, these positions are not affected by the annual neon injection, so the assumption that hydrogen
is emitted predominantly is justified.
 
Only \vasco results are available for the distant pressure bump locations. The accuracy of these can be estimated by comparing
\vasco calculations with the more detailed \molflow simulation results for the nearby bumps.
Figure\,\ref{fig:beamGasRates} shows that \bgr values obtained from \vasco are rather consistent with \molflow
predictions when $\meffneg\approx \mexpfor{5}{-3}$ is assumed for the latter; for the 19\,m pressure bumps \vasco gives an only 8\% higher \bgr
while for the 22\,m position it is 23\% lower. This good agreement seen for the nearby bumps provides confidence
in the pressure profiles produced by \vasco for the distant bumps.
 
Since the maximum pressure of the \vasco-calculated profile is scaled to the local gauge reading, an uncertainty arises from how
well the measured pressure represents the maximum at the beam line.
At both distant locations two nearby vacuum gauges are available. One is close to the NEG cartridge used for
the gas injection while the other is about 25\,cm away on the other side of a sector valve.
Both gauges are on chimneys, recessed from the beam line. In the absence of the pressure bump they report very consistent values.
When the gas is injected the pressure measurements differ significantly.
At both locations the gauge next to the NEG is used and can be assumed to provide an upper estimate for the pressure at the beam line.
At 58\,m the neighbouring gauge remains reasonably consistent with the steep edge of the pressure profile but
for the 151\,m bump, shown in Figure\,\ref{fig:bumpNeqProfiles151}, the pressures reported by the two adjacent gauges differ
by more than an order of magnitude. This effect is systematically seen for all fills and both
beams at the 151\,m location.
 
In the extreme case of using VGPB938 as the absolute pressure reference at 151\,m, \bgr would increase by an order
of magnitude. The VGPB935 gauge, however, is assumed to provide a better estimate of the pressure at the beam line, although it
could slightly overestimate \bgr:
the \molflow results shown in Figure\,\ref{fig:bumpNeqNewProfiles} suggest
that the gauge on the same chimney as the NEG cartridge overestimates the beam-line pressure by 30--150\%, depending on \effneg.
The equipment configuration at the nearby locations, however, is very different from that at 151\,m. In particular all chimneys have
a NEG coating, which is not the case at 151\,m. Thus the relevant difference should be that seen for the smallest \effneg-values, i.e.
$\sim$50\% at most.
 
It is evident from Figure\,\ref{fig:bumpNeqProfiles151} that such a 50\% correction is not nearly sufficient to establish agreement with the VGPB938 pressure measurement.
The origin of this discrepancy is most likely in the local configuration of the vacuum equipment:
the two gauges are located on different chimneys and the one housing VGPB938 has a much smaller diameter than the beam pipe and
therefore lower conductance. Thus the pressure drop along the chimney is larger than along the beam line. This is also consistent
with the difference appearing only when the gas is injected. This explanation implies that the pressure measured by VGPB938 is
not relevant, i.e. no additional uncertainty is derived from it.
 
An alternative explanation for the discrepancy between the two gauges is that it is due to an overestimate of the profile width:
if the slope is underestimated on the side of the cold magnet, the pressure at VGPB938 is overestimated. As the pressure
scale in the plot is logarithmic, a steeper slope, passing through the two measurements, would reduce \bgr by only $\sim$30\%.
 
Both uncertainties, discussed above, are asymmetric and tend in the direction of \bgr being overestimated. Although the discrepancy
between the two neighbouring gauges is significantly smaller at the 58\,m location, it seems prudent to assign the same uncertainties there
as at 151\,m.
 
Table\,\ref{tab:sysErrSummary} summarises the measurement-related uncertainties discussed above and in Section\,\ref{sect:sysnear}.
For the nearby pressure bumps the unknown neon contribution dominates all other effects.
For the distant bumps it is not obvious how the uncertainties should be combined. The gauge accuracy can be assumed to represent a Gaussian
uncertainty distribution, centred at the measured value. The asymmetric uncertainties, however, indicate the possible bias by which
the amount of gas might be overestimated. All these uncertainty components are independent.
Due to their different natures it appears justified to add the upper and lower limits separately in quadrature. This yields
a combined uncertainty range from $-30$\% to 70\%, which means that the analysis is slightly biased towards underestimating the
BIB detection efficiency.
 
\begin{table}
\centering
\caption{Summary of uncertainties affecting the determined values of \effbcm{z} and \effjet{z}. The first two
contributions apply also to the \omegf{z_i} values discussed on Section\,\protect\ref{sect:BkgdVsP}. The fill-to-fill variation
is included in the statistical errors of Tables\,\ref{tab:averageomegas},~\ref{tab:bcmSimEfficiency} and \ref{tab:jetSimEfficiency}.
The gauge accuracy is best at 19\,m where a VGI-type gauge is used. For the effective \effneg no central value is determined; instead
the uncertainty is expressed as a range of \effbcm{z} with the lower limit corresponding to the largest realistic contribution of neon.
For the nearby bumps the last two uncertainties, which are related to the bump shape, are considered to be part of the effective \effneg
uncertainty.}
\begin{tabular}{l|c|c|c}
$|z$|-position              &  19\,m        & 22\,m      &  58\,m \& 151\,m   \\ \hline
Proton-H$_2$ cross section  &  5\%          & 5\%        &  5\%               \\
Fill-to-fill                &  10\%         & 10\%       &  10\%            \\
Gauge uncertainty           &  10\%         & 30\%       &  30\%            \\
Effective \effneg           &  $0.15-1.0$   & $0.15-1.0$ &  n.a.              \\
Gauge to profile maximum    &  ---          & ---        &  $0-50$\%    \\
Bump shape                  &  ---          & ---        &  $0-30$\%    \\ \hline
\end{tabular}
\label{tab:sysErrSummary}
\end{table}
 
\subsection{Uncertainties of the \fluka simulations}
 
Systematic uncertainties affecting the \fluka simulations arise from the accuracy of modelling the beam-gas interactions, the geometry
description, the shower transport and the signal reconstruction in the detectors. All of these are difficult to estimate and disentangle
since they depend very much on the particular situation.
 
The formation of the BCM background trigger signature is based on a simple coincidence of two energy depositions in
thin diamond sensors. The main uncertainty is in the threshold ionisation to record a hit. Previous simulation studies\,\cite{DAPR-2017-01}
indicate that within a reasonable range of thresholds the rate varies only by a few percent, which is negligible compared with
the overall uncertainties of the shower simulations.
 
The jet reconstruction is approximated in the \fluka simulations by an evaluation of local energy
deposition\,\cite{DAPR-2017-01}. It was shown that the energy sum depends only weakly on the volume over which the integration
of deposited energy is performed. It is, however, not possible to determine the overall accuracy of this method, since it is merged with
all other simulation uncertainties, which are discussed below.
 
Extensive comparisons\,\cite{GENR-2019-02} have indicated that the simulations agree with
radiation background measurements within 20--30\% when the primary source is $p\!p$ collisions at the IP. It is not evident to what extent
this applies to BIB simulations, where the showers extend over a much larger distance and have quite different characteristics.
 
The BCM records BIB very close to the beam pipe, where the particle fluences originate mainly from interactions in the
beam-pipe material and other accelerator elements. Thus it can be assumed that the dominant uncertainty arises from the
accuracy with which the beam line is described in the simulation models.
Secondary particle yields and angular distributions
of the primary beam-gas interaction are another significant source of uncertainty for the BCM rates.
For the nearby pressure bumps the uncertainties evaluated for $p\!p$ collisions at the IP should give an indication of the error margin.
Based on this a 30\% uncertainty is assigned to the simulated \effbcm{19\,{\mathrm m}} and \effbcm{22\,{\mathrm m}} values.
 
A minor asymmetric uncertainty, affecting the nearby bumps, arises from the difference of the $p$-H and $p$-Ne interactions. The latter
produce about 15\% more BCM background triggers. The simulations, assuming only $p$-H collisions, might underestimate the
BCM background rate by up to 15\%, depending on the amount of neon.
 
For the two distant pressure bumps the geometry traversed before reaching the BCM is more complicated, and this must be reflected
as a larger uncertainty.
The simulated fake jet rates depend on the yield of high-energy muons in the very forward direction, which is a minor
part of the entire event topology. The muon transport, however, is almost insensitive to small differences in material
descriptions.
 
Previous BIB studies\,\cite{DAPR-2017-01, PhysRevAccelBeams.22.021004} show agreement between data and simulation within a factor of two
for the BCM and fake-jet rates.
In those earlier studies, as in the present one, the dominant uncertainties were in the estimated loss rate of primary beam
protons, which provides the absolute scale for the simulations.
In Ref.\,\cite{DAPR-2017-01} the loss rate was determined from \vasco pressure simulations for the entire LSS, in Ref.\,\cite{PhysRevAccelBeams.22.021004}
from the loss rate on the primary collimators of the LHC and in the present work from the gas density in the locally introduced bumps. These normalisations are
totally independent while the \fluka simulations are very similar in all three cases.
The fact that all three studies find agreement within a factor of about two indicates that the uncertainties of the \fluka simulations, themselves,
should be lower. Although this provides no rigorous argument to quantify the uncertainty associated with the \fluka simulations alone,
it suggests that a factor of 1.5 is reasonable.
 
\section{Comparison with simulations}
 
\subsection{BIB detection efficiencies}
\label{sect:bibdeteff}
 
The comparison of the simulated efficiencies with the averages of the values given in Tables\,\ref{tab:bcmBkgdEff} and \ref{tab:jetBkgdEff}
is the primary objective of this study.
The localised pressure bumps provide much more detailed benchmarking of the simulations than could be obtained in Ref.\,\cite{DAPR-2017-01}.
These comparisons are shown in Tables\,\ref{tab:bcmSimEfficiency} and \ref{tab:jetSimEfficiency} for BCM and fake-jet
backgrounds, respectively. In order to avoid the \ET-range where the jet trigger is not yet fully efficient, the jet rates are
compared for \ET$>$20\,\GeV. Therefore the values given in Table\,\ref{tab:jetSimEfficiency} are considerably lower than the
averages over the individual fills shown in Table\,\ref{tab:jetBkgdEff}.
 
\begin{table}
\centering
\caption{BCM efficiency to detect a beam-gas event in the various pressure bump locations.
The results from all fills are averaged and compared with the simulations, in which the two
beams are assumed identical.
As explained in the text, exact values of \effbcm{19\,{\mathrm m}} and \effbcm{22\,{\mathrm m}} cannot be determined due to the
possible but not quantifiable contribution of neon. The lower limit of the given range corresponds to the
maximum reasonable neon contribution, as discussed in Section\,\protect\ref{sect:sysnear}.
The fill-to-fill variation, derived from the per-fill data given in Table\,\protect\ref{tab:bcmBkgdEff}, is included in
the errors. The systematic effects discussed in Section\,\ref{sect:sysfar} are not included for the distant bumps.
The simulations assume $p$-H collisions and for the 19\,m and 22\,m locations a gas density profile corresponding to
$\meffneg=\mexpfor{5}{-3}$ is assumed. The uncertainties on the simulations are statistical only.
}
\begin{tabular}{r|c|c|c}
& \multicolumn{2}{c|}{Measured} & Simulated   \\
& \Beam{1}                   & \Beam{2}                   & Beams 1 or 2         \\ \hline
\effbcm{19\,{\mathrm m}}     &  $0.15-1.0$                & $0.15-1.0$                 & $0.424\pm 0.006$     \\
\effbcm{22\,{\mathrm m}}     &  $0.15-1.0$                & $0.15-1.0$                 & $0.432\pm 0.006$     \\
\effbcm{58\,{\mathrm m}}     &  $0.029\pm 0.001$          & $0.043\pm 0.005$           & $0.0284\pm 0.0014$   \\
\effbcm{151\,{\mathrm m}}    &  $0.0071\pm 0.0003$        & $0.0056\pm 0.0008$         & $0.0075\pm 0.0002$   \\ \hline
\end{tabular}
\label{tab:bcmSimEfficiency}
\end{table}
 
The simulations predict that the BCM detects about 40\% of $p$-H collisions taking place at 19\,m and 22\,m.
As discussed before, a significant neon contribution would increase the simulated efficiency by up to 15\%.
This increase, however, is anti-correlated with the effect neon has on the measurement, in which
it pushes the efficiency towards the lower limit of the range given in Table\,\ref{tab:bcmSimEfficiency}.
Assuming $\meffneg=\mexpfor{5}{-3}$ for H$_2$, the amount of neon needed to bring the measurement and simulation into
consistency can be estimated: for the \beam{1} pressure bump at 19\,m about \expfor{1.2}{13} neon atoms, which is $\sim$25\% of
the amount extracted in the laboratory tests, in the beam vacuum
would result in a measured $\meffbcm{19\,{\mathrm m}} \approx 0.5$. With that neon density about 90\% of the beam-gas interactions
would be $p$-Ne collisions, i.e. the simulated \effbcm{19\,{\mathrm m}} would rise by almost 15\% with respect to Table\,\ref{tab:bcmSimEfficiency},
reaching the same $\sim$0.5 as the measurement. For the other pressure bumps the amount of neon needed to lower the measured \effbcm{z} to
$\sim$0.5 would be less -- around \expfor{5}{12} atoms. Even with this amount, $p$-Ne collisions would dominate over $p$-H.
It is, actually, very difficult to conceive that the measurement and simulations could be consistent if hydrogen is the only gas contributing, while
the two-gas model presented in Section\,\ref{sect:neon} could establish consistency.
 
\begin{table}
\centering
\caption{Efficiency to detect fake jets with \ET$>20$\,\GeV from beam-gas events in the various pressure bump locations. The results from
all fills are  averaged and compared with simulation in which the two beams are assumed identical.
The upper table shows results for the full $|\eta|$-range considered in this study.
The lower table compares the data and simulations for $\eta|<1.5$, which roughly corresponds to the barrel calorimeters.
This reduced range is consistent with earlier work where similar comparisons are made\,\protect\cite{PhysRevAccelBeams.22.021004,DAPR-2017-01}.
The fill-to-fill variation, derived from the per-fill data given in Table\,\protect\ref{tab:jetBkgdEff}, is included in
the errors. The systematic effects discussed in Section\,\ref{sect:sysfar} are not included.
The uncertainties on the simulations are statistical only.}
\begin{tabular}{r|c|c|c}
& \multicolumn{2}{c|}{Measured ($|\eta|<3.2$)} & Simulated ($|\eta|<3.2$)  \\
& \Beam{1}                      & \Beam{2}                    & Beams 1 or 2 \\ \hline
\effjet{58\,{\mathrm m}}     & \expfor{(1.3\pm 0.2)}{-5}     & \expfor{(3.1\pm 0.2)}{-5}   &  \expfor{(1.5\pm 0.1)}{-5} \\
\effjet{151\,{\mathrm m}}    & \expfor{(8.7\pm 0.5)}{-5}     & \expfor{(8.3\pm 0.4)}{-5}   &  \expfor{(6.9\pm 0.1)}{-5} \\ \hline
\multicolumn{4}{c}{}  \\
& \multicolumn{2}{c|}{Measured ($|\eta|<1.5$)} & Simulated ($|\eta|<1.5$)  \\
& \Beam{1}                      & \Beam{2}                    & Beams 1 or 2 \\ \hline
\effjet{58\,{\mathrm m}}   & \expfor{(2.5\pm 0.2)}{-6}     & \expfor{(8.7\pm 0.1)}{-6}   & \expfor{(5.4\pm 0.6)}{-6} \\
\effjet{151\,{\mathrm m}}  & \expfor{(3.3\pm 0.4)}{-5}     & \expfor{(3.3\pm 0.3)}{-5}   & \expfor{(2.25\pm 0.08)}{-5} \\ \hline
\end{tabular}
\label{tab:jetSimEfficiency}
\end{table}
 
For the 151\,m bump the simulations agree almost perfectly with the data for the BCM backgrounds, and
underestimate the fake jets by about 20\%. For the \beam{1} bumps at 58\,m the simulations agree rather well with
the BCM backgrounds and the fake-jet measurements, especially if the latter are considered over the entire $|\eta|$-range.
For both, BCM and fake-jet backgrounds, the difference between data and simulation exceeds significantly
the statistical uncertainties for the 58\,m bump in \beam{2}.
This is consistent with Table\,\ref{tab:averageomegas}, which is based on pure data and reveals a substantial difference between \beam{1} and \beam{2}
for the 58\,m bump:
the BCM background per unit measured pressure is 50\% higher for \beam{2} than \beam{1}. For fake jets the difference
seems even larger -- about a factor of two. Since the pedestal background and pressure are subtracted prior to forming these ratios,
an additional source, not related to the pressure bump, can be excluded.
This could imply that either the gauge efficiency on the \beam{2} side was lower or the bump width at 58\,m was wider than predicted by
the \vasco calculations.\footnote{A comparison of the ratios between the two neighbouring gauges at 58\,m reveals that the preferred
gauge was giving up to 40\% lower readings on the \beam{2} side, especially in fills 5331 and 6239.
This could explain
the higher $\varepsilon$-values seen in Tables\,\ref{tab:bcmSimEfficiency} and \ref{tab:jetSimEfficiency} as well as the larger fill-to-fill
variation seen in Table\,\ref{tab:bcmBkgdEff}.}
 
\begin{figure}
\centering
\includegraphics[width=0.95\textwidth]{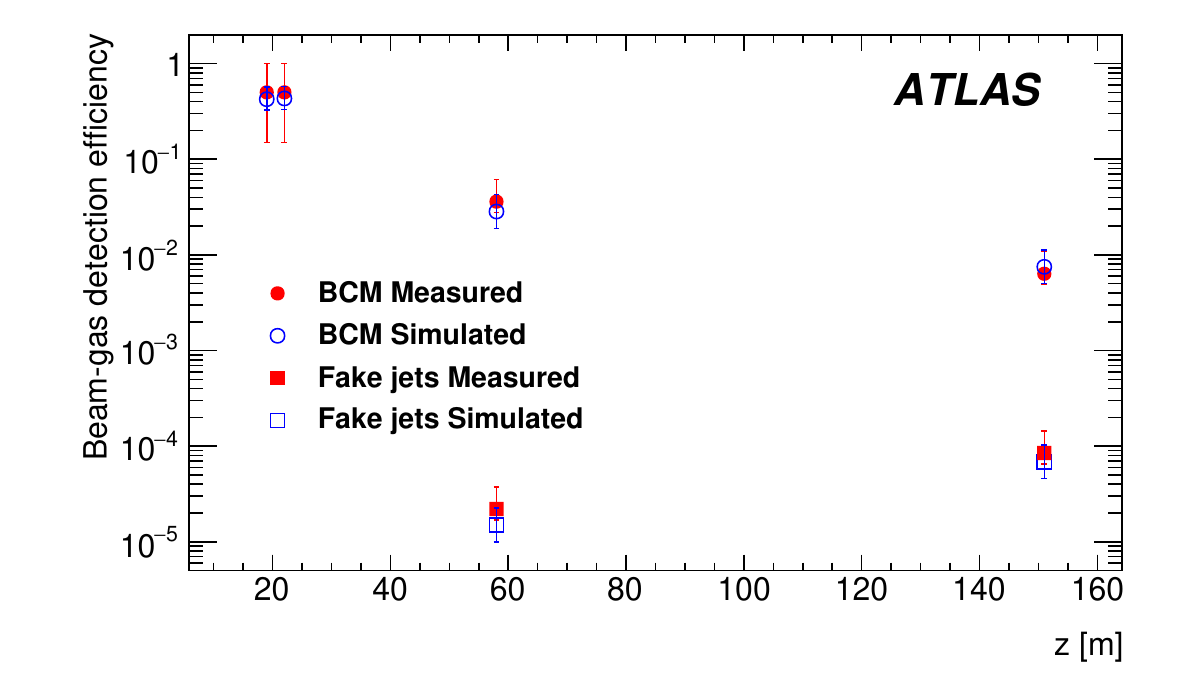}
\caption{Detection efficiency of beam-gas events as a function of $z$. The measurements correspond to the averages of both beams
in Tables\,\protect\ref{tab:bcmSimEfficiency} and \protect\ref{tab:jetSimEfficiency}. The error bars are dominated by the estimated systematic
uncertainties, as discussed in Section\,\protect\ref{sect:syserrors}. The average of the measured value for the two nearby points is set to 0.5,
which corresponds to assuming a significant neon contribution, as discussed in the text. The error bars display the range indicated in
Table\,\protect\ref{tab:bcmSimEfficiency}. The error bars on the measurements at the two distant points are dominated by the 50\% uncertainty
on how well the pressure measured by a gauge reflects the amount of gas in the bump region.}
\label{fig:ratePlot}
\end{figure}
 
Figure\,\ref{fig:ratePlot} shows \effbcm{z} and \effjet{z} as a function of $z$. The error bars in the plot include the estimated
systematic uncertainties, as discussed in Section\,\ref{sect:syserrors}. For the two nearby points it is assumed that neon plays a
significant role in lowering \effbcm{z} to $\sim0.5$, as discussed above. This summary plot illustrates the opposite trends of the
two BIB observables as a function of $z$: the \effbcm{z} decreases with distance from the IP while \effjet{z} increases.
It is worth emphasising  that the data shown in Figure\,\ref{fig:ratePlot} span over four orders of magnitude.
Even though the uncertainty estimates are large, they play no significant role in the principal outcome of this study,
which is a demonstration of the strong $z$-dependence of the beam-gas detection efficiency with opposite trends for the BCM and fake-jet
backgrounds.
 
A comparison in the reduced $|\eta|$-range is presented in Table\,\ref{tab:jetSimEfficiency} in order to maintain consistency with
earlier work\,\cite{PhysRevAccelBeams.22.021004,DAPR-2017-01}.
This ensures that the trigger is fully efficient over the $\eta$-range considered. Another reason for the restriction is that
the simulations assume that all fake jets are produced by high-energy muons. While this assumption is
certainly justified for the barrel calorimeters, the possibility of hadronic showers, initiated in the beam-pipe walls and reconstructed
as jets in the endcaps, cannot be excluded\,\cite{DAPR-2017-01}. A comparison of the results for the two $|\eta|$-ranges in Table\,\ref{tab:jetSimEfficiency}, however,
reveals that considering only central rapidities leads to a slightly larger reduction of the simulated
than the observed fake-jet rates. This confirms that radiative energy losses of high-energy muons are, by far, the dominant origin of the
fake-jet background even in the endcap calorimeters, but it also suggests that either the efficiency drop of the jet trigger is
significant in the endcap region or the simulations underestimate the radial spread of high-energy muons.
 
When the BIB detection efficiencies reported in Tables\,\ref{tab:bcmSimEfficiency} and \ref{tab:jetSimEfficiency} are compared with those for
BIB resulting from beam-halo losses on the TCT\,\cite{PhysRevAccelBeams.22.021004},
it is seen that they are significantly higher, especially for the BCM. This indicates that the higher BCM/Jet ratio reported in
Table\,\ref{tab:bcmPerJet}, compared to Ref.\,\cite{PhysRevAccelBeams.22.021004}, is due to more secondaries from beam-gas events staying inside the
beam-pipe aperture and reaching the BCM. If protons interact in the TCT jaws, a larger fraction of the produced particles is absorbed
locally or in the accelerator structures between the collision point and the IP.
Because fake jets are the type of BIB that can affect some physics searches, this large difference is potentially relevant: it indicates
that not only the $z$-position where the proton is lost matters, but also the material or accelerator element on which it is lost. Thus a monitoring
of BIB with the BCM alone might severely underestimate the fake-jet rates if beam halo losses on the TCT become more
important with evolving LHC beam optics and collimator settings.
 
\subsection{Jet distributions}
\label{sect:jet-dist}
 
Based on earlier simulations\,\cite{DAPR-2012-01}, the $\phi$-distribution of the muon background originating from $|z|<60$\,m is
expected to exhibit maxima in both the horizontal and vertical planes, while muons from farther away are predominantly in the
plane of the LHC ring. This azimuthal distribution should be reflected in the fake-jet rates.
Fake-jet data collected in the absence of a pressure bump, however, show pronounced maxima only at $\phi=0$ and $\phi=\pi$ and
only very minor humps at $\phi=\pm\pi/2$\,\cite{DAPR-2017-01}.
 
With the clear excess of fake jets observed from both, the 58\,m and the 151\,m locations in the pressure bump data, it is possible to
verify whether the azimuthal jet distribution depends on the $|z|$-location of the beam-gas event. In particular, it is possible to
investigate how the azimuthal BIB distribution is affected by the D1 dipole, which is situated between these two bump locations.
 
In data the additional fake jets due to
the pressure bumps are merged with the distribution from elsewhere around the ring. Especially for the 58\,m location this pedestal
is large dilutes the $\phi$-distribution. As it is impossible to determine the $|z|$-origin of an individual fake jet seen
in data, the pedestal can only be subtracted on average. For this the jet counts per azimuthal bin, collected
in periods (see, e.g. Figure\,\ref{fig:fits}) with a normal pressure distribution ($N_{\textrm{Base}}(\Delta\phi)$), are subtracted from those
during the gas injection ($N_{\textrm{Bump}}(\Delta\phi)$), after weighting with the ratio of wall-clock times
($t_{\textrm{Bump}}/t_{\textrm{Base}}$) over which the two data sets are collected:
\begin{equation}
N_z(\Delta\phi) = N_{\textrm{Bump}}(\Delta\phi) - \frac{t_{\text{Bump}}}{t_{\textrm{Base}}}N_{\textrm{Base}}(\Delta\phi).
\end{equation}
 
The resulting $N_z(\Delta\phi)$ is the estimated additional number of counts due to the pressure bump at $z$.
Table\,\ref{tab:jetBkgdEff} shows that the fake-jet rate per unit measured pressure varies significantly from fill to fill and between
the two beams, especially for the 58\,m location.
Including this variation would inflate the uncertainties and render a comparison of the $\phi$-distributions less conclusive.
In order to limit the comparison to the shape only, the different fills are made comparable by normalising all
distributions to unity before averaging $N_z(\Delta\phi)$ over the fills.
Due to this normalisation the \ET-cut becomes irrelevant so, unlike in Table\,\ref{tab:jetSimEfficiency}, all jets with \ET$>$12\,\GeV have
been retained in order to maximise the sizes of the samples.\footnote{Raising the cut threshold to 20\,\GeV has no effect on the shapes
of the distribution.}
Even with the relaxed \ET-cut, the data suffer from a low count rate; after subtracting the estimated pedestal, only slightly more
than 1100 fake jets remain for the 58\,m pressure bump. These are distributed among the 36 $\phi$-bins in the plot, and most entries
fall into the peaks. Thus the minima have just a few counts per bin. For the 151\,m pressure bump the count rate is higher:
about 5000 fake jets remain after pedestal subtraction.
 
The count rate in the \fluka simulations has no pedestal but is limited by the number of beam-gas events available from the
LHC simulations.
The number of counts is of similar magnitude to that in the data: slightly more than 3000 jets from both the 58\,m and the 151\,m locations.
For the 151\,m position the events are selected according to the pressure bump profile.
At 58\,m, where the probability to obtain a jet is lower, all beam-gas events with a flat distribution in
the range 53--63\,m had to be used in order to reach this number of fake jets. The probability for a high-energy
muon to generate a fake jet when traversing the ATLAS calorimeters is at the percent level. Thus the
number of simulated events could be doubled by using twice each muon produced by the LHC simulations.
This introduced a minor correlation, i.e. about 1\% of the fake jets originated from the same muon and, thereby, fall into the same $\phi$-bin.

\begin{figure}
\centering
\mbox{
\begin{subfigure}{0.49\textwidth}
\centering\captionsetup{width=0.9\textwidth}
\includegraphics[width=0.95\textwidth]{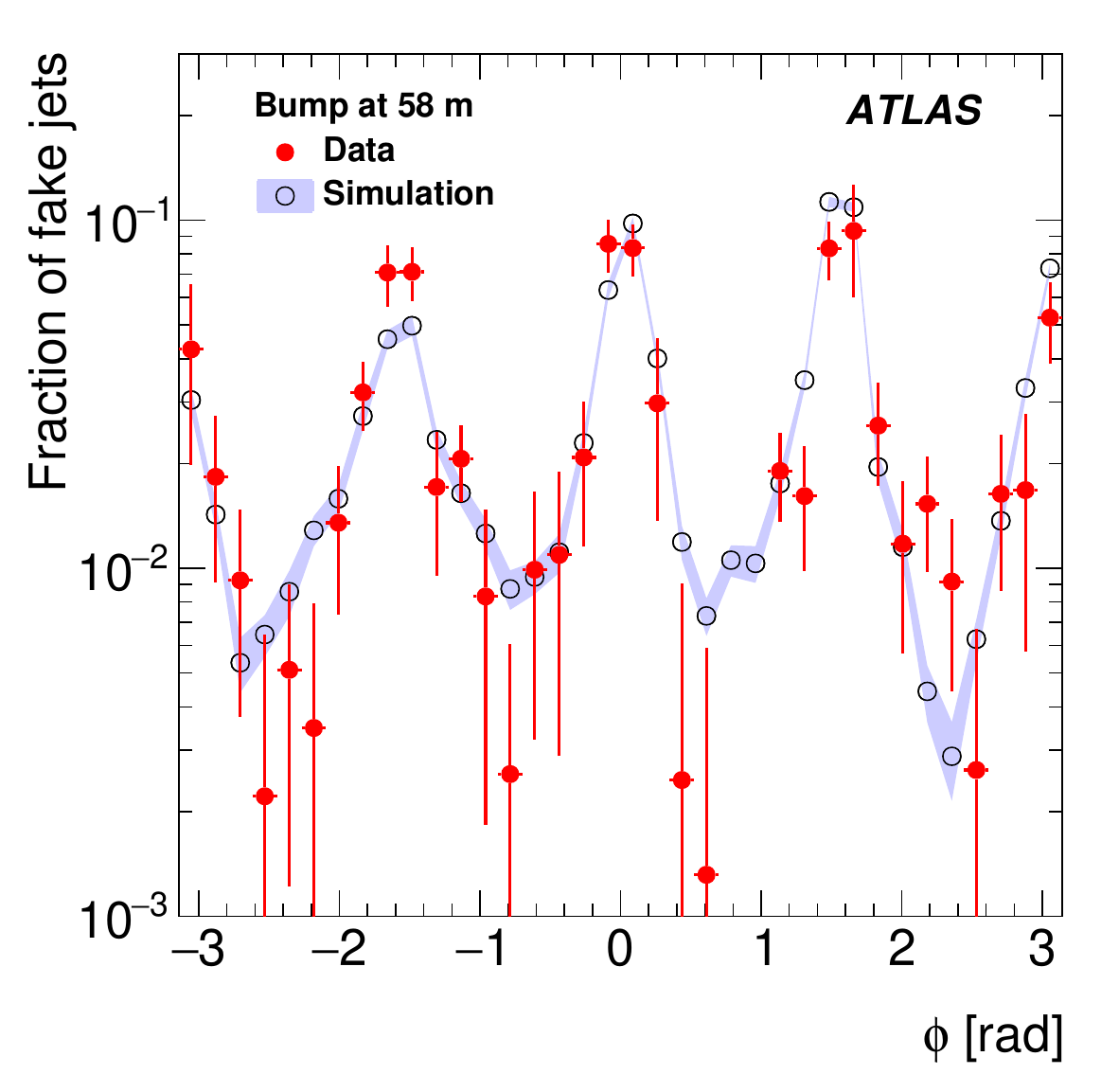}
\caption{}
\label{fig:AvePhiDist58}
\end{subfigure}
\begin{subfigure}{0.49\textwidth}
\centering\captionsetup{width=0.9\textwidth}
\includegraphics[width=0.95\textwidth]{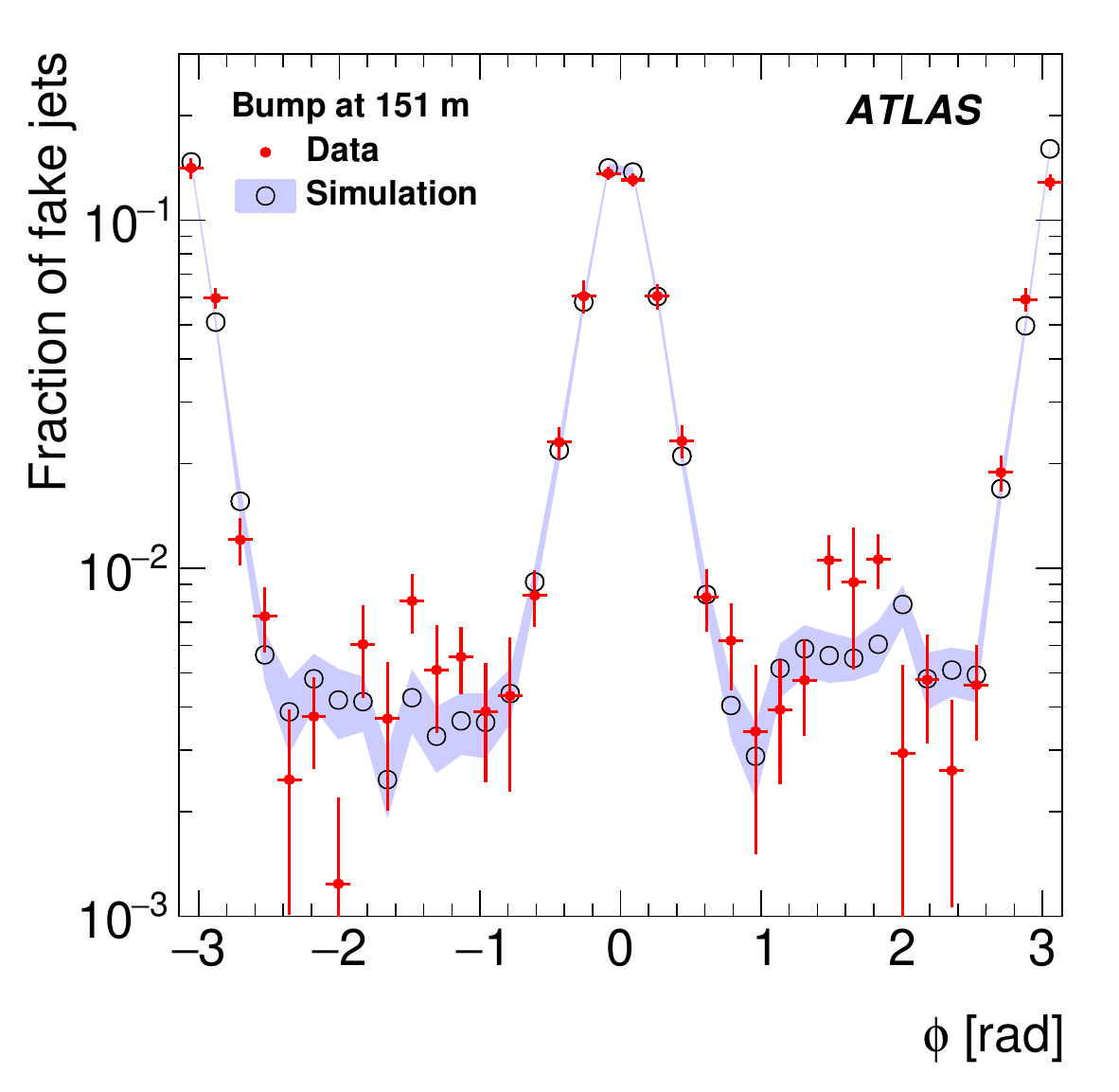}
\caption{}
\label{fig:AvePhiDist151}
\end{subfigure}
}
\caption{Comparison of azimuthal distributions of fake jets seen in data and predicted by simulations for the
\subref{fig:AvePhiDist58} 58\,m and \subref{fig:AvePhiDist151} 151\,m pressure bumps. All distributions are normalised to unity, thus only the shapes are compared, not the
absolute rates. The data of both beams are added.}
\label{fig:AvePhiDist}
\end{figure}
 
\begin{figure}
\centering
\mbox{
\begin{subfigure}{0.49\textwidth}
\centering\captionsetup{width=0.9\textwidth}
\includegraphics[width=0.95\textwidth]{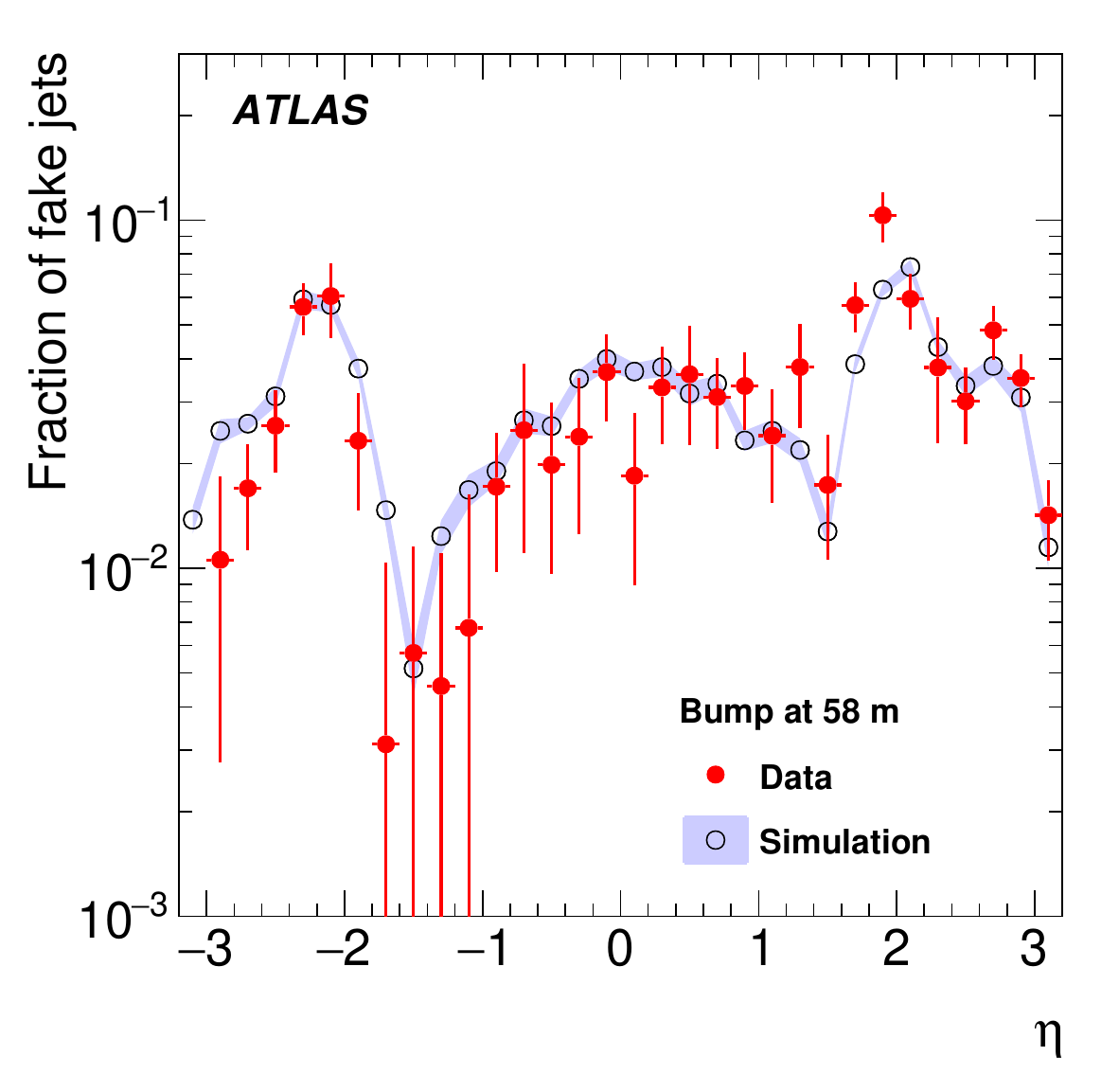}
\caption{}
\label{fig:AveEtaDist58}
\end{subfigure}
\begin{subfigure}{0.49\textwidth}
\centering\captionsetup{width=0.9\textwidth}
\includegraphics[width=0.95\textwidth]{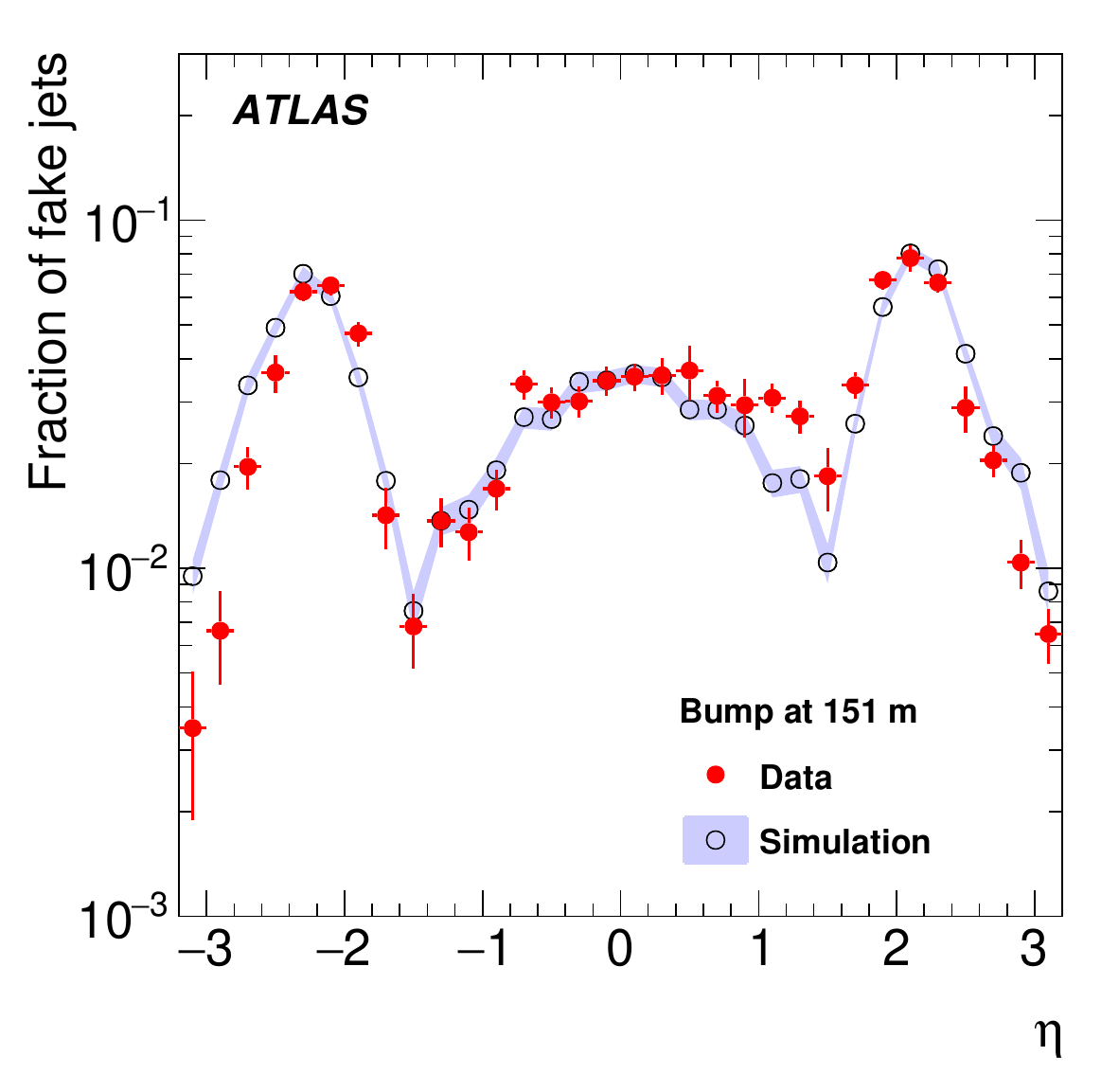}
\caption{}
\label{fig:AveEtaDist151}
\end{subfigure}
}
\caption{Comparison of pseudorapidity distribution of fake jets seen in data and predicted by simulations for the
\subref{fig:AveEtaDist58} 58\,m and \subref{fig:AveEtaDist151} 151\,m pressure bumps. The data of both beams are added after inverting $z$ for \beam{1} unpaired bunches, i.e. the
beam direction is always from negative to positive $\eta$. All distributions are normalised to unity, thus only the shapes are
compared, not the absolute rates.}
\label{fig:AveEtaDist}
\end{figure}

The comparisons, shown in Figure\,\ref{fig:AvePhiDist}, confirm that the data and simulations agree on the expected
feature: the $\phi$-distribution of fake jets from 58\,m exhibits four peaks while those from 151\,m are predominantly in the
plane of the accelerator. The agreement on the shape of the azimuthal distributions is remarkably good for both locations.
The only significant elements between 58\,m and the IP, which are not $\phi$-symmetric, are the quadrupoles of the inner
triplet. Between 58\,m and 151\,m the D1 dipole introduces an additional breaking of $\phi$-symmetry. Thus Figure\,\ref{fig:AvePhiDist}
confirms the assumption that the four-peak structure arises from the bending of muon trajectories in the quadrupoles, while the
more significant peaks in the accelerator plane are caused by the dipole fields.
 
Figure\,\ref{fig:AveEtaDist} shows the pseudorapidity distributions of the fake jets. The measured distributions are rather
similar for both distant pressure bump locations. There are slightly more fake jets in the downstream half, i.e. positive $z$.
The \fluka simulations reproduce these shapes reasonably well, although there are some differences, especially at high $|\eta|$.
These are, at least qualitatively, consistent with the drop of the jet trigger efficiency at $|\eta|>2.8$, which is not accounted for
in the simulations.
 
The distributions from both beams were inspected separately and found to agree within statistical uncertainties. Therefore the
contributions of both beams are added together in Figures\,\ref{fig:AvePhiDist} and \ref{fig:AveEtaDist}, in the latter case
after inverting the \beam{1} direction.
 
\section{Backgrounds excess following a pressure test}
\label{sect:aftertest}
 
\begin{figure}
\centering
\mbox{
\begin{subfigure}{0.49\textwidth}
\centering\captionsetup{width=0.9\textwidth}
\includegraphics[width=0.95\textwidth]{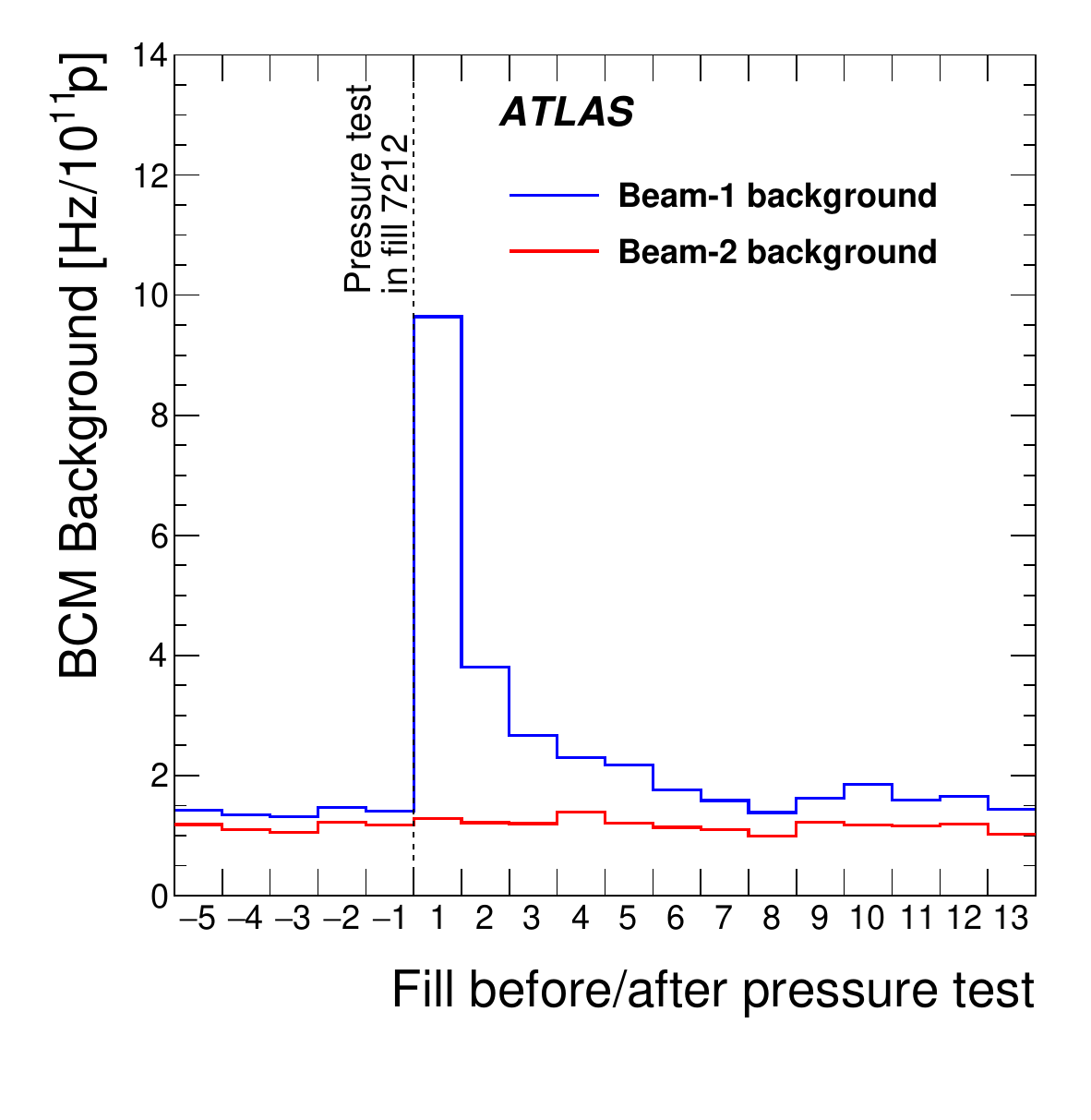}
\caption{}
\label{fig:backgroundAfter}
\end{subfigure}
\begin{subfigure}{0.49\textwidth}
\centering\captionsetup{width=0.9\textwidth}
\includegraphics[width=0.95\textwidth]{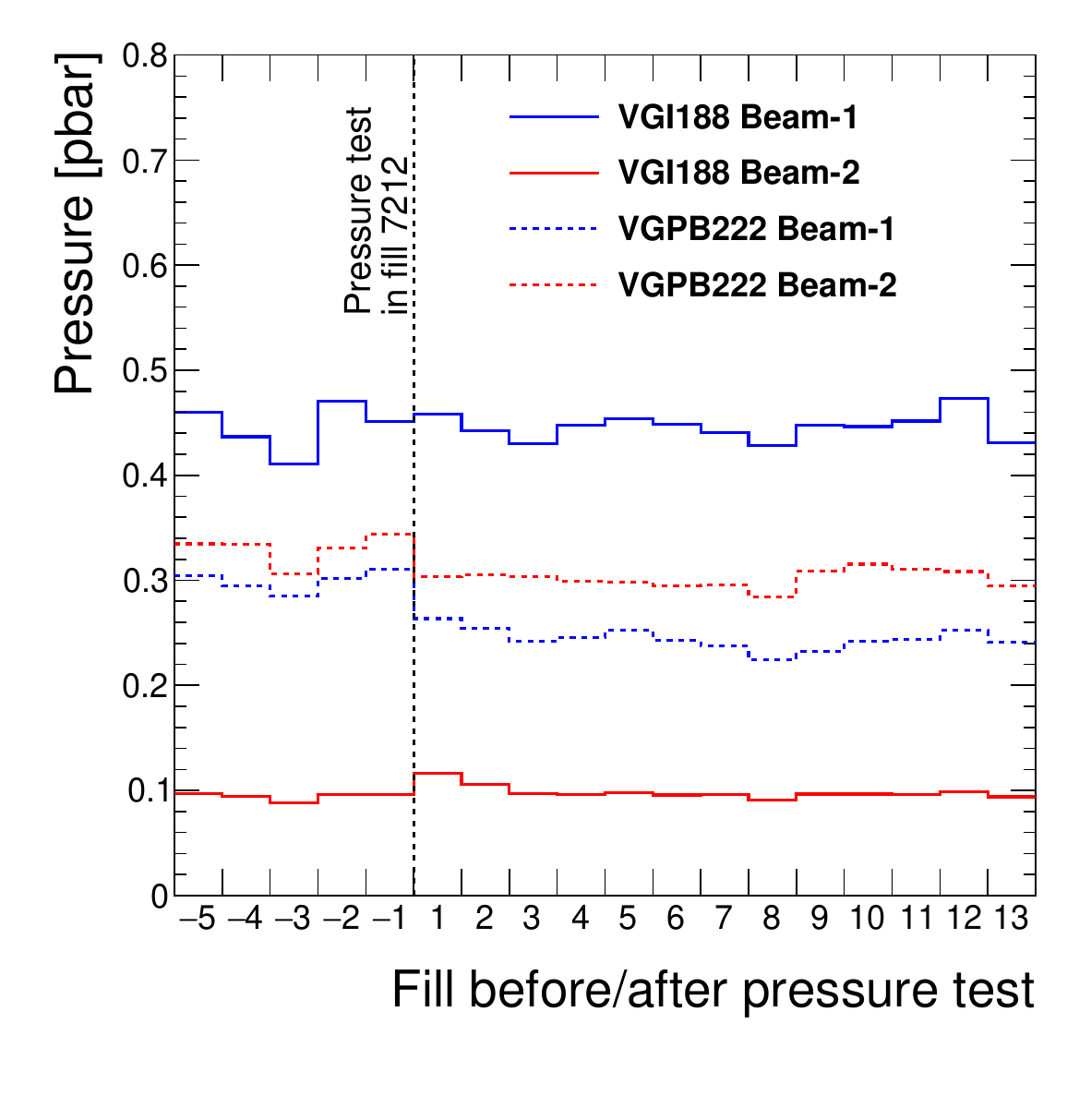}
\caption{}
\label{fig:pressuresAfter}
\end{subfigure}
}
\caption{\subref{fig:backgroundAfter} BCM backgrounds and \subref{fig:pressuresAfter} residual pressures for the two beams observed in normal
physics fills just before the pressure test and in several fills thereafter. Each bin represents the
average over the first 30\,minutes of data-taking at high luminosity.
The dashed vertical line indicates the pressure test, which itself is excluded from the
plot.}
\end{figure}

In the normal physics fills, following the pressure test in fill 7212, unusually high \beam{1} backgrounds were recorded by the BCM.
In the first fill after the test, the background jumped to about 10 times the normal level and then gradually decreased over the course
of the fill. In subsequent fills the same was observed, although the magnitude of the initial jump was reduced from fill to fill.
It took several normal physics fills for the \beam{1} start-of-collision background levels to recover to their pre-test levels.
This is illustrated in Figure\,\ref{fig:backgroundAfter} where the
start-of-collision backgrounds in five fills before the pressure test are compared with those in 13 fills following the test.
Only fills starting with high luminosity are included in the plot, and the backgrounds are averaged over the first 30 minutes of
data-taking, provided the average number of collisions per bunch crossing was higher than 45. This requirement
was imposed in order to prevent start-of-fill emittance scans\,\cite{bib:eScans} from biasing the results, since the background jump is
obviously luminosity-related.
The background increase seems to affect only \beam{1}; no sign of a similar increase is seen for BIB from \beam{2}.
As shown in Figure\,\ref{fig:pressuresAfter}, at the same time the vacuum gauges at the 19\,m and 22\,m positions,
where the pressure bumps had been introduced, showed no anomalies. The reason for this excess is not entirely clear.
Possibly some of the injected gas had condensed on the BS and was gradually desorbed by the particle debris hitting the
BS during high-luminosity operation.
 
Although the increase of the BCM background trigger rate for \beam{1} was an order of magnitude, it had no impact on the
operation or data quality of ATLAS. This can be taken as an indication of the margin that ATLAS at least has with respect
to BIB from nearby beam-gas collisions, before data-taking efficiency or quality of physics data start to be affected.
 
\section{Conclusions}
\label{sect:concl}
 
Several dedicated pressure bump tests were undertaken in five different fills during LHC \run{2} in order to determine the
sensitivity of the ATLAS beam-induced background (BIB) monitoring system to beam-gas events at different distances from the
interaction point (IP). In these tests individual non-evaporative getter (NEG) cartridges in the vicinity of the ATLAS experimental
area were heated in order to reverse the pumping process. In the bumps the gas density was up to four orders of magnitude higher
than elsewhere.
 
A very good correlation between the pressure measured at the bump location and the BIB rates seen in ATLAS is observed, i.e. the
BIB rate can be fitted with a linear combination of a constant term and the local pressure changes, weighted with $z$-dependent
proportionality constants.
 
The ratio of BIB from the different bump locations, seen by the beam conditions monitor (BCM) and as fake-jet backgrounds in
the calorimeters, is evaluated. It confirms previous simulation studies, which predict that beam-gas events very close to the detector
do not produce fake jets, while the BCM has high sensitivity to BIB originating from that region.
With increasing distance from the IP the sensitivity of the BCM drops while fake jets become more likely.
At $|z|=58$\,m the BCM/jet BIB ratio is $\sim$800 and at $|z|=151$\,m it is only $\sim$45. This latter
value is much larger than the ratio of about six, which has been measured for BIB from proton halo impacting on
the tertiary collimators, only a few metres closer to the IP.
These large differences imply that BCM and fake-jet backgrounds are very weakly correlated
and BIB monitoring by the BCM alone does not provide a reliable measure of the fake-jet rate.
 
In order to derive absolute background detection efficiencies for the different $z$-locations, the amount of gas traversed
by the beam was evaluated from the calculated pressure profile of each bump.
Although the calculations were adjusted to the measured pressure, they are subject to significant uncertainties,
in particular the pumping efficiency of the NEG-coating in the room-temperature sections of the beam pipe.
With a typical NEG efficiency of $\meffneg = \mexpfor{5}{-3}$ the results for the two
nearby pressure bumps indicate that more beam-gas events are detected than are produced.
A possible explanation is that upon heating the NEG cartridge releases some neon in addition to the expected hydrogen. The neon could
be a residue from filling the beam pipe with an ultra-pure noble gas to atmospheric pressure during access to the experiment.
Neon, which is not pumped by the NEG coating ($\meffneg = 0$), would have a wider distribution and longer residence time in the
beam pipe than hydrogen. It also has an about ten times higher cross section for proton scattering.
 
Even though systematic uncertainties are estimated to be of the order of a factor of two, they are rather minor compared to the
differences of the observed efficiencies, which extend over several orders of magnitude, depending on the $z$-location.
The \effbcm{z} values range from just below unity for pressure bumps at 19\,m and 22\,m to $\sim$3\% at 58\,m and $\sim$0.6\% at 151\,m.
No fake-jet backgrounds are detected from the two nearby bump positions. For the 58\,m and 151\,m bumps the \effjet{z} values, for \ET$>$12\,\GeV,
are $\sim\mexpfor{4}{-5}$ and $\sim\mexpfor{1.4}{-4}$, respectively.
 
For the nearby pressure bumps the \fluka simulations agree with the data at $\meffbcm{z}\sim 0.5$, but only if neon plays a
significant role. The simulation results for the two distant bumps are in good agreement with the measurements. The latter
are based on pressure profile calculations that are consistent with $\meffneg \approx \mexpfor{5}{-3}$.
In these regions no neon filling is applied, so the assumption of hydrogen dominance should be valid.
 
The shapes of the fake-jet distributions in the azimuthal plane and as a function of pseudorapidity have been compared with simulations,
and good agreement is found. The data show, in particular, that the azimuthal distribution of jets from the 58\,m
position has four about equally high maxima at $n \frac{\pi}{2}$, while jets from 151\,m are predominantly in the plane
of the accelerator ring, i.e. their distribution peaks at $n\pi$.
 
All tests were performed parasitically during normal physics data taking. No adverse effects for the performance of ATLAS or the LHC
were seen. Even the unexplained contamination of the beam vacuum, observed after the last test, had no impact on the operation of the
accelerator or the experiment.
 
This study completes a set of dedicated measurements and simulations that comprise also Refs.\,\cite{PhysRevAccelBeams.22.021004, DAPR-2017-01}.
Together these cover the dominant sources of BIB in the ATLAS detector and show that the sources of these backgrounds are well understood and
simulations can reproduce them within a factor of about two. These results, and the successful benchmarking of the simulation methods, provide
important information for the study and optimisation of future accelerator configurations and BIB mitigation methods.

\section*{Acknowledgements}


We thank CERN for the very successful operation of the LHC and its injectors, as well as the support staff at
CERN and at our institutions worldwide without whom ATLAS could not be operated efficiently.
 
The crucial computing support from all WLCG partners is acknowledged gratefully, in particular from CERN, the ATLAS Tier-1 facilities at TRIUMF/SFU (Canada), NDGF (Denmark, Norway, Sweden), CC-IN2P3 (France), KIT/GridKA (Germany), INFN-CNAF (Italy), NL-T1 (Netherlands), PIC (Spain), RAL (UK) and BNL (USA), the Tier-2 facilities worldwide and large non-WLCG resource providers. Major contributors of computing resources are listed in Ref.~\cite{ATL-SOFT-PUB-2023-001}.
 
We gratefully acknowledge the support of ANPCyT, Argentina; YerPhI, Armenia; ARC, Australia; BMWFW and FWF, Austria; ANAS, Azerbaijan; CNPq and FAPESP, Brazil; NSERC, NRC and CFI, Canada; CERN; ANID, Chile; CAS, MOST and NSFC, China; Minciencias, Colombia; MEYS CR, Czech Republic; DNRF and DNSRC, Denmark; IN2P3-CNRS and CEA-DRF/IRFU, France; SRNSFG, Georgia; BMBF, HGF and MPG, Germany; GSRI, Greece; RGC and Hong Kong SAR, China; ISF and Benoziyo Center, Israel; INFN, Italy; MEXT and JSPS, Japan; CNRST, Morocco; NWO, Netherlands; RCN, Norway; MNiSW, Poland; FCT, Portugal; MNE/IFA, Romania; MESTD, Serbia; MSSR, Slovakia; ARRS and MIZ\v{S}, Slovenia; DSI/NRF, South Africa; MICINN, Spain; SRC and Wallenberg Foundation, Sweden; SERI, SNSF and Cantons of Bern and Geneva, Switzerland; MOST, Taipei; TENMAK, T\"urkiye; STFC, United Kingdom; DOE and NSF, United States of America.
 
Individual groups and members have received support from BCKDF, CANARIE, CRC and DRAC, Canada; CERN-CZ, FORTE and PRIMUS, Czech Republic; COST, ERC, ERDF, Horizon 2020, ICSC-NextGenerationEU and Marie Sk{\l}odowska-Curie Actions, European Union; Investissements d'Avenir Labex, Investissements d'Avenir Idex and ANR, France; DFG and AvH Foundation, Germany; Herakleitos, Thales and Aristeia programmes co-financed by EU-ESF and the Greek NSRF, Greece; BSF-NSF and MINERVA, Israel; Norwegian Financial Mechanism 2014-2021, Norway; NCN and NAWA, Poland; La Caixa Banking Foundation, CERCA Programme Generalitat de Catalunya and PROMETEO and GenT Programmes Generalitat Valenciana, Spain; G\"{o}ran Gustafssons Stiftelse, Sweden; The Royal Society and Leverhulme Trust, United Kingdom.
 
In addition, individual members wish to acknowledge support from CERN: European Organization for Nuclear Research (CERN PJAS); Chile: Agencia Nacional de Investigaci\'on y Desarrollo (FONDECYT 1190886, FONDECYT 1230987); China: National Natural Science Foundation of China (NSFC - 12175119, NSFC 12275265); Czech Republic: Czech Science Foundation (GACR - 24-11373S), Ministry of Education Youth and Sports (FORTE CZ.02.01.01/00/22\_008/0004632); European Union: European Research Council (ERC - 948254, ERC 101089007), Horizon 2020 Framework Programme (MUCCA - CHIST-ERA-19-XAI-00), Italian Center for High Performance Computing, Big Data and Quantum Computing (ICSC, NextGenerationEU); France: Agence Nationale de la Recherche (ANR-20-CE31-0013, ANR-21-CE31-0013, ANR-21-CE31-0022), Investissements d'Avenir Labex (ANR-11-LABX-0012); Germany: Baden-Württemberg Stiftung (BW Stiftung-Postdoc Eliteprogramme), Deutsche Forschungsgemeinschaft (DFG - 469666862, DFG - CR 312/5-2); Italy: Istituto Nazionale di Fisica Nucleare (ICSC, NextGenerationEU); Japan: Japan Society for the Promotion of Science (JSPS KAKENHI 22H01227, JSPS KAKENHI 22KK0227, JSPS KAKENHI JP21H05085, JSPS KAKENHI JP22H04944); Netherlands: Netherlands Organisation for Scientific Research (NWO Veni 2020 - VI.Veni.202.179); Norway: Research Council of Norway (RCN-314472); Poland: Polish National Agency for Academic Exchange (PPN/PPO/2020/1/00002/U/00001), Polish National Science Centre (NCN 2021/42/E/ST2/00350, NCN OPUS nr 2022/47/B/ST2/03059, NCN UMO-2019/34/E/ST2/00393, UMO-2020/37/B/ST2/01043, UMO-2021/40/C/ST2/00187, UMO-2022/47/O/ST2/00148, UMO-2023/49/B/ST2/04085); Slovenia: Slovenian Research Agency (ARIS grant J1-3010); Spain: Generalitat Valenciana (Artemisa, FEDER, IDIFEDER/2018/048), Ministry of Science and Innovation (RYC2019-028510-I, RYC2020-030254-I), PROMETEO and GenT Programmes Generalitat Valenciana (CIDEGENT/2019/023, CIDEGENT/2019/027); Sweden: Swedish Research Council (VR 2018-00482, VR 2022-03845, VR 2022-04683, VR 2023-03403), Knut and Alice Wallenberg Foundation (KAW 2018.0157, KAW 2018.0458, KAW 2019.0447, KAW 2022.0358); Switzerland: Swiss National Science Foundation (SNSF - PCEFP2\_194658); United Kingdom: Leverhulme Trust (Leverhulme Trust RPG-2020-004), Royal Society (NIF-R1-231091); United States of America: Neubauer Family Foundation.


Discussions with Giuseppe Bregliozzi and Vincent Baglin from the CERN vacuum group were invaluable for the interpretation
of the results. Alessio Galloro kindly provided the results of the neon emission laboratory measurements.

\printbibliography
 
\clearpage

\begin{flushleft} 
\hypersetup{urlcolor=black} 
{\Large The ATLAS Collaboration}

\bigskip

\AtlasOrcid[0000-0002-6665-4934]{G.~Aad}$^\textrm{\scriptsize 104}$,
\AtlasOrcid[0000-0001-7616-1554]{E.~Aakvaag}$^\textrm{\scriptsize 17}$,
\AtlasOrcid[0000-0002-5888-2734]{B.~Abbott}$^\textrm{\scriptsize 123}$,
\AtlasOrcid[0000-0002-0287-5869]{S.~Abdelhameed}$^\textrm{\scriptsize 119a}$,
\AtlasOrcid[0000-0002-1002-1652]{K.~Abeling}$^\textrm{\scriptsize 56}$,
\AtlasOrcid[0000-0001-5763-2760]{N.J.~Abicht}$^\textrm{\scriptsize 50}$,
\AtlasOrcid[0000-0002-8496-9294]{S.H.~Abidi}$^\textrm{\scriptsize 30}$,
\AtlasOrcid[0009-0003-6578-220X]{M.~Aboelela}$^\textrm{\scriptsize 45}$,
\AtlasOrcid[0000-0002-9987-2292]{A.~Aboulhorma}$^\textrm{\scriptsize 36e}$,
\AtlasOrcid[0000-0001-5329-6640]{H.~Abramowicz}$^\textrm{\scriptsize 154}$
% ...
\bigskip
\\

% ...
$^{17}$Department for Physics and Technology, University of Bergen, Bergen; Norway.\\
% ...
$^{30}$Physics Department, Brookhaven National Laboratory, Upton NY; United States of America.\\
% ...
$^{36}$$^{(a)}$Facult\'e des Sciences Ain Chock, R\'eseau Universitaire de Physique des Hautes Energies - Universit\'e Hassan II, Casablanca;$^{(b)}$Facult\'{e} des Sciences, Universit\'{e} Ibn-Tofail, K\'{e}nitra;$^{(c)}$Facult\'e des Sciences Semlalia, Universit\'e Cadi Ayyad, LPHEA-Marrakech;$^{(d)}$LPMR, Facult\'e des Sciences, Universit\'e Mohamed Premier, Oujda;$^{(e)}$Facult\'e des sciences, Universit\'e Mohammed V, Rabat;$^{(f)}$Institute of Applied Physics, Mohammed VI Polytechnic University, Ben Guerir; Morocco.\\
% ...
$^{45}$Physics Department, Southern Methodist University, Dallas TX; United States of America.\\
% ...
$^{50}$Fakult\"{a}t Physik , Technische Universit{\"a}t Dortmund, Dortmund; Germany.\\
% ...
$^{56}$II. Physikalisches Institut, Georg-August-Universit\"{a}t G\"ottingen, G\"ottingen; Germany.\\
% ...
$^{104}$CPPM, Aix-Marseille Universit\'e, CNRS/IN2P3, Marseille; France.\\
% ...
$^{119}$$^{(a)}$New York University Abu Dhabi, Abu Dhabi;$^{(b)}$United Arab Emirates University, Al Ain; United Arab Emirates.\\
% ...
$^{123}$Homer L. Dodge Department of Physics and Astronomy, University of Oklahoma, Norman OK; United States of America.\\
% ...
$^{154}$Raymond and Beverly Sackler School of Physics and Astronomy, Tel Aviv University, Tel Aviv; Israel.\\
% ...
$^{a}$ Also Affiliated with an institute covered by a cooperation agreement with CERN.\\
$^{b}$ Also at An-Najah National University, Nablus; Palestine.\\
$^{c}$ Also at Borough of Manhattan Community College, City University of New York, New York NY; United States of America.\\
$^{d}$ Also at Center for Interdisciplinary Research and Innovation (CIRI-AUTH), Thessaloniki; Greece.\\
$^{e}$ Also at Centro Studi e Ricerche Enrico Fermi; Italy.\\
$^{f}$ Also at CERN, Geneva; Switzerland.\\
% ...

\end{flushleft}


\end{document}